\documentclass[a4paper]{book}
\usepackage[eng,doktor]{KTHEEtitlepage}
\usepackage[cp1252]{inputenc}
\usepackage[final]{pdfpages}
\usepackage{hyperref}
\usepackage{amsmath}
\usepackage{float}
\usepackage{epigraph}
\usepackage{xcolor}
\usepackage{tocloft}

\newif\ifpapers
\paperstrue

\newcommand{\kperp}{\mathbf{k_{\perp}}}

\hypersetup{
  colorlinks=true,
  urlcolor=teal,
  anchorcolor=black,
  linkcolor=black,
  citecolor=teal}%
  
\begin{document}

                \ititle{Quantum Optics in Dense Atomic Media: \\
From Optical Memories to Fluids of Light}
                \idate{}
                \irefnr{Laboratoire Kastler Brossel\\
                Paris, France\\  D\'ecembre 2018\\}
                \iauthor{Quentin Glorieux}

\includepdf[scale=1, pages=-]{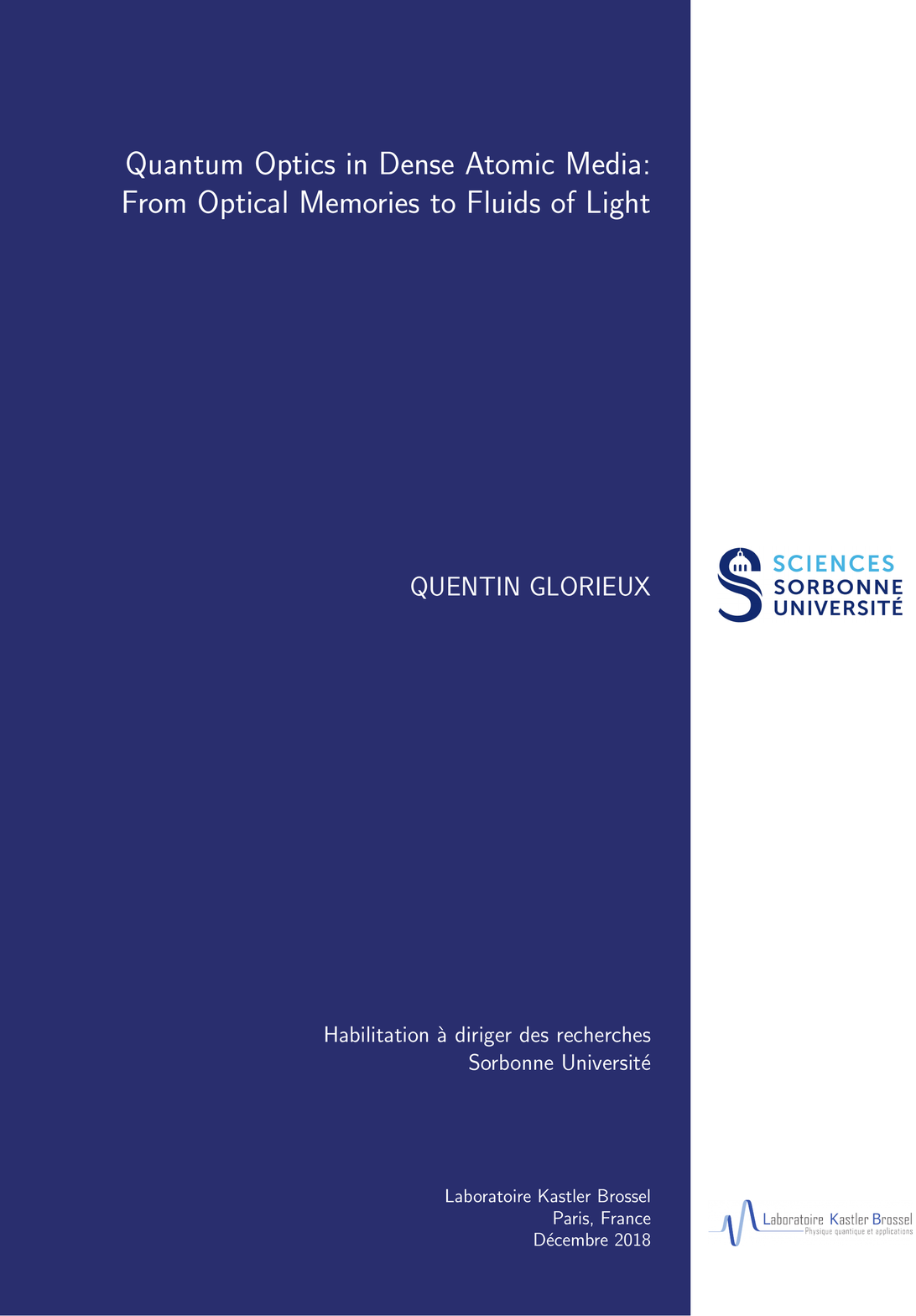}

\newpage
\strut 
\newpage
\tableofcontents

\chapter*{Introduction}
\addcontentsline{toc}{chapter}{Introduction}  

\hspace{0.33cm} Looking back over almost a decade of research after I graduated in 2010, I realized that what fascinates me the most in physics are the analogies between apparently unrelated phenomena.
Connecting microscopic descriptions and macroscopic quantities as does thermodynamics, or the general framework of spin 1/2 particle from NMR to quantum information... These are just two examples but one can find this everywhere in modern physics.\\

I cover in this work the links between what is  known today as quantum technologies and the more fundamental topic of quantum fluids.
I have always envisioned myself as a fundamental physicist, remotely interested in real-life applications. 
However, it is funny to note that part of what I believed to be fundamental science 10 years ago (quantum memories, squeezing, entanglement...) is today entering in the phase of technological development and applications with the major effort of the EU towards quantum technologies.  
This is exactly what should motivate basic science: sometimes it does bring applications to make our life easier and a better society and sometimes it just provides knowledge itself without any immediate  application (which also pushes toward a better society by making it smarter).
I am curious to see what quantum fluids will bring in 10 years.
This work summarizes my (modest) contribution to this field.\\

The main goal of this manuscript is to provide the tools to connect non-linear and quantum optics to quantum fluids of light.
Because the physics of matter quantum fluids and fluids of light has long been the territory of condensed matter physicists, it is uncommon to find textbooks which draw the analogies with quantum optics.
I have taken the reverse trajectory, being trained as a quantum optician  and moving progressively to quantum gases and quantum fluids of light.
Naturally, I try to use the concepts and resources developed by the quantum optics community to improve experiment about fluids of light, but I also reverse the approach and ask a very simple question: what does the concept of fluid of light bring to our understanding of non-linear and quantum optics ? 
It can be rephrased as an operational question: which new effects can we predict (and possibly observe) using the photon fluid formalism ?\\

This is actually a very exciting time, as we have recently demonstrated the validity of this approach with an experiment about the dispersion relation in a fluid of light \cite{fontaine}.
In this experiment, described in details in the chapter \ref{chap:4}, we have shown that light in a non-linear medium follows the Bogoliubov dispersion: a constant group velocity at small wavevectors and a linear increase with $k$ at large wavevectors.
If you trust the formal analogy between the non-linear Schr\"odinger equation and the paraxial propagation of light in a non-linear medium, this result is not surprising.
However, what is amazing is that, this description allowed us to design a non-linear optics experiment and helped us to understand it.
This experiment can also be thought as a correction to the Snell-Descartes law in a non-linear medium.
In this analogy,the group velocity is directly linked to the position of the beam in the transverse plane at the output of the medium. 
A linear increase with $k$ at large wavevectors translates to the standard refraction law: increasing the incidence angle will increase the distance from the optical axis after the medium.
However, at small angle the constant (non-zero) group velocity tells us that, whatever the angle of incidence, the beam will exit the medium at the same position !
The refraction law is independent of the incident angle.
Moreover, another idea can be extracted from quantum gases formalism: the group velocity depends on the medium density (at small k). Therefore the refraction law does not depend on the incident angle but on the light intensity !
Can we do something with this ? I don't know. Maybe we can use this novel understanding to image through non-linear medium. Or maybe we just understand a bit more of non-linear optics now.\\

This time is a very exciting one also because now we know that this approach works and we have tens of ideas for novel non-linear optics experiments testing the cancellation of drag force due to superfluidity, observing the Hawking radiation, or the Zel'dhovich effect... 
The next challenge is to bring this description to the field of quantum optics, where quantum noise and entanglement are of primary importance. What is the hydrodynamic analogue of squeezing or of an homodyne detection ? 
Understanding the effects of interactions in complex quantum systems beyond the mean-field paradigm constitutes a fundamental problem in physics. This manuscript just briefly ventures in this territory... But this will be, for sure, the next direction of my scientific career.\\

In fact I did not venture in this territory, but this manuscript intends to provide the tools to do so in the future.
Chapter 1 is a brief summary of what are the tools needed for quantum optics in a warm atomic medium: the two-level atoms model, electromagnetically induced transparency, four-wave-mixing, cooperative effects and decoherence.
In chapter 2, I describe one type of optical quantum memory based on the gradient echo memory protocol.
I present two implementations: in a warm vapor and in high optical depth cold atomic cloud and cover  how these two implementations are complementary with their specific strengths and weaknesses.
In chapter 3, I move to quantum optics with the study of imaging using the noise properties of light and propagation of quantum noise in a fast light medium. 
I complete the description of my previous works in chapter 4 including more recent experiments about fluid of light in an exciton-polariton microcavity and in an atomic vapor.
The final chapter is devoted to describe several outlooks and collaborative projects I have recently initiated.


\chapter{Hot atomic vapor}

\section{Atomic ensembles}
When learning about light-atom interaction, various approaches can be discussed. 
Light can be described as a classical electromagnetic field  or as quantum elementary excitations: the photons. 
Similarly an atom can be seen as a classical oscillating dipole or treated using quantum mechanics. 
The description of the interaction can then take any form mixing these 4 different perspectives.
While for most experiments a classical approach is sufficient, it is sometimes needed to invoke $\hbar$ and its \textit{friends} to explain a specific behaviour.
In this work, I will try, as much as possible, to avoid an artificial distinction between quantum and classical phenomena, as it does not bring  much to the understanding of the effects. Sometimes, a purely classical description is even more intuitive.\\

In this first section, I  remind readers of the basic tools to appreciate atomic ensemble physics, without entering into the details and complexities of genuine alkali atomic structures.
I  describe the differences in light-matter interaction between the case of one single atom and an atomic ensemble, and explain what I call a dense atomic medium. 
A short discussion will illustrate the relationship between collective excitations and Dicke states \cite{dicke1954coherence}.
This work is mainly focused on warm atomic ensembles (with the notable exception of section \ref{coldatoms}) and therefore a discussion on the role of Doppler broadening and the assumptions linked to it will conclude this part.

\subsection{Light--matter interaction}
Let us start with the simplest case: one atom interacting with light in free space.
Two processes, time-reversals of each other, can happen: absorption and stimulated emission\footnote{Obviously, a third process, called spontaneous emission, is also possible, but for this specific discussion  we do not need it.}.
An important question is: what is the \textit{shadow} of an atom~?
Or more precisely what is the dipole cross-section $\sigma_0$ of one atom interacting with light in free space ?
It is interesting to note that the answer to this question does not depend on the description of the atom as a classical oscillating dipole or a two-level atom. 
With $\lambda$ being the light wavelength, the resonant cross section for a two-level atom\footnote{More generally, the cross section for an atom with $F_g$ and $F_e$ being the total angular momenta for the ground and the excited states respectively, has the form $\sigma = \frac{2F_e+1}{2F_g+1}\frac{\lambda^2}{2\pi}$.} can be written as: \cite{steck2001rubidium}:
\begin{equation}\label{sigma0}
    \sigma_0=\frac{3\lambda^2}{2\pi}.
\end{equation}

From this relation, we can immediately conclude that it is a \textit{difficult} task to make light interact with a single atom.
Indeed, focusing light in free-space is limited by diffraction to a spot of area $\sim  \pi\lambda^2/4$.
Therefore, even a fully optimized optical system, will not reach a focusing area of $\sigma_0$.
What is surprising here, is that it does not depend on the choice of your favorite atom. 

Nevertheless various tricks can be tested to improve this coupling.
A common criteria to discuss the interaction strength is called cooperativity $C$ and is given by the ratio between the dipole cross section and the mode area of the light:
\begin{equation}
    C=\frac{\sigma_0}{\text{Area}_{\text{mode}}}.
\end{equation}
The cooperativity gives the ratio between the photons into the targeted mode and those emitted to other modes.

A simple way of increasing $C$ is to put your single atom in an optical cavity. By doing so you will approximately multiply C by the number of round trips inside the cavity. The cooperativity will be modified to:
\begin{equation}
    C=\frac{\sigma_0}{\text{Area}_{\text{mode}}}\frac 1T,
\end{equation}
with $T$ being the output mirror transmission.
A second approach would be to reduce the area of the mode below diffraction limit. It is obviously impossible with a propagating field but it has been observed that using the evanescent field near a nano-structure can indeed provide a mode area well below $ \pi\lambda^2/4$.
In this work we will use another way of improving coupling: by adding more than one atom we can increase the cooperativity by the number of atoms $N$ as 
\begin{equation}
    C=\frac{\sigma_0}{\text{Area}_{\text{mode}}}N.
\end{equation}
The cooperativity is therefore equal to the inverse of the number of atoms needed to observe non-linear effects.
Unfortunately, adding more atoms comes with a long list of associated problems that we will discuss throughout this work. 

\subsection{Dicke states and collective excitation}
Getting a large cooperativity means having a large probability for a photon to be absorbed by the atomic medium. However, we find here an important conceptual question: if a photon is absorbed by the atomic ensemble, in what direction should I expect the photon to be re-emitted ? 
If there is an optical cavity around the atomic ensemble, it is natural to expect the photon to be more likely emitted in the cavity mode, since it is most strongly coupled to the atoms. The coupling efficiency $\beta$ is the ratio between the decay rate in the cavity mode to the total decay rate and is given by: 
\begin{equation}
    C=\frac{\beta}{1-\beta}.
\end{equation}
But, what if we do not place the atoms in a cavity ? Why shouldn't we expect the recovered field in any arbitrary direction ? 
This would make the job of an experimental physicist way trickier...
\subsubsection{Dicke states}
We can get a physical intuition of why light is mainly re-emitted in the forward direction by introducing Dicke states. 
In 1954, Robert Dicke predicted in Ref. \cite{dicke1954coherence} that the behaviour of a cloud of excited atoms would change dramatically above a density of 1 atom per $\lambda^3$.
Indeed, when the inter-atomic distance becomes smaller than the wavelength of the emitted photons, it becomes impossible to distinguish which atoms are responsible for the emission of individual photons.
This indistinguishability lead to a spontaneous phase-locking of all atomic dipoles everywhere in the medium, and therefore a short and directional burst of light is emitted in the forward direction.
The anisotropic nature of the emission can be understood simply as a constructive interference due to the alignment of atomic dipoles thanks to the dipole-dipole interaction \cite{gross1982superradiance,scully2009super}.

\subsubsection{Collective excitation}\label{collective}
When the atomic density is lower than $\lambda^{-3}$, the dipole-dipole interaction becomes negligible but, fortunately, directional emission can still occur. 
The mechanism involved here is similar to the Dicke prediction. 
When one photon interacts with the atomic ensemble, the created excitation is delocalized in the entire cloud.
\textit{Every} atom participates in the absorption process and each of them retains the phase of the incoming field in the coherence between the ground and the excited states.
If the Fock state with only one photon is sent into the medium, the collective state of the atomic cloud $|\underline{e}\rangle$ will be written as a coherent superposition of all possible combinations of a single atom excited $|e\rangle$ and all the other ones in the ground state $|g\rangle$:
\begin{equation}
    |\underline{e}\rangle=\frac{1}{\sqrt N} \sum_{i=1}^N|g_1,g_2,...,g_{i-1},e_i,g_{i+1},...,g_N\rangle.
\end{equation}
Once again the collective enhancement comes from a constructive interference effect \cite{fleischhauer2002quantum}.
During the scattering process, an atom at position $\vec{r}_i$ in the cloud interacts with an incoming photon of wavevector $\vec k$ and therefore acquires (in the rotating frame)   a static phase term $e^{i\vec{k}\vec{r}_i}$ in the collective state superposition $|\underline{e}\rangle$.
After the scattering\footnote{We assume here an elastic Rayleigh scattering process and therefore the norm of $\vec k'$ is fixed, only its direction is a free parameter.} of a photon in the direction of a wavevector $\vec k'$, the atomic ensemble is back in the collective ground state $|\underline{g}\rangle$ but with an additional phase. The cloud state is:
\begin{equation}
  \left( \frac{1}{\sqrt N} \sum_{i=1}^N e^{i(\vec k-\vec k').\vec r_i}\right) |\underline{g}\rangle.
\end{equation}
The global factor before $|\underline{g}\rangle$ gives the square root of the probability of this process.
To benefit from the $\frac 1 N$ enhancement the phase term must be equal to zero and therefore we must have $\vec k =\vec k ' $.
The probability of scattering is then maximum in the forward direction with the same wavevector $\vec k $ as the incident photon.

\subsubsection{Decoherence}
With these two configurations (Dicke states and collective enhancement) we have understood qualitatively why emission will be in the forward direction after scattering through atomic vapor. 
Obviously there are many restrictions to these simple explanations, because the phase coherence is not always conserved.
These troublemakers are grouped under the term: decoherence.

One main reason for the decoherence will come from the fact that atoms are moving (and moving quite fast in a hot atomic vapor). 
Taking this into account, the position of re-emission  for atom $i$ will not be $\vec r_i$  and cancellation of the phase $e^{i(\vec k-\vec k').\vec r_i}$ will not be perfect anymore.
This will be discussed in more detail in paragraph \ref{doppler}.

Another drawback with \textit{real} atoms is that they often have more than one excited state with slightly different energies.
If light couples to these states (even with different coupling constants), the atomic ensemble state will acquire a temporal phase of the order of $e^{i\Delta E  t/\hbar}$ with $\Delta E$ being the characteristic  energy difference between excited states.
We will come back to this point when we will discuss 2-level atoms and rubidium D-lines (see paragraph \ref{2level}).

\subsection{What is a dense atomic medium ?}
As an associate professor, I am member of a CNU (Conseil National des Universit\'es) thematic section called \textit{dilute media and optics.} 
Then why do I entitle this work "Quantum Optics in dense atomic medium".   
What does dense mean  in this context ?
I do not mean that I have been working with dense media in condensed matter sense (with the notable exception of section \ref{polariton}) but rather with an \textbf{optically dense} cloud.
In the continuation I describe the relationship between temperature and atomic density in a warm vapor and the method to provide a good measure of the optical density: the optical depth.

\subsubsection{Atomic density}\label{atomicdensity}
A strong advantage of atomic vapors is that their density can be tuned at will by simply changing their temperature.
As alkali vapors are not perfect gases, there is a correction to the Boyle law given by C. Alcock, V. Itkin and M. Horrigan in Ref. \cite{alcock1984vapour}.
The vapor pressure $p$ in Pascal is given by: 
\begin{equation}
    p(\text{Pa})=10^{A+\frac{B}{T}},
\end{equation}
with  $A=9.318$ and $B-4040$~K.
The atomic density $n_{at}$ is then given by dividing the vapor pressure by $k_B$ the Boltzmann constant times the temperature $T$:
\begin{equation}
    n_{at}=\frac{p(\text{Pa})}{k_B T}.
\end{equation}

\begin{figure}
    \centering
     \includegraphics[width=12cm]{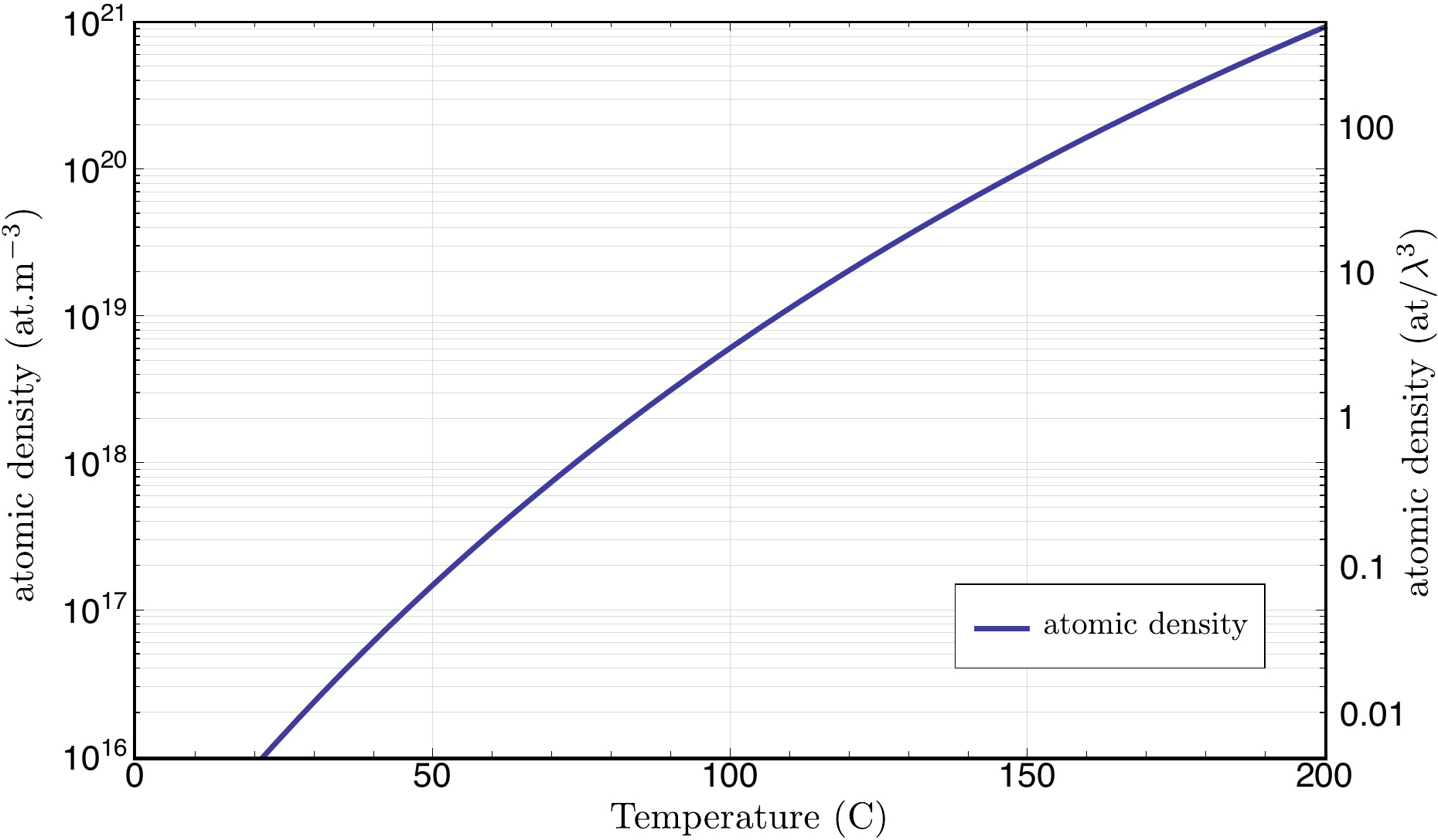}
        \caption{Atomic density of rubidium as function of temperature in Celsius. Right scale is given in unit of $\lambda^3$ for the D2 line.}
    \label{fig:density}
\end{figure}

\subsubsection{Optical depth}
Taking into account the density, light propagating in an atomic medium  will have a linear absorption $\alpha$ given by:
\begin{equation}
\alpha=n_{at}\sigma_0.
\end{equation}
The optical depth is simply defined as the product of $\alpha$ by $L$, the length of the atomic medium. 
Based on this definition we can directly write the Beer law of absorption for a beam of initial intensity $I_0$:
\begin{equation}
I=I_0\exp{[-\alpha L]}.
\end{equation}
For the D2 line of rubidium $\sigma_0=1.25\times 10^{-9}$~cm$^{-2}$.
We can note that the value is slightly different from Eq. \ref{sigma0} due to the multiplicity of atomic states.\\

\subsubsection{Saturation intensity}
This derivation is only valid on resonance and with a weak electromagnetic field. 
In fact, saturation can modify the behaviour of the medium.
To quantify this effect, we introduce the saturation intensity $I_{sat}$ which is defined as the value of intensity for which the cross section is reduced by half compared to the low intensity case.
This leads to the redefinition of the atomic cross section $\sigma$ as \cite{van2016atoms}:
\begin{equation}\label{sigma1}
\sigma= \frac{\sigma_0}{1+I/I_{sat}}.
\end{equation}
We can refine even more this model by adding a detuning $\Delta = \omega_L - \omega_0$ between the excitation (laser) field frequency and the atomic transition.
In this case we have:
\begin{equation}\label{sigmaDelta}
    \sigma(\Delta)=\frac{\sigma_0}{1+4 (\Delta/\Gamma)^2+I/I_{sat}}.
\end{equation}
The saturation intensity is defined for a resonant excitation. 
However, in practice, it is often useful to know if the atomic medium is saturated or not at a large detuning from the resonance.
A simple way to obtain a quick insight on this question is defining an \textit{off-resonance} saturation intensity $I_{sat}(\Delta)$.
It is given by \cite{van2016atoms}:
\begin{equation}\label{Isatdelta}
    I_{sat}(\Delta)=\left[1+4\frac{\Delta^2}{\Gamma^2}\right] I_{sat}^0,
\end{equation}
with $I_{sat}^0$ the resonant saturation intensity.
We have then reformulated Eq. \ref{sigma1} with $I_{sat}(\Delta)$ instead of $I_{sat}$.
We can remember that, when $I=I_{sat}(\Delta)$, the cross section is reduced by half compared to low intensity.

\subsection{Doppler broadening}\label{doppler}
Unfortunately, in an atomic cloud, every atom does not contribute equally to the light-matter interaction. This is a consequence of the Doppler effect.
Indeed, atoms moving at the velocity $\vec v$ will result to a frequency shift $\sim \vec k \vec v$.
In a warm atomic vapor this effect is crucial to understand the dynamics of the system, as Doppler broadening can be much greater than the natural linewidth $\Gamma$. For an atom of a mass $m$ at temperature $T$ the Doppler linewidth $\Gamma_D$ is given by:
\begin{equation}
    \Gamma_D=\sqrt{\frac{k_B T}{m\lambda^2}}.
\end{equation}
At 100$^{\circ}$C, this gives $\Gamma_D\approx 250$~MHz for Rb D2 line, compared to $\Gamma\approx 6$~MHz.\\

An important consequence is that the atomic response, derived in the next section, will have to be integrated over the atomic velocity distribution to obtain quantitative predictions.
We will not include this calculation here, as it adds unnecessary complexity but no immediate intelligibility. We refer the interested reader to \cite{glorieux2010etude,mishina2011electromagnetically,sheremet2010quantum}

\begin{figure}[]
    \centering
     \includegraphics[width=0.9\textwidth]{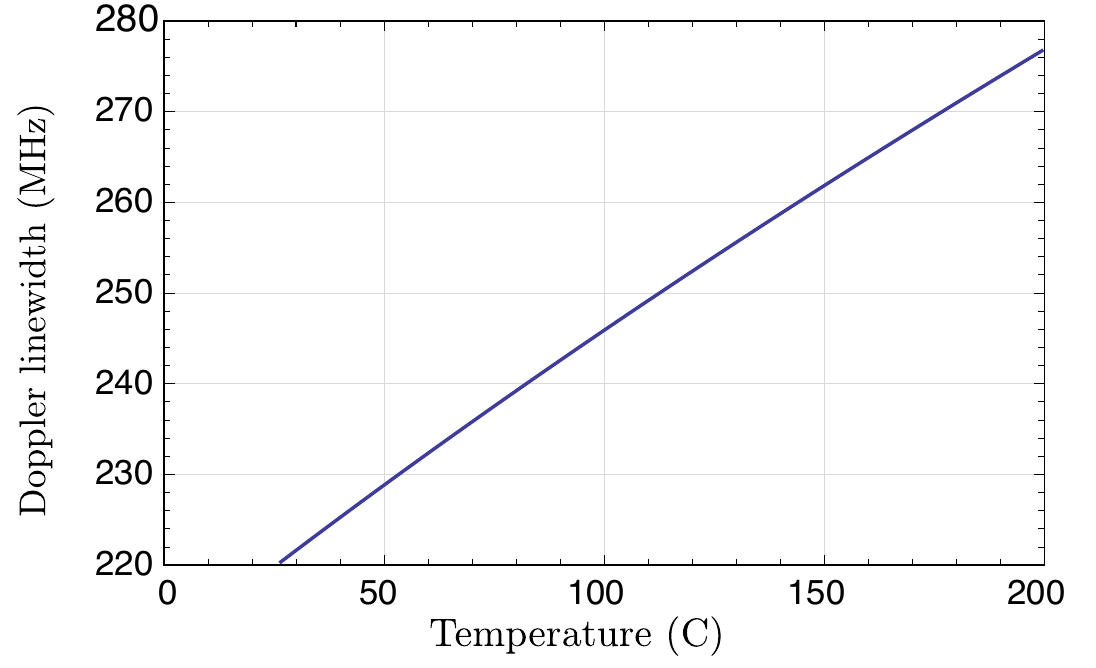}
        \caption{Doppler linewidth of rubidium D2 line as function of temperature. }
    \label{fig:doppler}
\end{figure}

\clearpage
\section{Non-linear optics in dense atomic media}

\epigraph{\textit{There is no two-level atom and rubidium is not one of them}}{William D. Phillips}

This famous quote from Bill Phillips reminds us, that how useful can  the two-level atom model be, it still remains simplistic compared to the complexity of alkali atomic structures.
In this section, I  briefly recall three important features of this model: linear absorption $\text{Im}(\chi^{(1)})$, linear phase shift $\text{Re}(\chi^{(1)})$ and non-linearity $\chi^{(3)}$ near a two-level atomic resonance.
This well known derivation is an important concept settling the background of this work.
Along this section I will highlight the theoretical tools one by one  to progressively cover more and  more complex situations.
Specifically, we are in this work focusing on non-linear effects with large intensities.
This means not only a large $\text{Re}(\chi^{(3)})$ but also a large $\text{Re}(\chi^{(3)})\times I$.\\

\noindent Therefore I introduce:
\begin{itemize}
    \item the saturation of non-linearity at larger intensity ($\chi^{(5)}$ correction term); 
    \item the concept of electromagnetically induced transparency (EIT) and possible extension when the probe beam is not perturbative anymore;
    \item the process of four-wave-mixing.
\end{itemize}

\subsection{2--level atoms}\label{2level}

Let us consider the interaction of a monochromatic electric field with a system of $N$ two-level atoms.
This interaction process can be described by the optical Bloch equation \cite{boyd2003nonlinear}:
\begin{equation}
\frac{d\hat{\rho}}{dt} = -i \frac{i}{\hbar}\left[\hat{H}, \hat{\rho}\right] - \hat{\Gamma}\hat{\rho},
\label{Bloch}
\end{equation}
where $\hat{\rho}$ is the density matrix of the atomic system, $\Gamma$ is the decay rate of the excited state, $\hat{H} = \hat{H}_0 + \hat{V}$ is the Hamiltonian of the system with the non-perturbative part $\hat{H}_0$ and the interaction $\hat{V}$ which can be written as $\hat{V} = - \mathbf{d}\cdot \mathbf{E} = -\hbar\Omega/2$ in the dipole approximation.
$\Omega$ is the Rabi frequency.

We can denote the ground and the excited state of the atom as $|g\rangle$ and $|e\rangle$ respectively with the resonant transition frequency $\omega_{eg}$. With these notations we can rewrite the Bloch equation \eqref{Bloch} for the slowly varying amplitudes $\sigma_{ij}(t)$ of the density matrix elements $\rho_{ij}(t) = \sigma_{ij}(t)\exp(-i\omega_{ij} t)$ in the following form:
\begin{eqnarray}
\dot{\rho}_{gg} &=& i\frac{\Omega}{2} \left(\sigma_{ge} - \sigma_{eg}\right) + \Gamma\rho_{ee}
\nonumber\\
\dot{\rho}_{ee} &=& -i\frac{\Omega}{2} \left(\sigma_{ge} - \sigma_{eg}\right) - \Gamma\rho_{ee}
\nonumber\\
\dot{\sigma}_{ge} &=& - i(\Delta - i\Gamma/2)\sigma_{ge} - i\frac{\Omega}{2} \left(\rho_{ee} - \rho_{gg}\right) ,
\label{Dmatrix}
\end{eqnarray}
where $\Omega$ is the Rabi frequency of the probe field and $\Delta = \omega - \omega_{eg}$ is the laser detuning from the excited state. 
To obtain this system of equations we have applied the Rotating Wave Approximation (RWA).
This approximation allows to eliminate the fast decaying terms and to rewrite the Bloch equation for slow-varying amplitudes \cite{boyd2003nonlinear}.
The elements $\rho_{gg}$ and $\rho_{ee}$ correspond to population of the ground and the excited states respectively, while the elements $\sigma_{eg} = \sigma_{ge}^{*}$ correspond to the atomic coherence.
We can rewrite the system of equations \eqref{Dmatrix} as:
\begin{eqnarray}
\dot{\rho}_{ee} - \dot{\rho}_{gg} &=& -i\Omega \left(\sigma_{ge} - \sigma_{eg}\right) - \Gamma\left(\rho_{ee} - \rho_{gg} + 1 \right)
\nonumber\\
\dot{\sigma}_{ge} &=& - i(\Delta - i\Gamma/2)\sigma_{ge} - i\frac{\Omega}{2} \left(\rho_{ee} - \rho_{gg}\right) ,
\label{Dmatrix2}
\end{eqnarray}
taking into account the condition that $\rho_{gg} + \rho_{ee} = 1$. 
Because the amplitudes $\rho_{gg}$, $\rho_{ee}$ and $\sigma_{eg}$ are slow-varying we can assume that $\dot{\rho}_{gg} = \dot{\rho}_{ee} = \dot{\sigma}_{ge}=0$ and the solution of \eqref{Dmatrix2} can be found in the following form:
\begin{eqnarray}
\sigma_{ge} &=& -\frac{\Omega/2\left(\rho_{ee} - \rho_{gg}\right)}{\Delta - i\Gamma/2},
\nonumber\\
\rho_{ee} - \rho_{gg} &=& -\frac{\Delta^2 + \Gamma^2/4}{\Delta^2 + \Gamma^2/4 + \Omega^2/2}.
\label{Bsolution}
\end{eqnarray}

The response of the medium on the interaction with light can be described in terms of the atomic polarization $\mathbf{P}$. The vector of polarization relates to the electric field with a proportional coefficient:
\begin{equation}
\mathbf{P} = \varepsilon_0\chi\mathbf{E},
\label{polarization}
\end{equation}
where $\chi$ is the atomic susceptibility.

In general, if we neglect frequency conversion processes, the polarization $\mathbf{P}$ can be written as an expansion in Taylor series in the electric field as:
\begin{equation}
P = \varepsilon_0\chi^{(1)}E + \varepsilon_0\chi^{(2)}|E|^2 + \varepsilon_0\chi^{(3)}|E|^2\cdot E + \varepsilon_0\chi^{(4)}|E|^4 + \varepsilon_0\chi^{(5)}|E|^4\cdot E + ...
\end{equation}
Here $\chi^{(1)}$ is known as a linear susceptibility, higher order terms are known as a second-order, a third-order, or n-order susceptibilities.
In general, for anisotropic materials the susceptibility is a $(n-1)$-order rank tensor. 
In our consideration we expand the Taylor series up to the rank $5$ to take into account the nonlinear response of the atomic medium. 
For a centro-symmetric medium, the even terms vanish and the polarization of the system can be written as \cite{boyd2003nonlinear}:
\begin{equation}\label{Pola1}
P = \varepsilon_0\chi^{(1)}E  + \varepsilon_0\chi^{(3)}|E|^2\cdot E  + \varepsilon_0\chi^{(5)}|E|^4\cdot E + ...
\end{equation}

\noindent The polarization can be found in terms of the density matrix elements:
\begin{equation}
P = N\ \mu_{eg}\sigma_{ge},
\label{polarization2}
\end{equation}
where $\mu_{eg}$ is the dipole moment of the transition.
From this expression we can find a full polarization of the atomic system:
\begin{equation}
P = -\frac{N|\mu_{eg}|^2(\Delta + i\Gamma/2)}{\hbar(\Delta^2 + \Gamma^2/4 + \Omega^2/2)}E = -\frac{4N|\mu_{eg}|^2}{\hbar\Gamma^2}\cdot\frac{\Delta + i\Gamma/2}{1 + 4\Delta^2/\Gamma^2 + 2\Omega^2/\Gamma^2}E.
\end{equation}
We can write $P$ as function of the saturation intensity $I_{\text{sat}}$ using: 
\begin{equation}
    \frac{I}{I_{\text{sat}}} = 2\left[\frac{\Omega}{\Gamma}\right]^2 \text{ with } I = \frac12 n_0 \varepsilon_0 c E^2.
\end{equation}
One obtains:
\begin{equation}
P = -\frac{4N|\mu_{eg}|^2}{\hbar\Gamma^2}\cdot\frac{\Delta + i\Gamma/2}{1 + 4\Delta^2/\Gamma^2 + I/I_{\text{sat}}}E.
\label{chiI}
\end{equation}
We should note that the real part of the susceptibility corresponds to the refractive index of the medium, while the imaginary part gives information about the absorption.

\begin{figure}
    \centering
    \includegraphics[width=\textwidth]{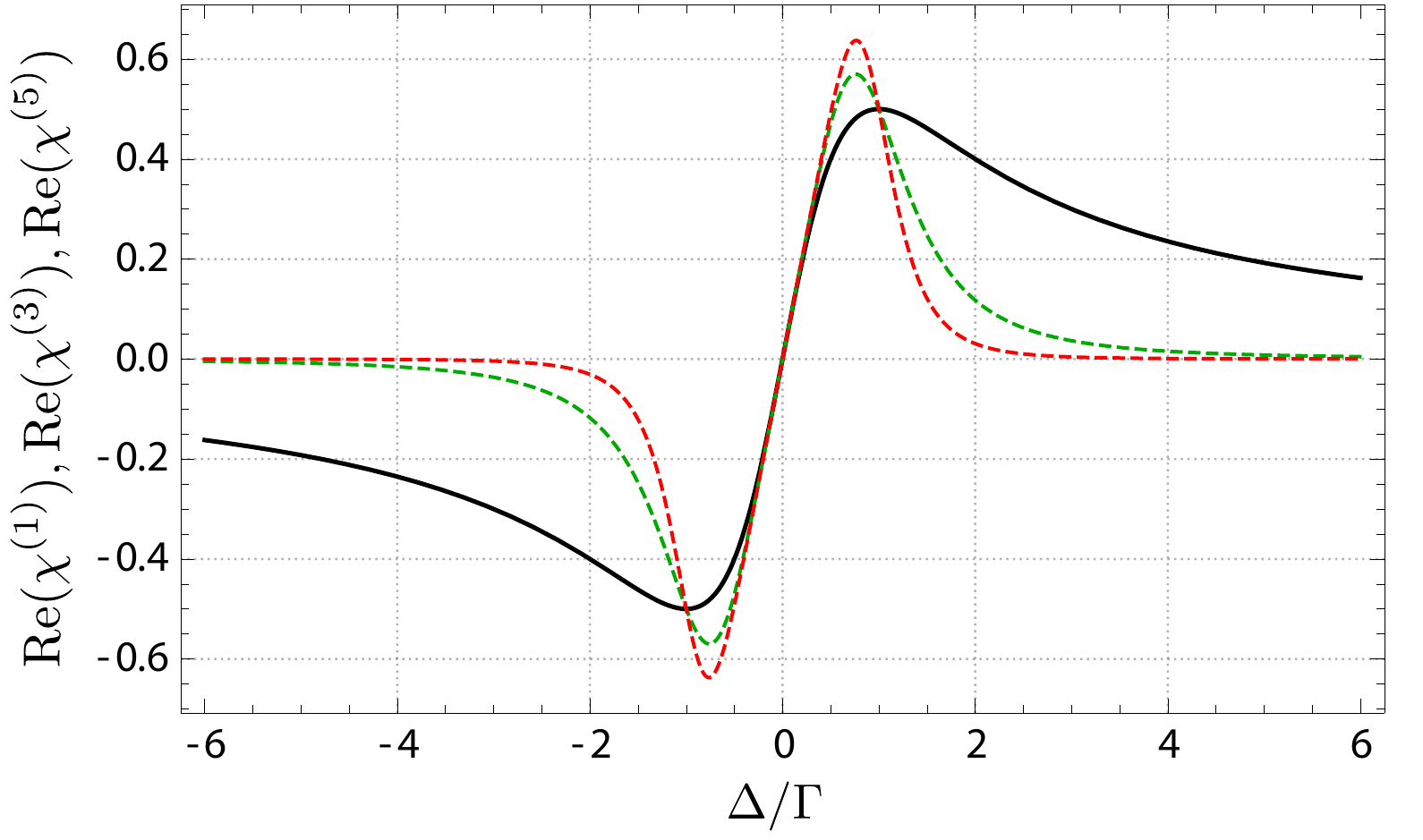}
    \caption{$\text{Re}(\chi^{(1)})$ in black ,$\text{Re}(\chi^{(3)})$ in dashed green,$\text{Re}(\chi^{(5)})$ in dashed red}
    \label{fig6}
\end{figure}

Next we calculate the zeroth, first and second contributions to the polarization of a collection of two-level atoms. 
By performing a power series expansion of Eq. \eqref{chiI} in the quantity $I/I_{\text{sat}}$:
\begin{equation}
P \approx  -\frac{4N|\mu_{eg}|^2}{\hbar\Gamma^2}\frac{\Delta + i\Gamma/2}{(1+4\Delta^2/\Gamma^2)}E\cdot \left[1 - \frac{I/I_{\text{sat}}}{1 + 4\Delta^2/\Gamma^2} + \frac{I^2/I_{\text{sat}}^2}{(1 + 4\Delta^2/\Gamma^2)^2} \right].
\label{expansion}
\end{equation}
We now equate this expression with Eq. \ref{Pola1} to find the three first orders of the atomic polarization:
\begin{eqnarray}
\varepsilon_0\chi^{(1)} &=&  -\frac{4N|\mu_{eg}|^2}{\hbar\Gamma^2} \cdot \frac{\Delta + i\Gamma/2}{(1+4\Delta^2/\Gamma^2)}
\nonumber\\
\varepsilon_0 \chi^{(3)}|E|^2 &=&\ \frac{4N|\mu_{eg}|^2}{\hbar\Gamma^2} \cdot \frac{\Delta + i\Gamma/2}{(1+4\Delta^2/\Gamma^2)}\cdot \frac{I/I_{\text{sat}}}{1 + 4\Delta^2/\Gamma^2}
\nonumber\\
\varepsilon_0\chi^{(5)}|E|^4 &=& -\frac{4N|\mu_{eg}|^2}{\hbar\Gamma^2} \cdot \frac{\Delta + i\Gamma/2}{(1+4\Delta^2/\Gamma^2)}\cdot \frac{I^2/I_{\text{sat}}^2}{(1 + 4\Delta^2/\Gamma^2)^2}.
\label{chinonl}
\end{eqnarray}
We introduce the usual\footnote{The coefficients $1,3$ and $10$ are used because we are only concern with the non-linear effects conserving the input frequency. 
For example the  $\chi^{(3)}$ term can lead to several frequency conversion, and we only keep triplets like:  $(+\omega,+\omega,-\omega)$, $(+\omega,-\omega,+\omega)$ and $(-\omega,+\omega,+\omega)$. 
This coefficient is given by the binomial coefficient $\tbinom{3}{2}$.}  power series expansion: 
\begin{equation}
    \chi_{\text{eff}} = \chi^{(1)} + 3  \chi^{(3)} |E|^2 + 10  \chi^{(5)} |E|^4.
\end{equation}
This notation allows us to write the effective refractive index $n_{\text{eff}}$ as:
\begin{equation}
    n_{\text{eff}}^2=1+\chi_{\text{eff}}.
\end{equation}
We use the standard definition \cite{boyd2003nonlinear} of the non-linear indices ($n_2$ and $n_3$) :  $n_{\text{eff}}=n_0+n_2I+n_3I^2$ and we can expand  $n_{\text{eff}}^2$ in:
\begin{equation}
    n_{\text{eff}}^2=n_0^2+2n_0n_2I+(2n_0n_3+n_2^2)I^2.
\end{equation}
We can then connect the indices to the expression of $\text{Re}(\chi)$.
We show the scaling of $\text{Re}(\chi)$ in Figure \ref{fig6}.
\begin{eqnarray}
n_0 &=& \sqrt{1 + Re\left[\chi^{(1)}\right]},
\nonumber\\
n_2 &=& \frac{3Re\left[\chi^{(3)}\right]}{n_0^2 \varepsilon_0 c},
\nonumber\\
n_3 &=&  \frac{20Re\left[\chi^{(5)}\right]}{n_0^3 \varepsilon_0^2 c^2}-\frac{n_2^2}{2n_0}.
\label{refractive}
\end{eqnarray}

\subsubsection{What to optimize ?}
When looking at the previous derivation of the non-linear susceptibility, we see that 3 different quantities can be optimized depending on the goal of your experiments.
\begin{itemize}
    \item If you want to observe non-linear effects at the single photon level, for example to create a photonic transistor,  you want to optimize  the value of Re$\left[\chi^{(3)}\right]$ in the limit $I\ll I_{sat}$;
     \item If you want to observe dispersive effects as slow and fast light propagation,  you will want to optimize the value of derivative of Re$\left[\chi\right]$ as function of the frequency.
     This is the configuration we will study in chapter 3;
     \item If you are interested in a large value for $\Delta n= n_2\times I$, you have access to 2 knobs: increasing $n_2$ or increasing $I$. However you must stay within the limit of $I< I_{sat}$ otherwise the Taylor expansion does not hold anymore. That is why we have done the calculation to the next order $I/I_{sat}$. This is the configuration we will study in chapter 4.
\end{itemize}
 
\subsubsection{Scaling for  effective 2-level atoms.}
We briefly review the dispersive limit. 
When the laser is detuned far enough from the atomic transition, we have a simplification of the atomic response.

First, the approximation of the 2-level atoms becomes more precise because the contribution from all the levels averages to an effective contribution. 
This is the case when the detuning $\Delta$ is much larger than the hyperfine splitting energy scale (typically around $\sim 500$~MHz for Rb).\\
The second consequence is that the absorptive part of the susceptibility Im$(\chi)$ becomes negligible with respect to the dispersive part Re$(\chi)$.
In this limit, the medium is virtually transparent but there is still a non-negligible phase shift (a linear and a non-linear one).
At $\Delta \gg \Gamma$, we can simplify Eq. \ref{chinonl} to obtain for the linear part of the absorption:
\begin{equation}
 \text{absorption: Im}(\chi^{(1)}) \propto  \frac{N}{\Delta^2},
\end{equation}
and for the non-linear dispersive part:
\begin{equation}
 \text{phase shift: Re}(\chi^{(3)}) \propto \frac{N}{\Delta^3}.
\end{equation}

The non-linear absorption is neglected as it scales with $\frac{1}{\Delta^4}$, as well as the linear phase shift which just results in a redefinition of the phase reference.\\

In the far detuned limit, an intuitive idea to improve the non-linear phase shift is to simply increase the atomic density, by rising the cloud temperature as described in section \ref{atomicdensity}. 
This is indeed true as Re$(\chi^{(3)})$ scales with $N$. 
However, this is often critical to conserve a large (fixed) transmission while increasing the non-linear phase shift.
This condition of fixed transmission means that $\frac{N}{\Delta^2}$ is a constant.
In other words, we can rewrite the phase shift Re$(\chi^{(3)})$ as this constant times $\frac{1}{\Delta}$. 

We see that the intuitive vision is no longer valid if we want to keep a fixed transmission: to maximize the phase shift at a given transmission it is therefore favorable to reduce $\Delta$, which in consequence leads to a lower temperature (to keep $\frac{N}{\Delta^2}$ constant).
This is obviously limited by the initial hypothesis of far detuned laser ($\Delta \gg \Gamma$).

In order to verify this model we have measured the non-linear phase shift for various temperatures and detuning and this is reported in chapter \ref{chap:4}.
In the next paragraph, I explain how to conduct this measurement.

\subsubsection{Measurement of the non-linear phase shift for  effective 2-level atoms.}

\begin{figure}
    \centering
    \includegraphics[width=\textwidth]{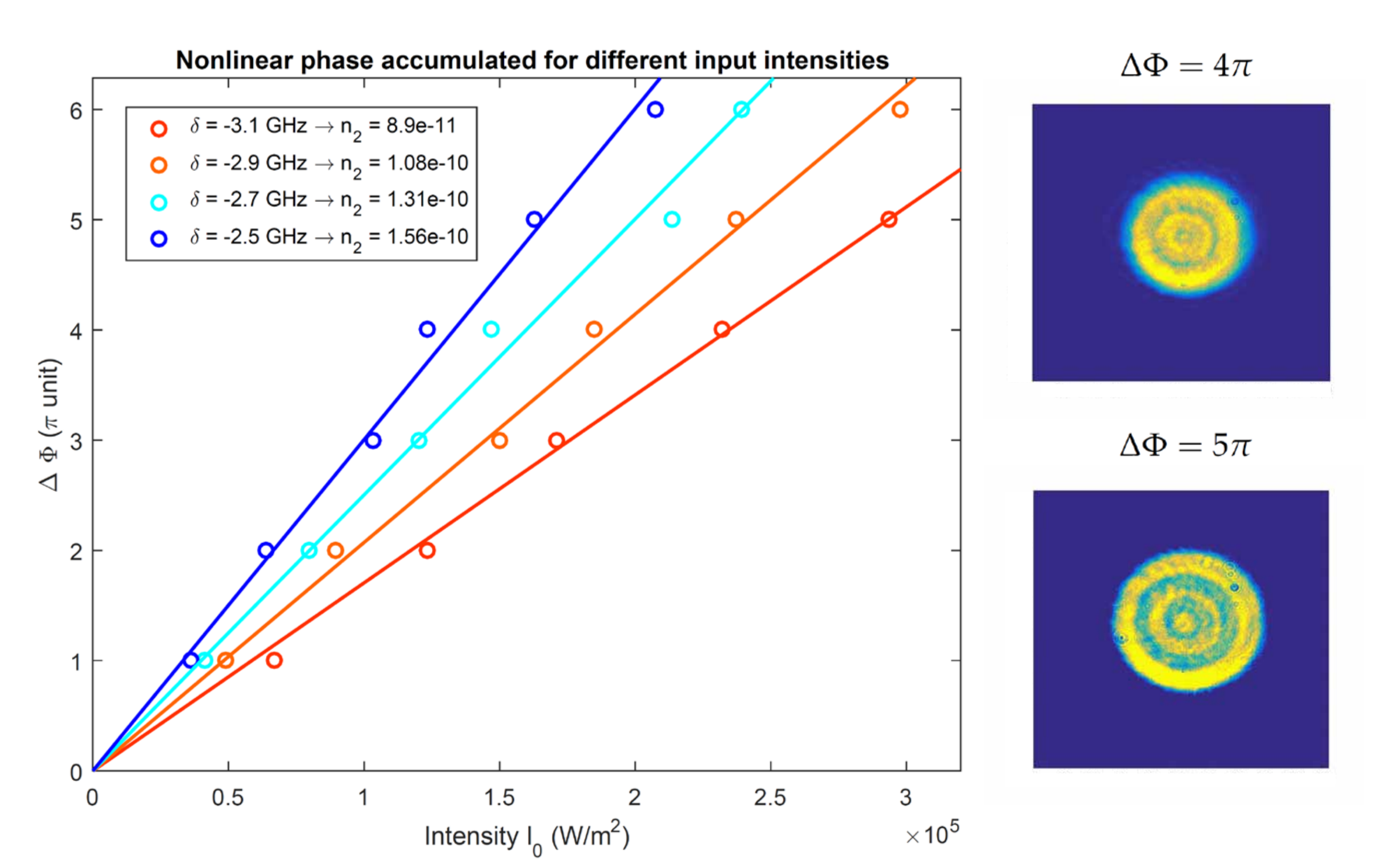}
    \caption{Typical non-linear phase shift measurement for effective 2-level atomic cloud. Detuning $\delta$ is given to D1 Rb85 line. Cell temperature is 130C. For larger intensity a clear deviation is observed due to $n_3$ term (not shown). Insets are examples of obtained far field measurements. }
    \label{fig:rings}
\end{figure}

\begin{figure}
        \centering
        \includegraphics[width=0.9\textwidth]{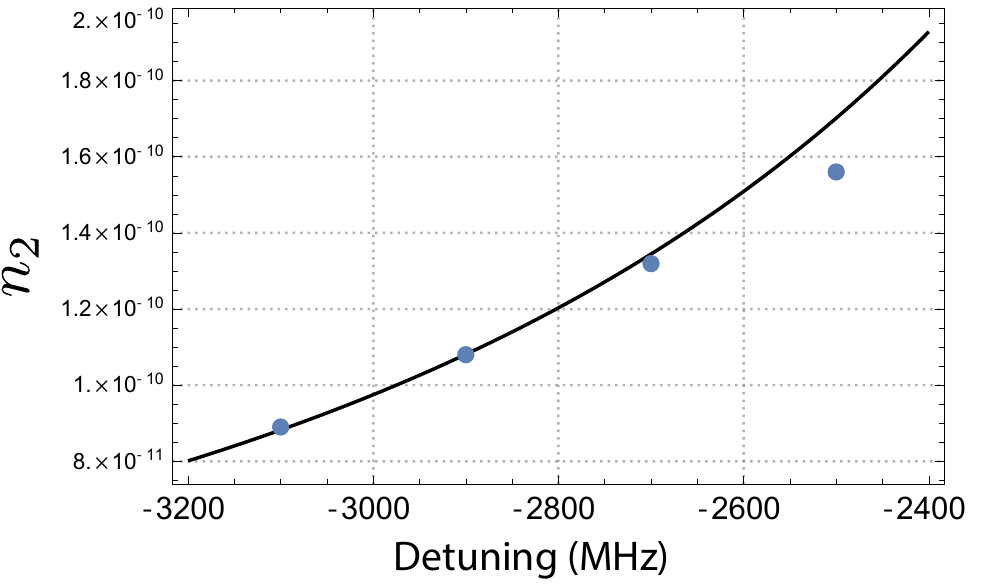}
        \caption{Comparison between 2-level atoms model and experimental measurement of $n_2$. Blue dots are experimental data presented in Fig. \ref{fig:rings}, and black line is the model of Eq. \ref{refractive}. The model is scaled by the atomic density which is a fitting parameter.}
        \label{fig:comparison2level}
    \end{figure}

The typical method to measure the non-linear phase shift of a sample is to realize a z-scan experiment \cite{zhao1993z,Rings_Focusing_Defocusing_Exp}.
However this technique works better with a thin layer of material. 
For thick samples, we can use a technique demonstrated in Ref. \cite{Rings_Rb}.
This allows to measure the accumulated phase $\Delta \phi$ along the propagation.
The phase accumulated can be written as:
\begin{equation}
    \Delta \phi (r)= k\int_{z_0}^{z_0+L}n_2I(r,z) dz,
\end{equation}
with $z_0$ is the coordinate of the front of the sample and $L$ is its length.
The intensity profile of the beam can then be replaced by the Gaussian profile of a TEM(0,0) beam at the input plane. We obtain:

\begin{equation}\label{phaseshift1}
    \Delta \phi (r) = k\int_{z_0}^{z_0+L}n_2 \Delta \phi (r) \frac{w_0^2}{w(z)^2}e^{-2r^2/w(z)^2}dz,
\end{equation}
with $w_0$ is the waist radius, $w(z)=w_0\sqrt{1+\left(\frac{z}{z_R}\right)^2}$ and $z_R=\frac{\pi w_0^2}{\lambda}$.
It is then possible  to derive the far field diffraction pattern in the Fraunhofer limit (see Ref. \cite{Rings_Rb} for details).
From Eq. \ref{phaseshift1}, it is clear that the phase shift in the center : $\Delta \phi (r=0)$ is directly proportional to $n_2$ and the central intensity $ I(0,0)$.
We will therefore observe a switch from a bright spot to dark spot in the center of the diffraction pattern when the phase is  modified by $\pi$ and back to a bright spot again for a phase change of $2\pi$.
By simply counting the number of rings $N_{Rings}$ that appear while slowly increasing $I(0,0)$, we can estimate the non-linear phase shift accumulated along $z$.
In the limit of long Rayleigh length ($z\ll z_R$) we can approximate $\Delta n$ to:
\begin{equation}
   \Delta n=n_2 I \simeq  \frac{\lambda}{L}N_{Rings}.
\end{equation}
In Figure \ref{fig:rings}, I present an experimental characterization of $n_2$ for rubidium 85.
From this figure, we have extracted the value of $n_2$ as function of the detuning from the atomic transition.
To validate the 2-level atoms model, I plot in Figure \ref{fig:comparison2level}, the results of the numerical model (Eq. \ref{refractive}) after integration over the Doppler profile and compare it to the experimental data.
We see that for large detuning the model is in excellent agreement. However as we get closer to the resonance the contribution of the $\chi^{(5)}$ term start to be not negligible anymore and $n_2$ is reduced\footnote{ $\chi^{(3)}$ and  $\chi^{(5)}$ are of opposite signs.} compared to the value predicted by  $\chi^{(3)}$. 
This is the main limitation to obtain a larger $\Delta n=n_2 I$ in experiments.\\

\begin{figure}[h!]
    \centering
    \includegraphics{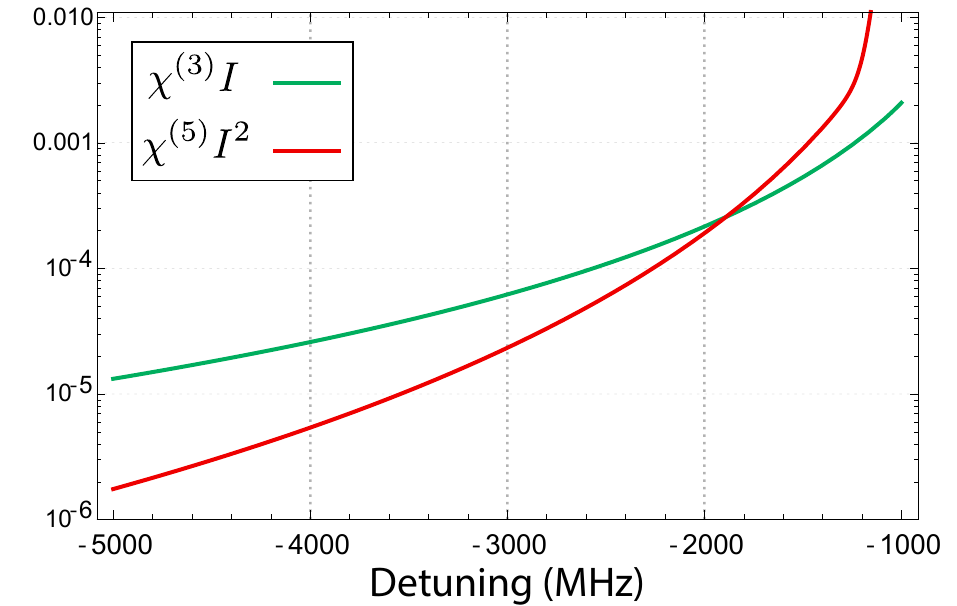}
    \caption{Comparison between $|\chi^{(5)}I^2|$ and $|\chi^{(3)}I|$. Here $I=2.10^5$~W/m$^2$ (which is still much smaller than the effective saturation intensity at this detuning).}
    \label{fig:chi3chi5}
\end{figure}

To get a better understanding of this effect we have compared $\chi^{(5)}I^2$ and $\chi^{(3)}I$ for the two-level model.
We see in Fig. \ref{fig:chi3chi5}, that the contribution of  $|\chi^{(5)}I^2|$  is huge when we get closer to resonance but at detuning larger than 2.5~GHz, it can be safely neglected at this intensity.

\clearpage

\subsection{Electromagnetically induced transparency}\label{sec:EIT}
In the previous section we have discussed non-linearity in a 2-level atoms model.
If we add one more level, more complex schemes using atomic coherences can be exploited. 
Here, I briefly describe the interaction of the system of $3$-level atoms with two electromagnetic fields in order to study the effect of \textit{electromagnetically induced transparency} (EIT) \cite{boller1991observation,fleischhauer2000dark}. 

The interaction process can be described by the evolution of the optical Bloch equations, as it was done in the section \ref{2level}. 
The three levels are noted:  $|g\rangle$ for ground, $|e\rangle$  for excited,  $|s\rangle$ for supplementary third level.
The optical Bloch equation (\ref{Bloch}) can be rewritten for slowly-varying amplitudes in the case of a $3$-level atom as following \cite{glorieux2010etude,boyd2003nonlinear}:
\begin{eqnarray}
\label{Bloch3}
\dot{\rho}_{gg} &=& i\frac{\Omega_p}{2}(\sigma_{ge} - \sigma_{eg}) + \Gamma_{eg}\rho_{ee}
\nonumber\\
\dot{\rho}_{ss} &=& i\frac{\Omega_c}{2}(\sigma_{se} - \sigma_{es}) + \Gamma_{es}\rho_{ee}
\nonumber\\
\dot{\rho}_{ee} &=& -i\frac{\Omega_p}{2}(\sigma_{ge} - \sigma_{eg}) - i\frac{\Omega_c}{2}(\sigma_{se} - \sigma_{es}) -\Gamma\rho_{ee}
\\
\dot{\sigma}_{ge} &=& -i(\Delta_p - i\Gamma/2)\sigma_{ge} - i\frac{\Omega_p}{2}(\rho_{ee} - \rho_{gg}) + i\frac{\Omega_c}{2}\sigma_{gs}
\nonumber\\
\dot{\sigma}_{se} &=& -i(\Delta_c - i\Gamma/2)\sigma_{se} - i\frac{\Omega_c}{2}(\rho_{ee} - \rho_{ss}) + i\frac{\Omega_p}{2}\sigma_{sg}
\nonumber\\
\dot{\sigma}_{gs} &=& -i(\Delta_p-\Delta_c -i\gamma_0)\sigma_{gs} - i\frac{\Omega_p}{2}\sigma_{es} + i\frac{\Omega_c}{2}\sigma_{ge}.
\nonumber
\end{eqnarray}
Here we characterize two electromagnetic fields: the probe field with the Rabi frequency $\Omega_p$ interacts between the initially populated ground state $|g\rangle$ and the excited state $|e\rangle$, while the control field with the Rabi frequency $\Omega_c$ couples the excited state $|e\rangle$ with the initially empty second ground state $|s\rangle$. 
The probe and the control fields are detuned from the corresponding atomic resonances with detunings $\Delta_p = \omega_p - \omega_{eg}$ and $\Delta_c = \omega_c - \omega_{es}$ respectively. Decay rates $\Gamma_{eg}$ and $\Gamma_{es}$ can be found with the Clebsch-Gordan coefficients of the corresponding transitions, and they satisfy to the condition $\Gamma_{eg} + \Gamma_{es} = \Gamma$. 
The decay rate $\gamma_0$ corresponds to the decay rate of the ground states coherence between $|g\rangle$  and $|s\rangle$.

In the RWA we assume $\rho_{ii}$ and $\sigma_{ij}$ as slowly varying amplitudes. With these conditions the system (\ref{Bloch3}) can be solved in the steady-state regime when $$\dot{\rho}_{gg} = \dot{\rho}_{ee} = \dot{\rho}_{ss} = \dot{\sigma}_{ge} = \dot{\sigma}_{se} = \dot{\sigma}_{gs} = 0.$$

In the following, we are interested in the nonlinear components of the atomic susceptibility $\chi$, which is the proportionality coefficient between the atomic polarization and the electric field, see Eq. \ref{polarization}.
We solve the system (\ref{Bloch3}) numerically.
The polarization induced by the probe field can be found in terms of the coherence at the corresponding atomic transition $\sigma_{ge}$, in the same way how it was done in Eq. \ref{polarization2}.

We can extract quantities similar to 2-level atoms: $\chi^{(1)},\chi^{(3)},\chi^{(5)}$ from a linear expansion of $\chi$ obtained numerically.
The code to implement these simulations in Mathematica is available \href{https://www.wolframcloud.com/objects/9fcca8a8-82e3-42fc-9966-327534f43d15}{here}.

\subsection{Four-wave-mixing}
\begin{figure}[h]
    \centering
    \includegraphics[width=0.70\textwidth]{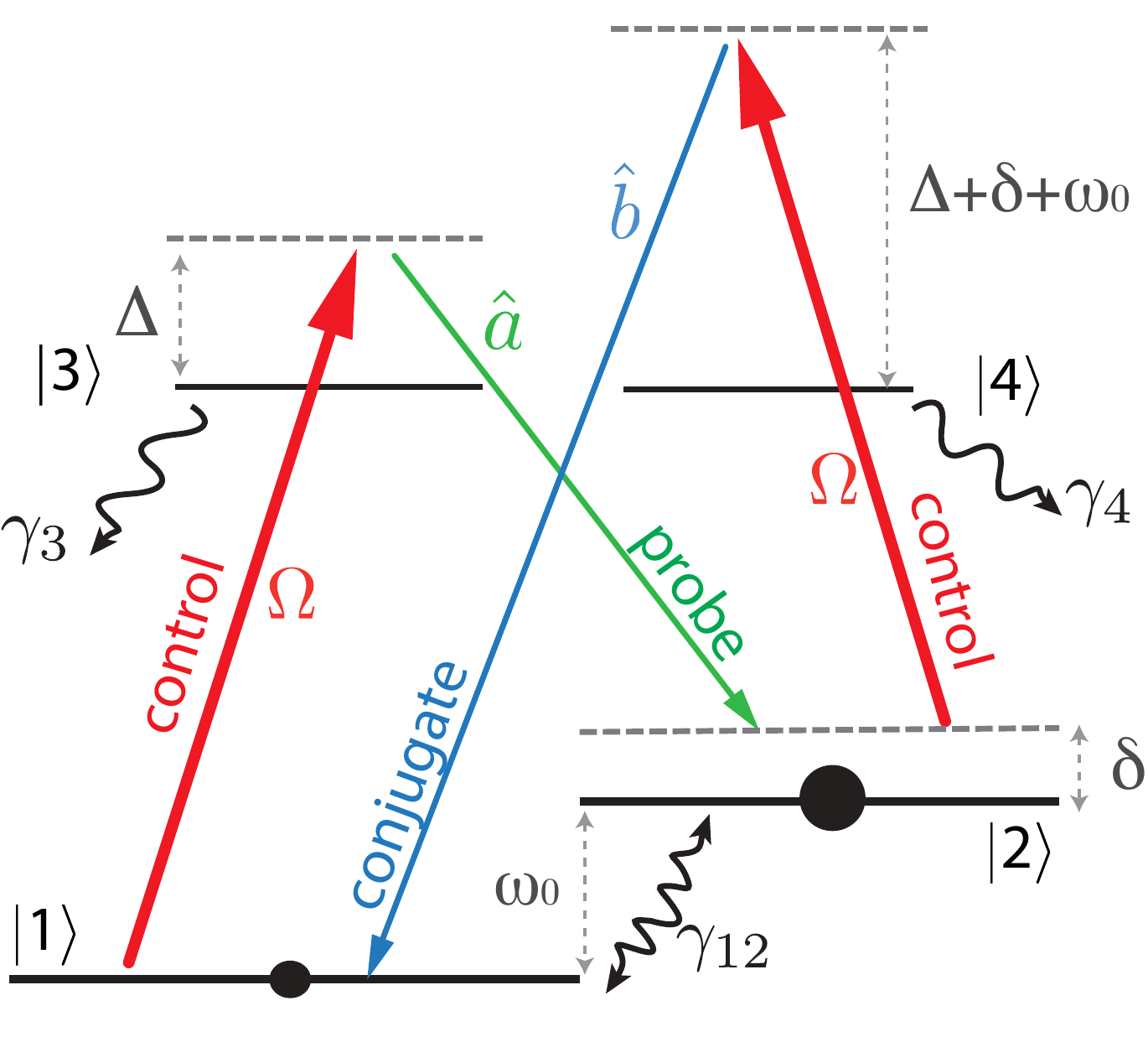}
    \caption{Levels scheme for four-wave-mixing in the double-$\Lambda$ configuration.}
    \label{fig:doublelambda}
\end{figure}

Adding one more level, one can add even more complexity (and also more fun) in the light-matter interactions.
I describe here, a typical configuration called double-$\Lambda$, where two intense pumps, one probe beam and one conjugate beam interact (see Figure \ref{fig:doublelambda}).
A lot of details about this configuration can be found in \cite{glorieux2010double,glorieux2011quantum,agha2011time,glorieux2012generation,corzo2013rotation,PhysRevA.87.010101,Vogl14}.

Here, I will just give an intuitive description of a few phenomenon. 
One important point to understand, is there is only one pump laser in real experimental configuration.
So if it is detuned by $\Delta$ from the transition $|1\rangle\rightarrow |3\rangle$ it has to be detuned by $\Delta+\omega_0$ from the transition $|2\rangle\rightarrow |4\rangle$. 
If $\Delta>0$ and $\Delta\gg \Gamma$, it directly implies a steady state for the population with a large amount of the atoms in state $|2\rangle$.\\
An interesting approach, to understand the quantum correlations which appears between the probe and conjugate is to think of four-wave-mixing as a DLCZ memory protocol \cite{duan2001long}.
Indeed, one starts with all the population in $|2\rangle$.
Sometimes (not often because of the large detuning) a $|2\rangle\rightarrow |4\rangle$ pump photon will \textit{write} his phase in the atomic coherence and induce the emission of an anti-Stokes (conjugate) photon.
In the DLCZ \cite{duan2001long} language, when this anti-Stokes (conjugate) photon is detected it implies that the memory has been loaded.
After a given time, the memory can be read (efficiently due to small detuning) by a pump photon on the $|1\rangle\rightarrow |3\rangle$ transition.
This process is accompanied by the emission of a probe photon, in a coherent manner (as describe in section \ref{collective}).\\

In chapter \ref{chap:3}, I use this technique to generate entangled pulses of light between the probe and the conjugate \cite{agha2011time}.
I also use another property of this system: to provide small group velocity and anomalous dispersion, in order to observe slow and fast light \cite{clark2014quantum}.

\subsection{Slow and fast light}
We begin by applying the curl operator to the Maxwell equations to obtain the Helmholtz equation
\begin{equation}
    (\nabla^2E+k^2)\mathbf{E}=0,
\end{equation}
where we have $k=\mu_0\epsilon_0\epsilon_r\omega^2$.
As usual we assume the beam to propagate along the $+z$ direction and we ignore polarization.
Solutions of the Helmholtz equation that fulfills these conditions are:
\begin{equation}
    E(z,t)=E_0(z)e^{i(kz-\omega t)}.
\end{equation}
The phase velocity is defined as the velocity at which the phase of this solution moves:
\begin{equation}
    v_{\phi}=\frac{\omega}{k}
\end{equation}
By replacing $k$ with its definition, it can be rewritten with the speed of light in vacuum $c$ and the index of refraction $n$:
\begin{equation}
    v_{\phi}=\frac{\omega}{\sqrt{\mu_0\epsilon_0\epsilon_r(\omega)\omega^2}}=\frac{c}{n(\omega)}.
\end{equation}
This is a well known result but it hides in the dependency on $\omega$ of $\epsilon_r(\omega)$ that different frequency will propagate at different velocities.
It has no consequence for monochromatic waves, but it implies that a pulse will distort while propagating in a dispersive media.
Slow and fast light terminology comes from this effect: a light pulse (basically a wave-packet) that  propagates slower than $c$ will be qualified as slow-light and reciprocally if it does propagate faster than $c$ it will be qualified as fast-light \cite{bigelow2003superluminal,milonni2004fast}.
Let us precise this terminology.\\

A wavepacket has the general form:
\begin{equation}
    E(z,t)=\int E_0(k)e^{i(kz-\omega(k) t) }\text{dk},
\end{equation}
where $E_0(k)$ is the Fourier transform of $E_0(z)$ at $t=0$. 
If the spectrum $E_0(k)$ is sufficiently narrowband (i.e. the pulse is not too short), we can call the central value $k_c$ and Taylor expand $\omega$ around $k_c$:
\begin{equation}\label{eq144}
    \omega(k)\approx\omega_c+\left.\frac{\partial \omega}{\partial k} \right\rvert_{k=k_c}(k-k_c).
\end{equation}
Using this approximation we can write:
\begin{equation}
    E(z,t)\approx  E\left(z-t\left.\frac{\partial \omega}{\partial k} \right\rvert_{k_c},0\right)\times e^{i(k_c\left.\frac{\partial \omega}{\partial k} \right\rvert_{k_c}-\omega_c)t}.
\end{equation}\label{eq145}
In this simple results we can see that the pulse will propagate largely undistorted (up to an overall factor phase and as long as Eq. \ref{eq144} is a good approximation) at the group velocity given by: 
\begin{equation}
    v_g=\left.\frac{\partial \omega}{\partial k} \right\rvert_{k_c}.
\end{equation}

As derived previously about 2-level atoms, it is often more convenient to express the group velocity as a function of the variation of $n$ with frequency.
We have 
\begin{eqnarray}
\nonumber \frac{\partial \omega}{\partial k}&=& \frac{c}{n}-\frac{ck}{n^2}\frac{\partial n}{\partial k}\\
&=&\frac{c}{n}-\frac{\omega}{n}\frac{\partial \omega}{\partial k}\frac{\partial n}{\partial \omega}.
\end{eqnarray}
We can then write the group velocity:
\begin{equation}
    v_g=\frac{c}{n(\omega_c)+\omega_c \left.\frac{\partial n}{\partial \omega}\right\rvert_{\omega_c}}.
\end{equation}
This equation gives us direct access of the group velocity if we know $n(\omega_c)$ (which is the case now that we master optical Bloch equations).
The group velocity is then given by the speed of light in vacuum divided by a term that includes both the index of refraction at the carrier frequency and the derivative of the index around the carrier frequency.
The denominator is commonly called the group index $n_g$ and it can take values larger or smaller than unity \cite{boyd2001slow}.

In the vast majority of dielectric media, far away from resonance, $\left.\frac{\partial n}{\partial \omega}\right\rvert_{\omega_c}$ is usually positive and $n_g>1$.
However, it is possible (using EIT for example) to obtain $\left.\frac{\partial n}{\partial \omega}\right\rvert_{\omega_c}<0$.
This type of medium is said to have \textit{anomalous dispersion}.
For sufficiently large negative value,  it is possible to reach $n_g<1$: a negative group velocity. 
I report on the use of this type of medium in chapter \ref{chap:3}.

\begin{figure}[h!]
    \centering
    \includegraphics[width=0.9\textwidth]{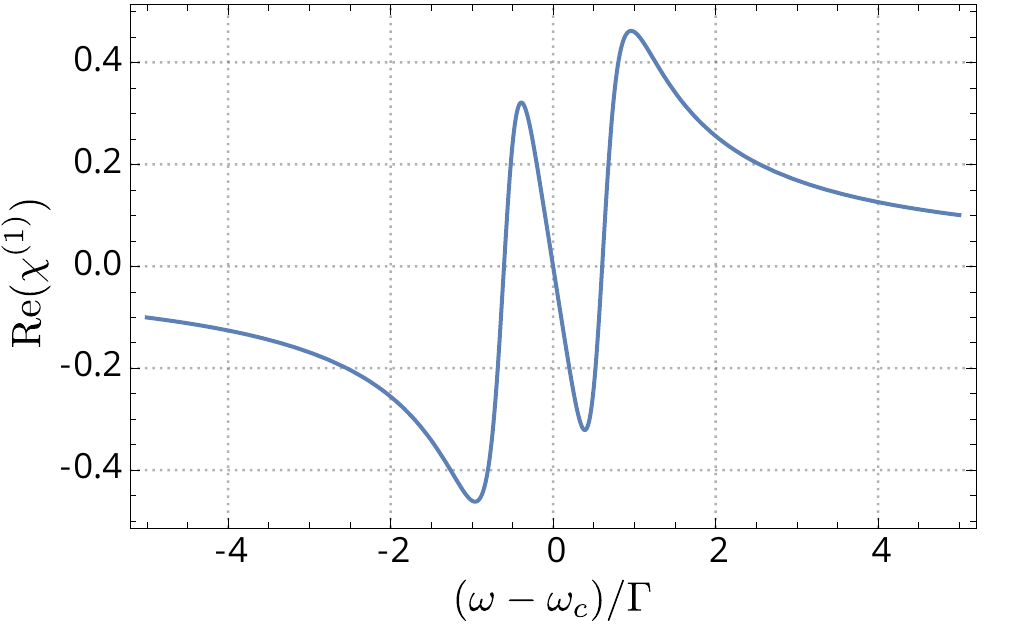}
    \caption{Real part of the the linear refractive index for the EIT configuration described in section \ref{sec:EIT}. We can see that around $\omega-\omega_c=0$ the slope of $\frac{\partial n}{\partial \omega}$ is large and negative. This is a perfect place of observe a negative group velocity.}
    \label{fig:my_label}
\end{figure}

\chapter{Multimode optical memory}
List of publications related to this chapter :
\begin{itemize}
    \item \href{https://doi.org/10.1364/OE.20.012350}{\textbf{Temporally multiplexed storage of images in a gradient echo memory}}.\\
\textbf{Q. Glorieux}, J. B. Clark, A. M. Marino,
Z. Zhou, and P. D. Lett.\\
\href{https://doi.org/10.1364/OE.20.012350}{ Optics Express \textbf{20}, 12350 (2012)}
 
     \item    \href{https://doi.org/10.1088/1367-2630/15/8/085027}{\textbf{Gradient echo memory in an ultra-high optical depth cold atomic ensemble}}.\\
    B. M. Sparkes, J. Bernu, M. Hosseini, J. Geng, \textbf{Q. Glorieux}, P. A. Altin, P. K. Lam, N. P. Robins, and B. C. Buchler. \\
     \href{https://doi.org/10.1088/1367-2630/15/8/085027}{New Journal of Physics \textbf{15}, 085027 (2013)  }

    \item \href{https://doi.org/10.1088/1367-2630/15/3/035005}{\textbf{Spatially addressable readout and erasure of an image in a gradient echo memory}}.\\
J. B. Clark, \textbf{Q. Glorieux} and P. D. Lett.\\
\href{https://doi.org/10.1088/1367-2630/15/3/035005}{New Journal of Physics \textbf{15}, 035005 (2013) }\\
Not included.

    \item \href{https://doi.org/10.1088/1742-6596/467/1/012009}{\textbf{An ultra-high optical depth cold atomic ensemble for quantum memories}}.\\
    B. M. Sparkes, J. Bernu, M. Hosseini, J. Geng, \textbf{Q. Glorieux}, P. A. Altin, P. K. Lam, N. P. Robins, and B. C. Buchler. \\ \href{https://doi.org/10.1088/1742-6596/467/1/012009}{Journal of Physics, \textbf{467}, 012009 (2013)} \\
    Not included.

\end{itemize}
\clearpage

\section{Gradient Echo Memory}

Light in vacuum propagates at $c$.
This statement is not only a fundamental principle of physics it is also the foundation of the definition of the meter in the international system of units.
Delaying \cite{hau1999light}, storing \cite{kozhekin2000quantum} or advancing \cite{boyd2009controlling} light are therefore only possible in a medium (i.e. not in vacuum).
In the next two chapters, I will show how to play with the speed of light in a optically dense atomic medium \cite{khurgin2008slow}.\\

The main motivation to delay or store photons in matter is the need to synchronize light-based communication protocols.
Quantum technologies promise an intrinsically secure network of long distance quantum communication.
However for a realistic implementation, the \textit{quantum internet} will need   quantum repeaters in order to compensate losses in long distance channels \cite{kimble2008quantum}.
The core element of a quantum repeater is a quantum memory \cite{lvovsky2009optical} which can store and release photons, coherently and on demand.
This chapter is covering this topic with the presentation of an important memory protocol: the gradient echo memory and two implementations of this protocol one in a warm vapor \cite{glorieux2012temporally} and the second in a cold atomic cloud \cite{sparkes2013gradient}.

\begin{figure}[h]
    \centering
    \includegraphics[width=1.05\textwidth]{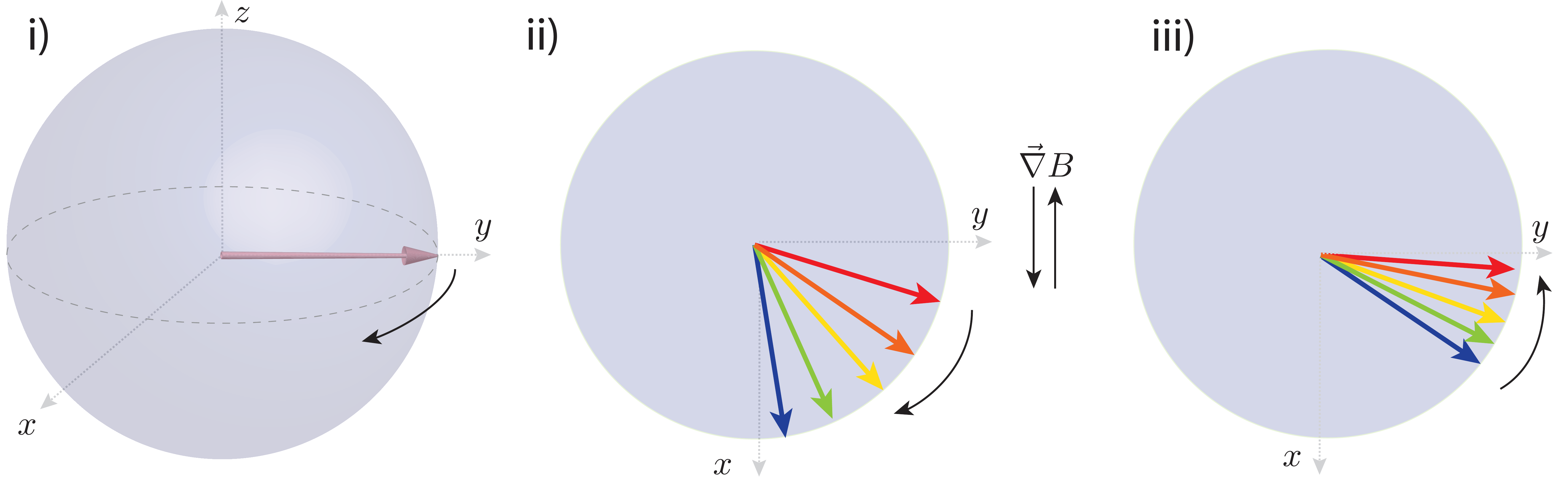}
    \caption{GEM protocol. i) Bloch sphere with the state initialized on the equator. ii) and iii) are views from top of the equatorial plan. ii) Dephasing with the gradient in one direction. iii) Rephasing with the gradient in the other direction.   The dephasing is shown in the rotating frame. }
    \label{figGem}
\end{figure}

If you are somewhat familiar with MRI, understanding qualitatively gradient echo memory (GEM) is  straightforward.
Imagine you want to encode quantum information in a light.
Various approaches are available (from polarization encoding, time-bin, orbital angular momentum...) but for qubit encoding (i.e. an Hilbert space of dimension 2), you always can map your information into a spin 1/2 system and represent it on the Bloch sphere.
Let's say your information is of the generic form $|\psi\rangle=\frac{1}{\sqrt2}\left(|0\rangle +e^{i\phi}|1\rangle\right)$, this means on the equator of the Bloch sphere with a latitude  $\phi$ as shown on Fig. \ref{figGem} i).
In GEM, during the storage process one applies a spatially dependant energy shift in order to enlarge the transition \cite{hetet2008electro,hetet2008photon}.
This technique enables the storage of photons with a broad spectrum (i.e. short in time).

Because $|\psi\rangle$ is not the energy ground state, the state will start to rotate in the equatorial plane of the Bloch sphere.
However, due to the spatially dependant energy shift, each frequency component will rotate at a different speed and dephase as shown in figure \ref{figGem} ii) 
The trick used in gradient echo techniques is to reverse the time evolution of the dephasing process.
By simply inverting the energy gradient after the evolution time $T$, we can see on figure \ref{figGem} iii) that the frequency components will start to precess in the opposite direction with opposite speed\footnote{This is true if not only the energy gradient is switched but also the sign of the energy. If only the energy gradient changes sign, the components will still precess in the same direction but the slowest components will become the fastest.}. 
At a time $2T$, all frequency components will be in phase again and the collective enhancement described in chapter 1 can occur.

Various approaches have been tested to produce  a spatially varying energy shift.
The two most successful techniques are using an AC-Stark shift in rare-earth crystal and using a Zeeman shift on atomic vapors by applying a magnetic field gradient.
In this chapter I  describe the Zeeman shift implementation in warm and cold atomic ensembles \cite{hosseini2011high,glorieux2012temporally,sparkes2013gradient}.

\subsection{Storage time and decoherence}
I should now describe  how to convert coherent light excitation into a matter excitation.
In the two-level-atom configuration, a quantum field $\hat{\mathcal{E}}(z,t)$ (carrying the information) is sent into the atomic vapor with ground state $|g\rangle$ and excited state $|e\rangle$. 

Here, we are focusing on the propagation of a light pulse into a medium, therefore the time derivative terms in the Maxwell-Bloch equations have to be conserved, but the transverse gradient is neglected\footnote{The opposite approximation is done in the Chapter \ref{chap:4}, where the time evolution is not considered but the diffraction in the transverse plan is.}.
Using the rotating wave approximation described in chapter 1, and in the presence of a spatially varying magnetic field, the evolution equations are given by:
\begin{eqnarray}\label{eq:coherence2}
\nonumber \frac{\partial}{\partial t}\hat{\sigma}_{ge}(z,t)&=&[-\Gamma+i\eta(t)z]\hat{\sigma}_{ge}(z,t)-ig\hat{\mathcal{E}}(z,t)\\
\left(\frac{\partial}{\partial t}+c\frac{\partial}{\partial z}\right) \hat{\mathcal{E}}(z,t)&=&igN\hat{\sigma}_{ge}(z,t).
\end{eqnarray}
We have introduced the term $i\eta(t)z$ which represents the energy shift form the spatially varying magnetic field. 
$\eta(t)$ is the gradient slope that can be inverted (or modified in a more complex manner) at the desired time.
From this coupled equations it is straightforward to point out the problem of the two-level-atom configuration.
The information is transferred from the field to the atomic coherence $\hat{\sigma}_{ge}(z,t)$. However this coherence will decay with the decay rate of the excited state $\Gamma$.
This is a terrible problem for a quantum memory !
The good news is: this problem can be solved using a third level.\\

\begin{figure}
    \centering
    \includegraphics[width=\textwidth]{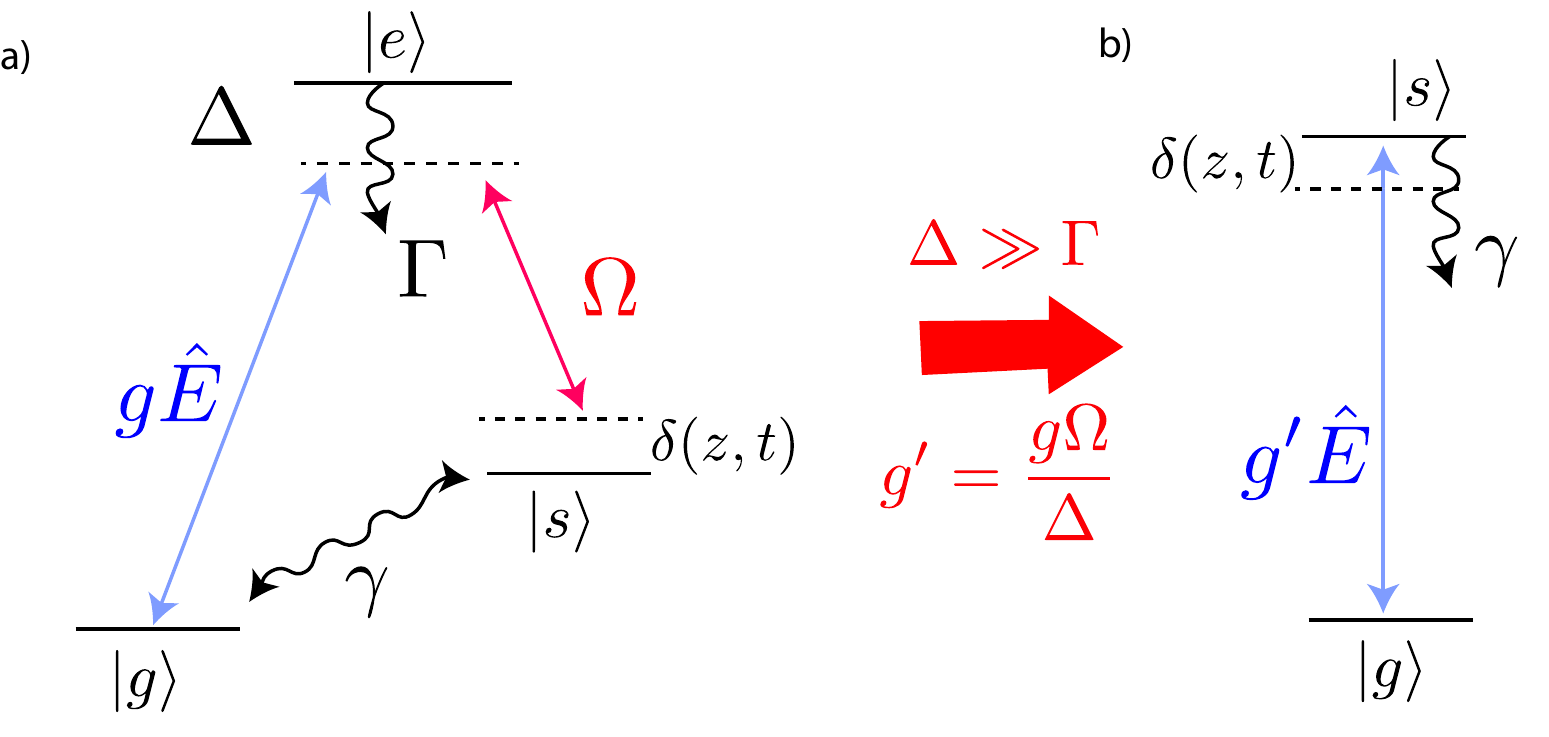}
    \caption{Adiabatic elimination notation summary.}
    \label{fig:adiabatic}
\end{figure}

The solution comes from mapping the photonic excitation into the ground state coherence of a three-level atom as discussed in section \ref{sec:EIT}.
Using a two-photon detuning equal to zero and a large one photon detuning compared to $\Gamma$, it is possible to apply the adiabatic elimination of the excited state and rewrite the evolution equation for the ground state coherence in a way similar to equations~\ref{eq:coherence2}:
\begin{eqnarray}\label{eq:coherence3}
\nonumber \frac{\partial}{\partial t}\hat{\sigma}_{gs}(z,t)&=&[-\gamma+i\eta(t)z-i\frac{\Omega^2}{\Delta}]\hat{\sigma}_{gs}(z,t)-i\frac{g\Omega}{\Delta}\hat{\mathcal{E}}(z,t)\\
\left(\frac{\partial}{\partial t}+c\frac{\partial}{\partial z}\right) \hat{\mathcal{E}}(z,t)&=&i\frac{g^2N}{\Delta}\hat{\mathcal{E}}(z,t)+i\frac{gN\Omega}{\Delta}\hat{\sigma}_{gs}(z,t).
\end{eqnarray}
We apply a transformation to this equations by going into the rotating frame at frequency $\frac{g^2N}{\Delta}$.
Moreover we neglect the ac-Stark shift induced by  the control field and we finally obtain \cite{hosseini2011high,clark2014timing}:
\begin{eqnarray}\label{eq:coherence3b}
\nonumber \frac{\partial}{\partial t}\hat{\sigma}_{gs}&=&[-\gamma+i\eta(t)z]\hat{\sigma}_{gs}-i\frac{g\Omega}{\Delta}\hat{\mathcal{E}}\\
\left(\frac{\partial}{\partial t}+c\frac{\partial}{\partial z}\right) \hat{\mathcal{E}}&=&i\frac{g \Omega N}{\Delta}\hat{\sigma}_{gs}.
\end{eqnarray}
We found an equation similar to Eq. \ref{eq:coherence2} but where the decoherence $\Gamma$ has been replaced by $\gamma$ which is the long-lived coherence between the two ground states. 
This is a major improvement as the ground state coherent is immune from spontaneous emission, and therefore much longer memory time can be envisioned (and demonstrated actually \cite{steger2012quantum}).
Another important improvement is obtained by going to three levels and modifying the coupling $g$ to $\frac{g \Omega}{\Delta}$, which can be modified and tuned with the Rabi frequency of the control field.

A final transformation \cite{clark2014quantum} is made to the equations \ref{eq:coherence3b} by changing  $z$ to the retarded frame $z'=z+ct$:
\begin{eqnarray}\label{eq:coherence3c}
\nonumber \frac{\partial}{\partial t}\hat{\sigma}_{gs}&=&[-\gamma+i\eta(t)z]\hat{\sigma}_{gs}-i\frac{g\Omega}{\Delta}\hat{\mathcal{E}}\\
\frac{\partial}{\partial z'} \hat{\mathcal{E}}&=&iN\frac{g\Omega }{\Delta}\hat{\sigma}_{gs}.
\end{eqnarray}

\subsection{GEM polaritons}
Equations \ref{eq:coherence3c} can be solved numerically but new insights emerge when writing them in the picture of normal modes called GEM-polaritons \cite{hetet2008photon,hosseini2012quantum}.
The polariton picture for light-matter interaction in a 3-level system has been pioneered by Lukin and Fleischhauer with the introduction of the dark-state polariton \cite{fleischhauer2002quantum}.
This pseudo particle is a mixed excitation of light and matter (atomic coherence) where the relative weight can be manipulated with the external control field.
This is a boson-like particle which propagates with no dispersion and a tunable group velocity.
It is possible to consider a similar approach for GEM.
Neglecting the decoherence and taking the spatial Fourier transform of Eq. \ref{eq:coherence3c}, we can write \cite{hosseini2012quantum}:
\begin{eqnarray}\label{eq:coherence3d}
\nonumber \frac{\partial}{\partial t}\hat{\sigma}_{gs}(t,k)&=&-\eta\frac{\partial}{\partial k}\hat{\sigma}_{gs}(t,k)+i\frac{g\Omega}{\Delta}\hat{\mathcal{E}}(t,k)\\
k \hat{\mathcal{E}}(t,k)&=&N\frac{g\Omega }{\Delta}\hat{\sigma}_{gs}(t,k).
\end{eqnarray}
In analogy to \cite{fleischhauer2002quantum}, we consider the polariton-like operator in k-space:
\begin{equation}
    \hat{\psi}(k,t)=k \hat{\mathcal{E}}(t,k)+N\frac{g\Omega }{\Delta}\hat{\sigma}_{gs}(t,k).
\end{equation}
This particle will follow an equation of motion:
\begin{equation}
    \left(\frac{\partial}{\partial t}+\eta(t)\frac{\partial}{\partial k}-iN\frac{g^2\Omega^2}{\Delta^2k}\right)\hat{\psi}(k,t)=0
\end{equation}
This equation indicates that $|\psi(k,t)|$ propagates un-distorted along the $k$-axis with a speed defined by the gradient $\eta$.
We are here in front of a strange polariton.

The first difference with dark-state-polariton is that it does propagate in the $k$-plane instead of the normal space.

The second difference, maybe more fundamental is that the weight of one of the two components depends on $k$.
If the polariton is normalized (and it has to be, in order to fulfill the bosonic commutation relation), lower spatial frequencies (short $k$) will be more an atomic excitation and higher spatial frequencies will be more a field excitation.
We see here the limit of the analogy with the dark-state polariton.
The GEM polariton picture is mainly useful to understand that the velocity of the pseudo particle is given by $\eta(t)$ and therefore will change sign when $\eta(t)$ will be flipped.
We have here another interpretation of the rephasing process explained with the Bloch sphere, earlier in the chapter.

\subsection{GEM in rubidium vapor}
\begin{figure}
    \centering
    \includegraphics[width=\textwidth]{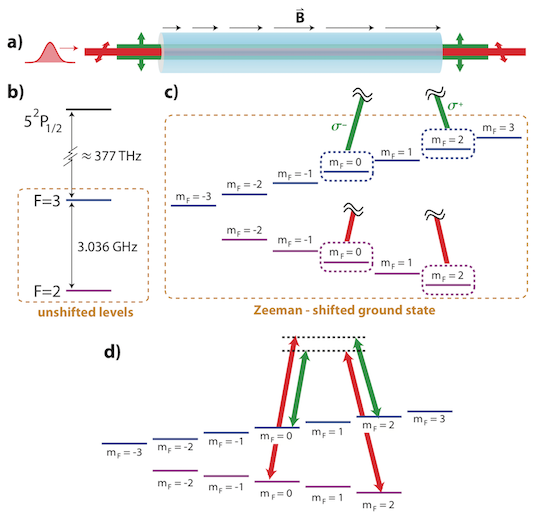}
    \caption{a) Orthogonal linearly polarized Raman beams used in the $\Lambda$ GEM experiments presented in this thesis. 
    When propagating along the direction of the magnetic field, a linear polarization can be viewed as a superposition of $\sigma+$ and $\sigma-$ polarizations. We therefore simultaneously drive two Raman transitions in any given echo experiment. 
    b) An unshifted hyperfine structure for the D1 transition in 85Rb. 
    c) A weak bias magnetic field to Zeeman shift the hyperfine structure of the ground state. In the weak field limit, the $m_F$ levels are linearly and oppositely shifted in the F = 2 and F = 3 manifolds.
    For the   GEM protocol, we couple $m_F = 2$ and $m_F = 0$ magnetic sublevels. 
    d) Complete depiction of the two Raman transitions. 
    Figure and caption are extracted from the J. Clark thesis \cite{clark2014quantum}, whom I supervised at NIST.}
    \label{gemCLARK}
\end{figure}
I give here a short description of the implementation of GEM in a hot Rubidium 85 vapor.
More details are given in the article: \textit{Temporally multiplexed storage of images in a gradient echo memory}, attached to this work \cite{glorieux2012temporally}.
The first task is to isolate a pair of ground states that can be individually addressed by light fields and whose energy splitting can be tuned by a linear Zeeman shift.
This is done by applying a bias magnetic field (about 50~G) to lift the degeneracy of the $m_F$ sub-levels for  F=2 and F=3 ground states (see figure \ref{gemCLARK}).\\
\noindent The second step is to apply a gradient magnetic field along the atomic vapor (a 20~cm long cell) to control the inhomogeneous broadening of the two-photon Raman transition.
The amplitude of this gradient field must be smaller than the bias field and large enough to accept the full bandwidth of the light pulse we expect to store.
Typically, in rubidium a rule of thumb is a Zeeman shift of 1~MHz per Gauss. So a gradient with 1 Gauss amplitude already allows for the storage of 1~microsecond Gaussian pulses.\\
An important part of the development of such setup consists in the precise control of the gradient and the ability to reverse it quickly.
Indeed in the echo description we have done so far, the inversion of the gradient is supposed to be instantaneous. It is clear that switching a large current in an inductor instantaneously is a tricky issue.
More details on the methods used are given in the article \textit{Gradient echo memory in an ultra-high optical depth cold atomic ensemble} attached to this work \cite{sparkes2013gradient}.\\

Finally, an important reason why GEM is a powerful tool for storing and retrieving light is that it is intrinsically temporally multimode: one can store multiple pulses at different times inside the memory and retrieve them independently \cite{hetet2008multimodal,glorieux2012temporally,clark2013spatially}.
But it is also spatially multimode, so images can be stored in the transverse plane (only limited by decoherence due to atomic diffusion).
We have leveraged these two advantages to demonstrate the first storage of a "movie" inside an atomic memory \cite{glorieux2012temporally}. Results are presented in the next pages.

\subsection{Press releases}
These results led to multiple press articles covering this topic. I reproduce here some excerpts.\\

\textit{The sequence of images that constitute Hollywood movies can be stored handily on solid-state media such as magnetic tape or compact diskettes. At the Joint Quantum Institute images can be stored in something as insubstantial as a gas of rubidium atoms. No one expects a vapor to compete with a solid in terms of density of storage. But when the "images" being stored are part of a quantum movie --the coherent sequential input to or output from a quantum computer --then the pinpoint control possible with vapor will be essential.} \textbf{Phys.org}\\

\textit{One of the enabling technologies for a quantum internet is the ability to store and retrieve quantum information in a reliable and repeatable way. 
One of the more promising ways to do this involves photons and clouds of rubidium gas. Rubidium atoms have an interesting property in that a magnetic field causes their electronic energy levels to split, creating a multitude of new levels.} \textbf{MIT Technology Review}\\

\textit{Having stored one image (the letter N), the JQI physicists then stored a second image, the letter T, before reading both letters back in quick succession. The two "frames" of this movie, about a microsecond apart, were played back successfully every time, although typically only about 8 percent of the original light was redeemed, a percentage that will improve with practice. According to Paul Lett, one of the great challenges in storing images this way is to keep the atoms embodying the image from diffusing away. The longer the storage time (measured so far to be about 20 microseconds) the more diffusion occurs. The result is a fuzzy image.} \textbf{Science Daily.}

\chapter*{Article 1: Temporally multiplexed storage of images in a gradient echo memory}

\section*{Preprint}
\noindent The preprint version of this article is accessible on \href{https://arxiv.org/abs/1205.1495}{ arXiv:1205.1495}.

\section*{Published version}
\href{https://doi.org/10.1364/OE.20.012350}{\textbf{Temporally multiplexed storage of images in a gradient echo memory}}.\\
\textbf{Q. Glorieux}, J. B. Clark, A. M. Marino,
Z. Zhou, and P. D. Lett.\\
\href{https://doi.org/10.1364/OE.20.012350}{ Optics Express \textbf{20}, 12350 (2012)}\\

\addcontentsline{toc}{section}{\textbf{Article 1}: Temporally multiplexed storage of images in a gradient echo memory. \textbf{Optics Express 20}, 12350 (2012)}
\ifpapers
\fi
\clearpage

\section{GEM in a cold atom ensemble}\label{coldatoms}
Shifting from warm atomic vapor (with the simplicity of high and tunable density) to laser-cooled atoms trapped in a magneto-optical trap (MOT) has many advantages.
The first of them being the reduction of  decoherence due to atomic motion.
However, in order to obtain the large optical depth required for an efficient gradient echo memory, a substantial optimization work has been needed.
In the paper attached to this work, I highlight the various techniques I developed and used during my stay at the Australian National University under the supervision of Ping Koy Lam.\\

For quantum memory applications we aim to reach low temperatures and the very large optical depths (OD) required for high efficiency.
Because OD is related to the integrated absorption of photons through a sample, there are a number of ways to increase the OD. 
These include: increase the atom number (for instance by increasing the length of the cloud) or increase the density.
Low temperature is also important as the thermal diffusion of atoms is a limiting factor for the memory lifetime.\\

In the attached paper, I present a MOT that achieves a peak OD of over 1000 at a temperature of 200~$\mu$K. 
We obtain this result by combining three techniques:
\begin{itemize}
    \item The optimization of the static loading of the MOT through geometry.
    The optimal shape for our atomic ensemble is a cylinder along the direction of the memory beams to allow for maximum absorption of the probe. This can be achieved by using rectangular, rather than circular, quadrupole coils or using a 2D MOT configuration.
    \item The use of a spatial dark spot. 
    The density in the trapped atomic state  is limited by re-absorption of fluorescence photons within the MOT (leading to an effective outwards radiation pressure). 
    By placing a dark spot of approximately 7.5 mm in diameter in the repump, atoms at the centre of the MOT are quickly pumped into the lower ground state and become immune to this unwanted effect, allowing for a higher density of atoms in the centre of the trap, as first demonstrated in \cite{ketterle1993high}.
    \item The use of optical de-pumping, followed by a compression phase with a temporal dark spot \cite{depue2000transient}.
\end{itemize}

With the setup built using these techniques, we have implemented the gradient echo memory scheme on the D1 line of Rubidium. 
The results shown here demonstrate a memory efficiency of 80~$\pm$~2\% and a coherence time up to 195~$\mu$s.
This coherence time is a factor of eight greater than  GEM experiments done in atomic vapor (described in the previous section).

\chapter*{Article 2: Gradient echo memory in an ultra-high optical depth cold atomic ensemble}

\section*{Preprint}
\noindent The preprint version of this article is accessible on \href{https://arxiv.org/abs/1211.7171}{  arXiv:1211.7171}.

\section*{Published version}

 \href{https://doi.org/10.1088/1367-2630/15/8/085027}{\textbf{Gradient echo memory in an ultra-high optical depth cold atomic ensemble}}.\\
    B. M. Sparkes, J. Bernu, M. Hosseini, J. Geng, \textbf{Q. Glorieux}, P. A. Altin, P. K. Lam, N. P. Robins, and B. C. Buchler. \\
     \href{https://doi.org/10.1088/1367-2630/15/8/085027}{New Journal of Physics \textbf{15}, 085027 (2013)  }

\addcontentsline{toc}{section}{\textbf{Article 2}: Gradient echo memory in an ultra-high optical depth cold atomic ensemble. \textbf{NJP 15}, 085027 (2013) }
\ifpapers
\fi

\chapter[Quantum optics in fast-light media]{Quantum optics in anomalous dispersion media}\label{chap:3}

List of publications related to this chapter :
\begin{itemize}

    \item    \href{https://doi.org/10.1364/OE.20.017050}{\textbf{Imaging using quantum noise properties of light}}.\\
    J. B. Clark, Z. Zhou, \textbf{Q. Glorieux}, A. M. Marino and P. D. Lett.\\
     \href{https://doi.org/10.1364/OE.20.017050}{Optics Express \textbf{20}, 17050 (2012)}
     
             \item  \href{https://doi.org/10.1038/NPHOTON.2014.112}{\textbf{Quantum mutual information of an entangled state propagating through a fast-light medium}}.\\
      J. B. Clark, R. T. Glasser, \textbf{Q. Glorieux}, U. Vogl, T. Li, K. M. Jones, and P. D. Lett.
     \href{https://doi.org/10.1038/NPHOTON.2014.112}{Nature Photonics \textbf{14}, 123024 (2014)}
     
          \item  \href{https://doi.org/10.1088/1367-2630/14/12/123024}{\textbf{Generation of pulsed bipartite entanglement using four-wave mixing}}.\\
      \textbf{Q. Glorieux}, J. B. Clark, N. V. Corzo, and P. D. Lett.\\
     \href{https://doi.org/10.1088/1367-2630/14/12/123024}{New Journal of Physics \textbf{14}, 123024 (2012)}. Not included.
     
      \item    \href{https://doi.org/10.1140/epjd/e2012-30037-1}{\textbf{Extracting spatial information from noise measurements of multi-spatial-mode quantum states}}.\\
    A. M. Marino, J. B. Clark,  \textbf{Q. Glorieux}  and P. D. Lett.\\
     \href{https://doi.org/10.1140/epjd/e2012-30037-1}{European Physical Journal D \textbf{66}, 288 (2012)}.  Not included.

       \item  \href{https://doi.org/10.1103/PhysRevA.88.043836}{\textbf{Rotation of the noise ellipse for squeezed vacuum light generated via four-wave mixing}}.\\
        N. V. Corzo,  \textbf{Q. Glorieux}, A. M. Marino, J. B. Clark, R. T. Glasser and P. D. Lett.
         \href{https://doi.org/10.1103/PhysRevA.88.043836}{Physical Review A \textbf{88}, 043836 (2013)}.  Not included.
         
         \item  \href{https://doi.org/10.1103/PhysRevA.87.010101}{\textbf{Experimental characterization of Gaussian quantum discord generated by four-wave mixing}}.\\
      U. Vogl, R. T. Glasser, \textbf{Q. Glorieux}, J. B. Clark, N. V. Corzo, and P. D. Lett.
     \href{https://doi.org/10.1103/PhysRevA.87.010101}{New Journal of Physics \textbf{87}, 010101 (2013)}. Not included.
     
        \item  \href{https://doi.org/10.1088/1367-2630/16/1/013011}{\textbf{Advanced quantum noise correlations}}.\\
      U. Vogl, R. T. Glasser, J. B. Clark, \textbf{Q. Glorieux},  T. Li, N. V. Corzo, and P. D. Lett.
     \href{https://doi.org/10.1088/1367-2630/16/1/013011}{New Journal of Physics \textbf{16}, 013011 (2014)}. Not included.

\end{itemize}

\section*{Introduction}
A large part of my work at NIST-JQI has been devoted to the study and use of noise properties of entangled beams generated using four-wave-mixing in warm atomic vapor.
It is not possible to describe here all these experiments but I will highlight two results that are valuable in relation to the other works presented in this manuscript.\\

We have seen in the previous chapter that GEM has the ability to store multiple spatial modes in the transverse direction.
In the first experiment that I describe in this chapter, we used the multimode properties of to quantum noise  \cite{Marino2012}   to be able to identify an object with a higher precision than would normally be achieved with a typical laser.\\

The second result I will comment concerns the study of the opposite process of a quantum memory: advancing quantum information \cite{boyd2009controlling}.
We shown that under certain conditions, an atomic vapor can exhibit an anomalous dispersion and give rise to a group velocity larger than $c$.
In the context of quantum communication, these results can be shocking as it is clear that no signal can travel faster than light \cite{schutz2009first}. 
However, we demonstrated the role of quantum noise to erase all information while  velocity becomes larger than $c$.

\section{Imaging with the noise of light}
Imagine that you have an object (an intensity or a phase mask) that you want to image but you are also limited in the number of photons you can send onto this object.
We propose two techniques (one \textit{classical} using thermal noise and one \textit{quantum} using quantum correlations) and compare the uncertainty in shape estimation between both.
To simplify the comparison we restricted the shape estimation to a 1D problem.\\
The resource needed for this experiment is a pair of squeezed vacuum beams generated using four-wave-mixing in a atomic vapor.
These beams are too weak to be measured directly with an amplified photodiode and need to be homodyned with local oscillators to be detected.
Taken independently both beams (also known as probe and conjugate) exhibit an excess noise compared to a coherent state.
They are indeed thermal states.
However, when the intensity difference or the phase sum is analyzed, we observed a noise below the shot noise indicating quantum correlations.\\
The parameter we are trying to estimate is the angle of a bow-tie mask (1D problem) placed on the conjugate path. To do so the local oscillator (LO) is designed with a bow-tie shape of same size and same center as the mask, using a spatial light modulator. However the orientation of the mask is unknown.\\

The \textit{classical} technique consists in overlapping the bow-tie shaped LO with the conjugate beam (after the mask) and measure the detected noise.
The LO is then rotated and when the noise is maximized, the search is finished and the LO angle is an estimation of the mask angle.
Indeed the fraction of the conjugate beam which passes through the mask contains spatial modes with extra-thermal noise compared to vacuum. 
In the limiting case of zero overlap between the LO and the mask, the LO will just beat with vacuum and detect shot noise because no light coming from the conjugate beams interfere with it.\\

The \textit{quantum} technique makes use of the second part of the entangled pair (the probe beam), which surprisingly does not have to interact directly with the mask.
For this measurement, we simply added a second LO with the same angle to beat with the probe beam and then measured the quadrature that minimized the detected noise.
This time, the search is terminated when the quadrature noise is minimal.
In the paper \textit{Imaging using the quantum noise properties of light}, we have demonstrated that the precision in the estimation of the mask angle is enhanced by a factor 6 with the quantum technique compared to the classical one, even though the second beam does not interact directly with the mask.\\

\begin{figure}[h]
    \centering
    \includegraphics[width=1\textwidth]{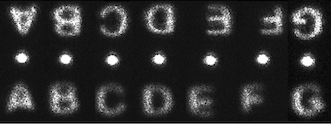}
    \caption{Example of probe and conjugate LO shape created using a spatial light modulator. The intense dot in the middle is the (filtered) pump beam. Each letter is an independent image and they are collated for presentation.}
    \label{fig:alphabet}
\end{figure}

The second experiment described in this paper is a direct application of this principle to the field of pattern recognition. 
We start by defining a pattern set (the alphabet letters).
We are able to imprint this set on the local oscillator with a spatial light modulator as shown in figure \ref{fig:alphabet}.
An unknown mask from the set is placed in the path of the conjugate beam and we compare the two techniques to see if pattern recognition is enhanced using quantum correlations.\\

On the presented results (mask of the letter Z) by taking into account the uncertainty on the noise measurements it is not possible to conclude on the pattern in the classical case because more than one pattern noise measurement lie within one standard deviation.
On the quantum case however, the letter Z is identified with more than 7 standard deviation certainty.\\

\chapter*{Article 3: Imaging using quantum noise properties of light}

\section*{Preprint}
\noindent The preprint version of this article is accessible on \href{https://arxiv.org/abs/1207.1713}{arXiv:1207.1713}.

\section*{Published version}

   \href{https://doi.org/10.1364/OE.20.017050}{\textbf{Imaging using quantum noise properties of light}}.\\
    J. B. Clark, Z. Zhou, \textbf{Q. Glorieux}, A. M. Marino and P. D. Lett.\\
     \href{https://doi.org/10.1364/OE.20.017050}{Optics Express \textbf{20}, 17050 (2012)}

    \addcontentsline{toc}{section}{\textbf{Article 3}: Imaging using the quantum noise properties of light. \textbf{Optics Express 20}, 17050 (2012)}
\ifpapers
\fi

\clearpage

\section[Quantum noise and fast-light medium]{How quantum noise affects the propagation of information. }

\begin{figure}[h]
    \centering
    \includegraphics[width=0.85\textwidth]{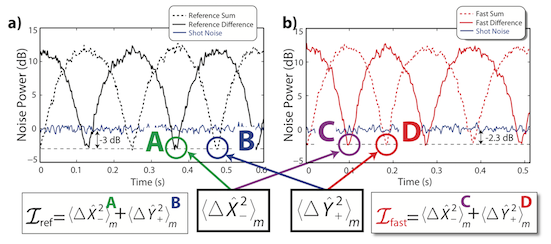}
    \caption{Determination of inseparability $\mathcal{I}$ using a spectrum analyzer.
    Part a) corresponds to the reference case while b) corresponds to data taken with fast light. 
    The minimum noise power of the homodyne difference signal corresponds to squeezing of the $\hat X$ quadrature difference (i.e. $\langle \hat X_-^2\rangle$) while the minimum noise power of the homodyne sum gives the squeezing of phase sum $\langle \hat Y_+^2\rangle$.
    With the fast-light cell inserted into the path of the conjugate, we were able to see $\mathcal{I}<2$. The figure shows how the inseparability can be calculated (from the data obtained from the spectrum analyzers) by adding together the values of the appropriate minima and averaging. Adapted from \cite{clark2014quantum}}
    \label{fig:insep}
\end{figure}

We have already discussed various approaches to delay light and quantum information carried by entanglement.
Although entanglement cannot be used to signal superluminally \cite{peres2004quantum}, it is thought to be an essential resource in quantum information science \cite{braunstein2005quantum,weedbrook2012gaussian}.
It is then of fundamental interest to study what happens to entanglement when part of an entangled system propagates through a fast light medium.
Much work has been done to understand fast-light phenomena associated with anomalous dispersion, which gave rise to group velocities that are greater than the speed of light in vacuum, as described in chapter 1 \cite{boyd2009slow}. 
For classical pulses propagating without the presence of noise, it has been well established theoretically that the "pulse front" propagates through a linear, causal medium at the speed of light in vacuum \cite{brillouin1914fortpflanzung}.
It is often argued that this part of the pulse carries the entirety of the pulse's classical information content since the remainder of the signal can in principle be inferred by measuring the pulse height and its derivatives, just after the point of non-analyticity has passed \cite{stenner2003speed,kuzmich2001signal}.
Experimentally, in the inevitable presence of quantum noise, pulse fronts may not convey the full story of what is readily observed in the laboratory.\\
In the paper \textit{Quantum mutual information of an entangled state
propagating through a fast-light medium}, we discuss in details the detrimental role of quantum noise.
Instead of using a pulse with a given shape, in this work we encode the information in  correlated quantum noise, and therefore the definition of a "pulse front" becomes impossible.
The fluctuations of the probe and conjugate electric fields are not externally imposed, and they present no obvious pulse fronts or non-analytic features to point to as defining the signal velocity. 
As such, classically-rooted approaches to defining the signal or information content of the individual modes are not readily applicable to this system.
To quantify the quantity of information usable in the entangled beams we use the criteria of inseparability (Fig. \ref{fig:insep} for details).
As a reminder, note that an inseparability lower than 2 guarantees the presence of entanglement.

The entangled beams are generated using four-wave-mixing in a hot atomic vapor, and one of the two beams then propagates through a tunable fast-light medium (with group velocity larger than $c$). 
This fast light medium can be tuned to slow-light by changing the experimental parameters and this allows for direct comparison.
Due to the Kramers-Kronig relation, a group velocity larger than $c$, is systematically associated with gain or loss through the medium. 
The main point of our study is to understand the role of this gain or loss on the quantum information encoded in our beams.\\

The first step consists in measuring the arrival time of the cross-correlation maximum (or minimum) when the conjugate beam propagates in vacuum (actually in air) and compare it to the case of propagation in a fast-light medium.
This approach is analogous to the classical technique of measuring the arrival time of a light pulse maximum.
After accumulating enough statistics, we have observed an advancement of the normalized correlation peak by $3.7$~ns $\pm~0.1$~ns. Compared to the 300~ns of correlation time, this corresponds to a un-ambiguous relative advancement of more than 1\%.

However this advancement is coming with a reduction of the quadrature squeezing from $-3$~dB below the shot noise in the case of free-space to $-2.3$~dB in the case of fast-light media (Fig. \ref{fig:insep}).
Obviously this reduction is not observed when comparing the cross-correlation maximum or normalized. 
Therefore, in a second set of measurements, we report the inseparability as function of time.
Similarly we see an advancement of the inseparability peak in the case of the fast-light medium.
This was expected as the inseparability is computed from normalized cross-correlation maximum and minimum (see figure \ref{fig:insep} for details).\\

More insights can be found in the behaviour of the leading edge of inseparability.
Indeed, the inseparability can be used to define the mutual information (using the covariance matrix) and therefore, a larger value of inseparability means more mutual information shared between the two sides of the entangled pair.
If at some moment in time, the inseparability has a larger value than later it  means that the amount of information is larger.
What we observe in the experiment is that the noise added by the fast light medium, consistently reduces the value of inseparability of the leading edge in order that it never outperforms the case of free-space.
In the opposite case of slow light, we also have shown clear evidence of delaying not only the maximum of inseparability but also the trailing edge.  

It is interesting to contrast this asymmetry in the fast- and slow-light behaviour of the mutual information with the results of previous experiments studying the velocity of classical information propagating through dispersive media. 
In refs \cite{stenner2003speed} and \cite{stenner2005fast}, new information associated with a 'non-analytic' point in the field was found to propagate at $c$ in the presence of slow- and fast-light media alike. 
Because of copyright only the first page of the paper is reproduced here. For full access please check on our webpage : \href{www.optiquequantique.fr}{www.optiquequantique.fr} or  \textbf{Nature Photonics 8}, 515 (2014).

\chapter*{Article 4: Quantum mutual information of an entangled state propagating through a fast-light medium}

\section*{Preprint}
\noindent The preprint version of this article is accessible on \href{https://arxiv.org/abs/1405.7726}{arXiv:1405.7726}.

\section*{Published version}

 \href{https://doi.org/10.1038/NPHOTON.2014.112}{\textbf{Quantum mutual information of an entangled state propagating through a fast-light medium}}.\\
      J. B. Clark, R. T. Glasser, \textbf{Q. Glorieux}, U. Vogl, T. Li, K. M. Jones, and P. D. Lett.
     \href{https://doi.org/10.1038/NPHOTON.2014.112}{Nature Photonics \textbf{14}, 123024 (2014)}

\addcontentsline{toc}{section}{\textbf{Article 4}: Quantum mutual information of an entangled state
propagating through a fast-light medium. \textbf{Nature Photonics 8}, 515 (2014) }
\ifpapers
\fi

\clearpage

\chapter{Hydrodynamics of light}\label{chap:4}

List of publications related to this chapter :
\begin{itemize}
   
    \item \href{http://journals.aps.org/prl/abstract/10.1103/PhysRevLett.116.116402}{\textbf{Injection of Orbital Angular Momentum and Storage of Quantized Vortices in Polariton Superfluids.}}\\
    T. Boulier, E. Cancellieri, N. D. Sangouard, \textbf{Q. Glorieux}, A. V. Kavokin, D. M. Whittaker, E. Giacobino, and A. Bramati. \\ \href{http://journals.aps.org/prl/abstract/10.1103/PhysRevLett.116.116402}{{\it Phys. Rev. Lett.} \textbf{116}, 116402} (2016)

   \item  \href{https://journals.aps.org/prl/accepted/dd076Yd5H9214c7c92b385f9f1b8d57b2c746dd87 }{\textbf{Observation of the Bogoliubov dispersion in a fluid of light}}.  \\
   \begingroup
    \fontsize{9.5pt}{9.5pt}\selectfont
    Q. Fontaine,  T. Bienaim\'e,  S. Pigeon, E. Giacobino,  A. Bramati,   \textbf{Q. Glorieux}.\\
\endgroup
   \href{https://journals.aps.org/prl/accepted/dd076Yd5H9214c7c92b385f9f1b8d57b2c746dd87 }{{\it Phys. Rev. Lett.}}, \textit{Accepted }, (2018).
     
     \item \href{https://doi.org/10.1103/PhysRevB.98.024503}{\textbf{Coherent merging of counterpropagating exciton-polariton superfluids.}}\\
    T. Boulier, S. Pigeon, E. Cancellieri, P. Robin, E. Giacobino, \textbf{Q. Glorieux} and A. Bramati.\\
\href{https://doi.org/10.1103/PhysRevB.98.024503}{\textit{Phys. Rev. B} \textbf{98}, 024503}  (2018).\\
    Not included.
    
    \item \href{http://www.sciencedirect.com/science/article/pii/S1631070516300330}{\textbf{Lattices of quantized vortices in polariton superfluids.}}\\
  T. Boulier, E. Cancellieri, N. D Sangouard, R. Hivet, \textbf{Q. Glorieux}, E. Giacobino, A. Bramati. \\ 
  \href{http://www.sciencedirect.com/science/article/pii/S1631070516300330}{{\it Comptes Rendus Acad\'{e}mie des Sciences} \textbf{17}, 893} (2016).\\
  Not included.

    \item \href{http://www.nature.com/articles/srep09230}{\textbf{Vortex chain in a resonantly pumped polariton superfluid.}}\\
    T. Boulier, H. Tercas, DD. Solnyshkov, \textbf{Q. Glorieux}, E. Giacobino, G. Malpuech, A. Bramati. \\
    \href{http://www.nature.com/articles/srep09230}{{\it Scientific Reports} \textbf{5}, 9230} (2015).\\
    Not included.
    
\end{itemize}

    \clearpage
    
\section{What is a fluid of light ?}
We know, since Einstein, that vacuum photons are well described as mass-less non-interacting particles, behaving as an ideal gas.
In this chapter we discuss experiments where photons acquire a sizeable effective mass and tunable interactions mediated by a coupling with matter, forming a \textbf{fluid of light.}
This approach relies on a description of these hybrid photons in terms of quasi-particle of light generically known as polariton \cite{fleischhauer2002quantum}.

\subsection{Hydrodynamic formulation of the non-linear Schr\"odinger equation}\label{section:hydro}

The non-linear Schr\"odinger equation is used to describe a large variety of phenomena.
This equation can be written in a mathematical form, with $t$ and $\nabla^2$ dimensionless:
\begin{equation}
    i \frac{\partial A}{\partial t} = \left( -\frac{1}{2 } \nabla^2 + g\vert A \vert^2 \right)A. \label{nlse_dimensionless}
\end{equation}
$A$ is an arbitrary quantity and $g$ is the non-linear coupling coefficient.\\

In this manuscript, we will use this equation (with some modifications) to study the time evolution of excitons polaritons in a microcavity (see \ref{section:polariton}) and the spatial evolution of an electric field propagating in a non-linear medium (see \ref{section:propagation}).
In chapter \ref{outlooks}, we will also discuss its relevance in slow light and quantum memory experiments.

In the context of fluid of light, it is useful to use the  hydrodynamic formulation of the non-linear Schr\"odinger equation \cite{wilhelm1970hydrodynamic,sulem2007nonlinear,coste1998nonlinear}.
To show that this approach is very general, I will do this derivation in the case of the Gross-Pitaevskii equation which describes the dynamics of an atomic Bose-Einstein condensate \cite{dalfovo1999theory}.
The Gross-Pitaevskii equation is a time dependant non-linear Schr\"odinger equation including a potential term:
\begin{equation}
    i \hbar \frac{\partial \psi(\mathbf{r},t)}{\partial t} = \left( -\frac{\hbar^2}{2m } \nabla^2 + V(\mathbf{r})+g\vert  \psi(\mathbf{r},t) \vert^2 \right) \psi(\mathbf{r},t) , \label{GPE1}
\end{equation}
where $ \psi(\mathbf{r},t)$ is the condensate wave-function, $V(\mathbf{r})$ is an external potential and $g$ is the non-linear coupling coefficient, which is real.
As usual the term $\vert  \psi(\mathbf{r},t) \vert^2$ gives the local density of the condensate.
To find an hydrodynamic formulation, we need to extract a quantity analogue to a local velocity.
This can be done combining the time evolution of $\psi$ and $\psi^*$ such as we find
\begin{equation}
     \frac{\partial |\psi|^2}{\partial t}+\nabla\cdot\left[\frac{\hbar}{2mi}\left(\psi^*\nabla\psi-\psi\nabla\psi^*\right)\right]=0.
     \label{GPE3}
\end{equation}

We can note that we would have obtained the same equation in the absence of external potential $V(\mathbf{r})$  and in the absence of the non-linear term  $g\vert  \psi(\mathbf{r},t) \vert^2$ as both these terms are real.
Indeed this derivation has first been made by Madelung for the usual (linear) Schr\"odinger equation \cite{fedele2002solitary,madelung1927quantentheorie}.

We note $\rho=\vert  \psi(\mathbf{r},t) \vert^2$ the local density. 
We can see that Eq. \ref{GPE3} has the form of a continuity equation for $\rho$, and may be written as:
\begin{equation}\label{eq:continuity}
    \frac{\partial \rho}{\partial t}+\nabla\cdot (\rho \mathbf{v})=0,
\end{equation}
where we have introduced a quantity $\mathbf{v}$ analogue to a velocity: 
\begin{equation}
     \mathbf{v}=\frac{\hbar}{2mi}\frac{\psi^*\nabla\psi-\psi\nabla\psi^*}{\vert  \psi(\mathbf{r},t) \vert^2}.
\end{equation}
It is possible to obtain a simpler expression for the velocity by applying the Madelung transformation to $\psi(\mathbf{r},t)$ and using the real part of the wavefunction $\sqrt{\rho}$ and its phase $\phi$:
\begin{equation}
   \psi(\mathbf{r},t)=\sqrt{\rho(\mathbf{r},t)}e^{i\phi(\mathbf{r},t)}.
\end{equation}
This allows to write the velocity as:
\begin{equation}
   \mathbf{v}=\frac{\hbar}{m}\nabla\phi.
\end{equation}

For example, this derivation can be done in a similar way for the time-evolution of an electromagnetic field in a cavity filled by a $\chi^{(3)}$ non-linear medium \cite{lugiato1994transverse,lugiato1999optical,brambilla1991transverse,staliunas1993laser,pomeau1993nonlinear}.
The equation is slightly modified (due to cavity losses \cite{staliunas1993laser}) but the main idea remains: the light intensity corresponds to the fluid density, the spatial gradient of its phase to the fluid velocity and the collective behaviour originates from the effective photon-photon interactions due to the non--linearity of the medium inside the cavity.

Historically, the first connection made between superfluid hydrodynamics and non-linear optics dates back to the 80's. Back then, P. Coullet \emph{et al.} related optical phase singularities to quantized vortices in superfluid \cite{coullet1989optical}.
A seminal attempt to experimental investigation of superfluid behaviour in a system based on non-linear medium in a macroscopic cavity \cite{vaupel1996hydrodynamic} has been followed by a series of theoretical articles by R. Chiao on superfluidity of light in an atomic medium within a Fabry--Perot cavity  \cite{chiao1999bogoliubov}. 
Surprisingly, no experiments were reported thereafter, possibly because large non-linearities and high-Q cavities were hardly available at the time.
In parallel, non-linear resonator filled with dye has been investigated and have allowed for the observation of photon BEC \cite{weitz}.

\section[Overview of exciton-polaritons in a microcavity]{General overview of exciton-polaritons in a semiconductor microcavity}\label{section:polariton}

Modern research on quantum fluids of light has shifted to exciton-polaritons in micro-cavities.
These semiconductor nanostructures, thanks to the progress in nano--fabrication, offer unprecedented control of light-matter interaction.
In these systems, photons entering a Fabry--Perot cavity strongly couple to excitonic dipolar transitions through quantum wells located at the cavity electric-field maxima. 
This leads to the creation of  massive interacting bosonic quasi-particles known as exciton-polaritons. 

\subsubsection{Where does the photon effective mass comes from ?}\label{exp:pol}
\hfill\textit{From the confinement inside a Fabry-Perot cavity.}\\

If you take an optical cavity of length $L$ and a refractive index $n_c$, then the resonance condition leads to  $p\lambda_p=2n_cL$,
with $p$ a positive integer and $\lambda_p$ the wavelength associated to the mode $p$.
The dispersion relation of light with wavevector $\mathbf{k}$ is:
\begin{equation}
E_c(\mathbf{k})=\frac{\hbar c}{n_c}|\mathbf{k}|=\frac{\hbar c}{n_c}\sqrt{k_x^2+k_y^2+k_z^2}.
\end{equation}

\noindent If we assume the cavity to be along the $z$ direction, the quantification of the modes inside the cavity leads to:
\begin{equation}
k_z=p\frac{\pi}{n_c L}.
\end{equation}

\noindent We restrict ourselves to small angle of incidence $\theta$ on the cavity plane and therefore $k_z=|\mathbf{k}|\cos(\theta)\gg k_x,k_y$.  
We note the in-plane wavevector $k_{\perp}=\sqrt{k_x^2+k_y^2}$. We obtain the general dispersion relation for a cavity:

\begin{equation}
\hbar \omega_c\simeq\frac{\hbar c k_z}{n_c}+\frac{\hbar^2 k_{\perp}^2}{2m^*},
\end{equation}
with the definition of an effective mass $m^*=n_c\hbar k_z c$.
In typical cavities we have $m^*\sim 5.10^{-7}m_e$, with $m_e$ the electron mass.
In this work, we will use microcavities made of GaAs and of length 2$\lambda$. 
The cavity finesse is $F=3000$.

\subsubsection{Where does the interactions comes from ?}
\hfill\textit{From the quantum wells excitons embedded within the microcavity.}\\

A quantum well in a semiconductor consist on thin layer (nano-scale) of small band gap semiconductor, \textit{sandwiched} between two other semiconductor with wider band gap. 
In our case, we are talking about an InGaAs layer embedded in GaAs. 
The material discontinuity in the growth direction leads to a confinement of the electronic excitations inside the well region and a discretization of their states inside the well.
In semiconductor, the absorption of a photon of adequate wavelength can promote an electron of the valence band (below the Fermi energy) to the conduction band (above the Fermi energy).
This will leave an empty spot in the valence band that can be treated as a quasi-particle of positive charge called hole.
Both excitations having opposite electronic charge can bound via  Coulombian interaction.
The energy of the electron-hole pair being reduced, they form a pseudo-particle named exciton. 
The exciton is then analogue to the 1s orbital of the hydrogen atom, with a positive charge of large mass (the hole) and a negative charge of light mass (the electron) bounded by the Coulombian interaction.
For Wannier-Mott exciton, the exciton size is larger than the typical crystal inter-atomic distance.\\

Excitons can be approximated as bosonic quasi-particles interacting mediated hard-core interaction \cite{ciuti1998role}. Defining a creation operator $\hat b_\mathbf{q}$, for an exciton with wavevector $\mathbf{q}$.
The interaction hamiltonian is then written as:
\begin{equation}
    \hat H_{XX}=\frac 12 \sum_{\mathbf{k},\mathbf{k'},\mathbf{q}}V_{XX}\hat b^{\dag}_{\mathbf{k}+\mathbf{q}}\hat b^{\dag}_{\mathbf{k'}-\mathbf{q}}\hat b_{\mathbf{k}}\hat b_{\mathbf{k'}}.
\end{equation}
This expression describes a scattering process with destruction of 2 excitons at wavevectors $\mathbf{k}$ and $\mathbf{k'}$ and the creation of 2 excitons at wavevectors $\mathbf{k+q}$ and $\mathbf{k'-q}$.
We assume the interaction potential $V_{XX}$ to be constant with $\mathbf{q}$ because we are interested at small wavevectors compared to the inverse of the exciton length scale.\\

I have briefly described the cavity part which gives an effective mass to the photons and the quantum well which will provide the interactions.
In the next paragraph, I bring together (strongly) these elements and I introduce the concept of exciton-polaritons.


\subsubsection{Exciton-polaritons}
The physics of coupling between these two systems is a fantastic example for undergraduate quantum mechanics class. 
One starts with two bosonic particles associated with the creation operators $\hat a_{\mathbf{k}}^{\dag}$ and $\hat b_{\mathbf{k}}^{\dag}$ for cavity photons and excitons respectively.
These particles have their own eigen-energies $E_c=\hbar \omega_c$ for the cavity photons and $E_X=\hbar \omega_X$ for the excitons. 
The hamiltonian of the uncoupled system is clearly diagonal.
One adds a coherent coupling energy $\hbar \Omega_R$ between them and the system is not diagonal anymore.
After a simple 2-by-2 matrix diagonalization, one obtains the two new eigen-states and eigen-energies, named upper and lower polaritons\footnote{Obviously, for this simple treatment we have neglected the exciton-exciton interactions described in the previous section. For a more detailed description please refer to \cite{yamamoto2003semiconductor}}.\\

We will briefly derive this. In the photon-exciton basis we can write the hamiltonian $ \hat H$ as:
\begin{equation}
 \hat H=\begin{bmatrix}
  E_c & \hbar\Omega_R  \\
  \hbar\Omega_R & E_X \\
 \end{bmatrix}.
\end{equation}

\begin{figure}
    \centering
    \includegraphics[width=13cm]{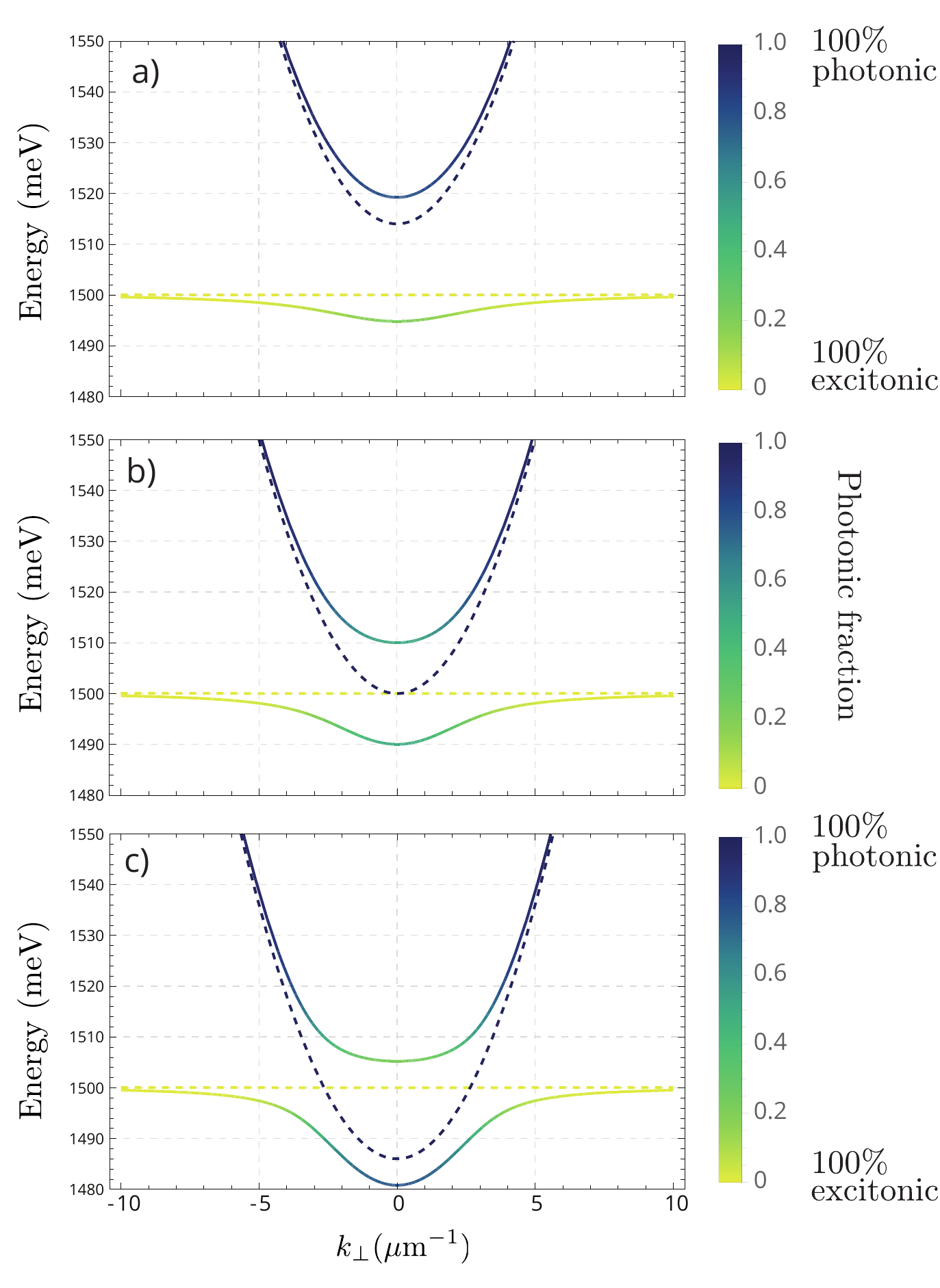}
    \caption{Dispersion of upper and lower polaritons. Yellow dashed is the bare exciton dispersion and blue dashed is the bare cavity dispersion. The relative photonic fraction is given in the colorbar. Exciton energy: $Ex=1500$~meV, Coupling frequency: $\Omega_R=10$~meV, a) Blue detuned cavity $Ec=1515$~meV (Marignan), b) Cavity on-resonance $Ec=1500$~meV, Red detuned cavity $Ec=1485$~meV.}
    \label{fig:pola}
\end{figure}

Diagonalization of this hamiltonian gives rise to two new bosonic particles associated with their creation operators $\hat p_{\mathbf{k}}^{\dag}$ and $\hat u_{\mathbf{k}}^{\dag}$ for the lower and upper polaritons respectively.
As bosonic particles, these operators follow the usual commutation relations.
The new eigen-energies are :
\begin{eqnarray}
\omega_{UP}(\mathbf{k})=\frac{\omega_C(\mathbf{k})+\omega_X(\mathbf{k})}{2}+\frac{1}{2}\sqrt{[\omega_C(\mathbf{k})-\omega_X(\mathbf{k})]^2+4\Omega_R^2},\\
\nonumber\omega_{LP}(\mathbf{k})=\frac{\omega_C(\mathbf{k})+\omega_X(\mathbf{k})}{2}-\frac{1}{2}\sqrt{[\omega_C(\mathbf{k})-\omega_X(\mathbf{k})]^2+4\Omega_R^2}.
\end{eqnarray}
$\omega_{LP}$ and $\omega_{UP}$ are associated to the lower and upper polaritons respectively.
The eigen-energy is then the mean of the exciton energy and photon energy, with a correction that is the root mean square of the coupling Rabi frequency $\Omega_R$ and half of the cavity-exciton detuning $\delta=\omega_C(\mathbf{k})-\omega_X(\mathbf{k})$.\\
The polariton eigen-states can be written using the Hopfield coefficients \cite{hopfield1958theory}:
\begin{equation}
\left\{
  \begin{array}{lr}
    \hat p_{\mathbf{k}}= X_{\mathbf{k}} \hat b_{\mathbf{k}} +C_{\mathbf{k}} \hat a_{\mathbf{k}}\\
    \hat u_{\mathbf{k}}= X_{\mathbf{k}} \hat a_{\mathbf{k}} -C_{\mathbf{k}} \hat b_{\mathbf{k}}
  \end{array}
\right.
\end{equation}
with
\begin{eqnarray}
    X_{\mathbf{k}}=\frac{1}{\sqrt{1+\left[\frac{\Omega_R}{\omega_{LP}(\mathbf{k})-\omega_C(\mathbf{k})}\right]^2}} \\
    \nonumber 
    C_{\mathbf{k}}=\frac{-1}{\sqrt{1+\left[\frac{\omega_{LP}(\mathbf{k})-\omega_C(\mathbf{k})}{\Omega_R}\right]^2}} .
\end{eqnarray}
To give an intuition of the behaviour of these pseudo particles we have plotted in Figure \ref{fig:pola} the dispersion relation for the upper and lower polaritons for different detunings $\delta=\omega_X(\mathbf{k})-\omega_C(\mathbf{k})$.
This situation is similar to what we realize experimentally \cite{boulier2016injection}.
The exciton mass is much larger than the cavity photon effective mass and therefore the dispersion of exciton is almost flat on this scale.
To tune the detuning $\delta$ we can only control the cavity energy. 
To do so, the cavity is slightly wedged in one direction, providing a large choice of cavity-exciton detunings by pumping at different positions in the sample.
From this figure we can understand intuitively the effect of detuning.
For positive detuning (Fig. \ref{fig:pola}-a), the lower polariton branch becomes more excitonic at $\mathbf{k}=0$ and therefore this enhances the polariton-polariton interactions.
On the opposite side, for negative detuning (Fig. \ref{fig:pola}-c), the lower polariton branch becomes more photonic. This can be used to reduce the effective mass and to get more light outside of the system.
An important characteristic of this system is that we are only concern on the 2D evolution that in the transverse (cavity) plane.

\subsubsection{Strong coupling condition}
In the previous paragraph, I have explained what are exciton-polaritons. However, I have skipped one important discussion about strong coupling.
Indeed, exciton and photon have a finite lifetime in this system.
Photon lifetime $\tau_C$ can be obtained from the quality factor $Q$ of the cavity as:
\begin{equation}
    \tau_C=\frac{Q}{\omega_C}.
\end{equation}
For the GaAs cavities we have used in this manuscript, the typical value of $ \tau_C$ is 20~ps and quality factor are in the order of $10^4$.\\
Exciton lifetime is far less under control and depends mainly of surface irregularities inside the wells.
These two quantities give rise to a broadening of the energy by a width $\hbar$ times the inverse of the lifetime.
Typical broadening for a GaAs cavity is $\hbar \Gamma_C\sim50$~$\mu$eV.\\

On the other hand the  Rabi frequency can be estimated from the oscillator strength of the exciton $f$, the size of the cavity $L$ and the exciton mass $M$:
\begin{equation}
    \Omega_r=\sqrt{\frac{fe^2}{2n_c^2LM\varepsilon_0}}.
\end{equation}
However this expression assumes a perfect overlap of the exciton and the photon wavefunctions.
Quantum wells must be placed carefully in the anti-nodes of the electromagnetic field in order to optimize the coupling.
It is common to increase the Rabi frequency by adding more wells as it scales with the square root of the number of wells (because the effective oscillator strength scales as the number of wells).\\

We can now give an \textit{graphical} understanding of the strong coupling. 
The strong coupling condition is achieved when the Rabi frequency is large enough so that polariton states can be distinguished form the uncoupled exciton and photon states.
This condition directly depends on the exciton and cavity photon linewidth and can be written:
\begin{equation}
    \Omega_R > \Gamma_C,\Gamma_X.
\end{equation}
If this condition is not satisfied, there is no need to talk about polariton states.

\subsection{Driven-dissipative Gross-Pitaevskii equation}
Remarkably, the time evolution of the exciton-polaritons wave-function follows dynamics similar to the Gross-Pitaevksii equation \ref{GPE1} describing the evolution of a dilute atomic Bose-Einstein condensate or a monochromatic light propagating in a $\chi^{(3)}$ non-linear media as presented in section \ref{section:hydro}.
However, for exciton-polaritons, it includes additional non-equilibrium features, streaming from their intrinsic \textbf{driven-dissipative nature}.
The evolution equation for the lower polaritons reads \cite{pigeon2011fluides}:
\begin{equation}
    i \hbar \frac{\partial \psi(\mathbf{r},t)}{\partial t} = \left( -\frac{\hbar^2}{2m^* } \nabla_{\perp}^2 + V(\mathbf{r})-i\frac{\hbar \gamma_{LP}}{2}+ g\vert  \psi(\mathbf{r},t) \vert^2 \right) \psi(\mathbf{r},t) +P(\mathbf{r},t) , \label{eq:GPE_DD}
\end{equation}
with $m^*$ the effective mass of lower polaritons. 

On one side, due to the finite reflectively of the cavity's mirrors, photons have a finite lifetime such that eventfully, they exit the cavity.
This, which can be seen as a drawback, has proved to be essential as it allows for the detection and measurement of polaritons (via their photonic component). 
This leads to a dissipation term in the equation proportional to the polariton linewidth: $\hbar \gamma_{LP}$.
On the other hand, if photons exit the cavity (after 20~ps typically), new particles must be added (continuously) to the system to maintain a constant number of particles.
This is done by pumping the cavity with an external field and leads to a driving term $P(\mathbf{r},t)$ in the equation.
Finally, we should restrict the evolution to the plane of the cavity and therefore the 3D spatial derivative is changed to 2D $\nabla_{\perp}$. \\

An important feature of polariton physics is therefore that it is out of equilibrium, as particles constantly leave the system.
However, with the seminal observation of \textbf{exciton-polaritons BEC} \cite{deng2002condensation,kasprzak2006bose} and the demonstration of \textbf{superfluidity} through resonant Rayleigh scattering \cite{carusotto2004probing,amo2009superfluidity}, this system has become a realistic tool for quantum hydrodynamics.
It is worthy to notice that the possibility to observe such fluid of light is allowed by the conjunction of relatively strong $\chi^{(3)}$ non-linearity and a very low effective mass.
The latter is essential as it provides for a very fast dynamics to form the fluid in a time scale smaller than the dissipation.
This allows for the formation of long-range correlations for example.

\section{Superfluidity}
To conclude this short introduction to light hydrodynamics, I will cover the concept of superfluidity and apply it to light.
Superfluidity is the property of a fluid to flow without viscosity below a certain temperature.
It has been discovered in Helium 4 cooled at 2~K by Kapitza \cite{kapitza1938viscosity}, Allen and Misener \cite{allen1938flow} in 1937, about 60 years before the first observation of an atomic BEC \cite{davis1995bose}.

However, in 1938 London highlights that superfluid transition temperature and BEC temperature for an ideal gas at the same density are similar and he proposes a link between the two phenomena \cite{nozieres2018theory}.
Unfortunately, this explanation is not relevant for liquid Helium 4, due to the strong interactions which break the ideal gas model.\\

In 1941, Landau introduced a novel model based on two fluids: one classical and one condensed which was much more accurate to describe the physics of liquid Helium \cite{feynman1954atomic,landau1941theory}.
He also proposed various criteria to define superfludity. 
These criteria are still used today to assess superfluidity and to measure the condensed fraction.

\subsubsection{Rotating fluid}\label{rotatingfluid}
These criteria are based on a conceptually simple experiment.
A fluid (presumably a superfluid) is placed in a bucket and the bucket is rotated.
For a superfluid, under the critical velocity, the fluid does not move with the bucket.
Above the critical velocity, quantized vortices  appear in the fluid.
This effect has been observed for the first time by Hess and Fairbank in 1967 \cite{hess1967measurements}.
I discuss this criteria in more details in the section \ref{polariton}, as we have used a similar technique to store quantized vortices in a polariton superfluid.

\subsubsection{Persisting current}
Another approach is to put in rotation a classical fluid and then cool it down below the critical point.
If now, one removes the rotating mechanism, a superfluid will  rotate for a virtually infinite time.
This effect has been observed in 1964 in liquid Helium \cite{reppy1964persistent} and in 2007 in a BEC with interactions \cite{ryu2007observation}.
We can note an interesting proposal to observe a similar mechanism in photon fluid proposed in Ref. \cite{silva2017persistent}.

\subsubsection{Landau criteria for superfluidity}
To derive these criteria we consider a uniform Bose gas  with interaction (a uniform quantum fluid) in its ground state.
Imagine that we move at velocity $\mathbf{v}$ a solid obstacle through the fluid.\\

We want to know at which critical velocity $v_c$, it starts to become possible to create an elementary excitation in the fluid.
It is more convenient to work in the reference frame of the obstacle, as the obstacle exerts a time-independent potential to the fluid in this frame.
Let us recall the standard Galilean transformations to write the energy $ E(\mathbf{v} )$ in the frame moving at velocity $\mathbf{v}$ as function of the energy $E$ and $\mathbf{p}$ the momentum of the fluid in one frame of reference:
\begin{equation}
    E(\mathbf{v} )=E-\mathbf{p} \cdot \mathbf{v} +\frac{1}{2}M \mathbf{v}^2,
\end{equation}
where $M$ is the total mass of the system.\\

If we apply it to our case, we can write the energy of the ground state as:
\begin{equation}\label{eq422}
    E(\mathbf{v} )=E_0 +\frac{1}{2}M \mathbf{v}^2,
\end{equation}
where $E_0$ is the ground state energy in the frame where the fluid is at rest, and $\mathbf{p}=0$.
We note $\epsilon_p$ the energy cost to add a single excitation of momentum $\mathbf{p}$ to the system.
In the original frame the energy of the state with a single excitation is:
\begin{equation}
    E_{ex}=E_0 +\epsilon_p,
\end{equation}
and therefore in the moving frame:
\begin{equation}\label{eq424}
    E_{ex}(\mathbf{v} )=E_0 +\epsilon_p-\mathbf{p}\cdot \mathbf{v}+\frac{1}{2}M \mathbf{v}^2.
\end{equation}
We can now make the difference between Eq. \ref{eq424} and Eq. \ref{eq422} to obtain the energy to create an excitation in the moving frame.
Thus at a velocity 
\begin{equation}
    v=\frac{\epsilon_p}{p},
\end{equation}
it becomes possible for the obstacle to create an excitation with momentum parallel to $\mathbf{v}$ and at higher velocities, the excitation momentum will have an angle to $\mathbf{v}$.
We can conclude with the Landau critical velocity $v_c$ defined as the minimum velocity at which it is possible to create excitation:
\begin{equation}\label{LandauCriteria}
    v_c=\text{min}\left(\frac{\epsilon_p}{p}\right).
\end{equation}
Below the critical velocity there is no mechanism to create elementary excitations and the liquid will exhibit superfluidity.
We can note that for a parabolic dispersion of the elementary excitations $\epsilon(p)\propto p^2$, the critical velocity is $0$. 
On the other hand, a linear dispersion $\epsilon(p)\propto p$ will allow for superfluidity.

\subsubsection{Suppression of Rayleigh scattering}
An important experiment, done in the LKB group before I joined, has demonstrated superfluidity of polaritons using the setup described in the section \ref{exp:pol}.
A polariton flow is sent towards a defect and depending on velocity and density of the fluid, elementary excitations are created or not  \cite{amo2009superfluidity}.
This experiment is really the foundation of most of the work done in polariton hydrodynamics afterwards and I will review it briefly. \\

\begin{figure}
    \centering
    \includegraphics[width=\textwidth]{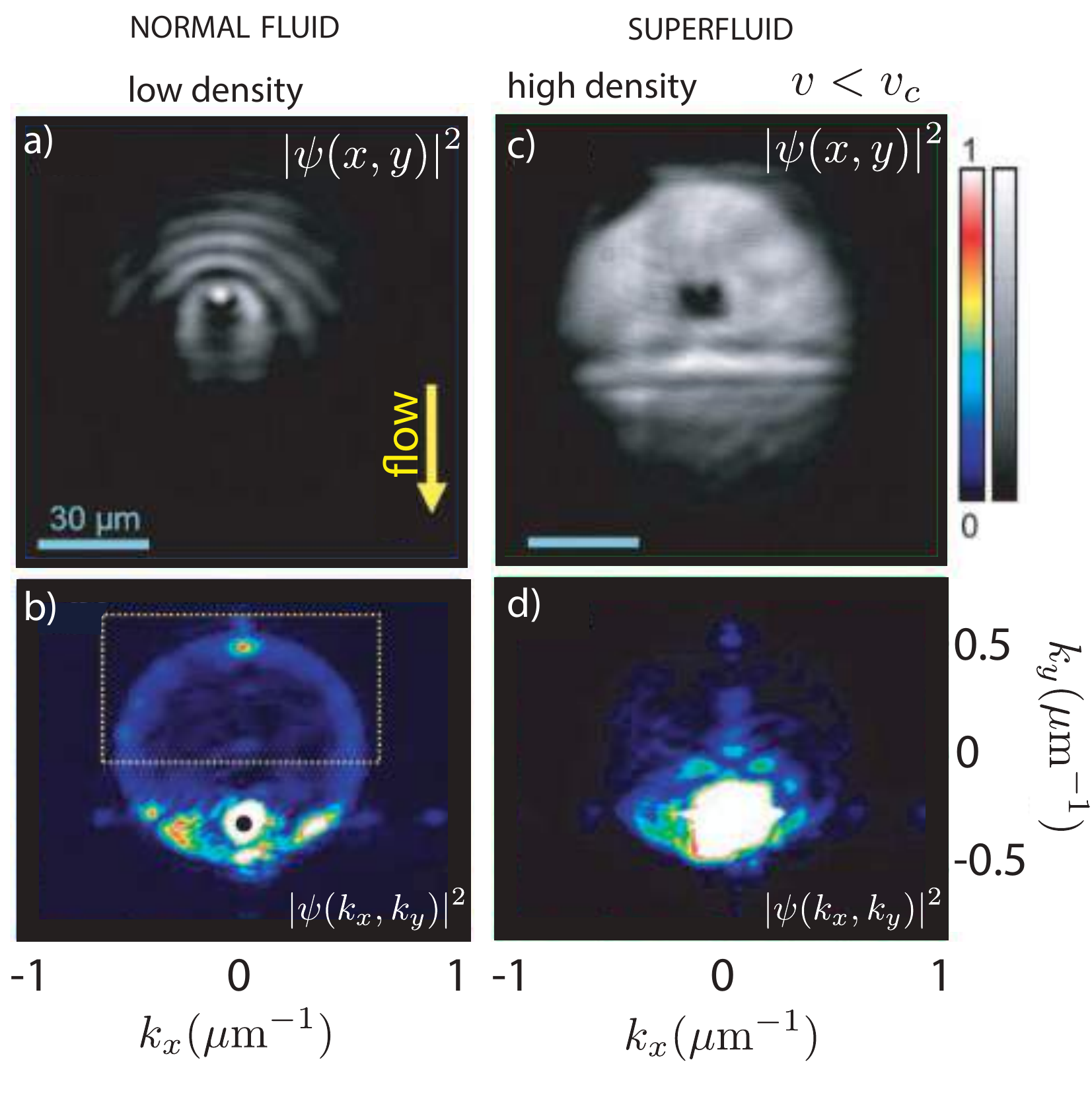}
    \caption{Comparison between a normal fluid (at low density a. and b.) and a superfluid (at high density c. and d.). The flow speed $v$ is downwards in real space images a) and c). Interference fringes in a) are due to scattering on the defect. They vanish for higher intensity due to superfluid flow.
    In k-space images, the Rayleigh scattering is clearly visible in b) and absent in d) due to superfluidity. For the superfluid case $v<v_c$. Adapted from \cite{amo2009superfluidity}}.
    \label{fig:amosuperfluid}
\end{figure}

A microcavity is pumped at a small angle by a resonant laser field.
The light enters the cavity and hybridizes with excitons to create polaritons with a momentum equal to the projection of the incident laser wavevector in the cavity plane.
Some defects can be found in microcavities due to scratches or imperfect growth, which can be used to study the scattering behaviour of polaritons and probe the superfluidity.
In Fig. \ref{fig:amosuperfluid} a) we see the polariton fluid hitting the defect and creating fringes upstream the defect. This situation is at low driving intensity, which is equivalent to a low density of polaritons inside the cavity. In this regime the non-linear interaction can be neglected.
Two interpretations can be given to these fringes: 
\begin{itemize}
    \item We can think as opticians and interpret this pattern as an interference pattern between incoming photons and photons scattered by the defect. The fringe distance will change when varying the projection of the incident laser wavevector in the cavity plane. This can be verified experimentally.
    \item The second interpretation comes from the hydrodynamic language. When a normal fluid propagates and hits a defect, characteristic waves appear upstream of the defect similar to water hitting a bridge pile.
\end{itemize}
In Fig. \ref{fig:amosuperfluid} b), we show the k-space (the emission angle of the cavity). 
The black spot at $k_x=0$ et $k_y=-0.3\mu$m$^{-1}$ is the pumping laser. 
We can observe a redistribution of the impulsion around a ring at constant $k$ which is Rayleigh scattering on the defect.\\

To switch to the superfluid regime, we do not have to cool down the fluid as in liquid Helium experiments, but we increase the density.
This regime is shown in Fig. \ref{fig:amosuperfluid} c) and d).
The disappearance of interference fringes is understood intuitively with the hydrodynamic interpretation because the fluid of light, now flows without viscosity. Correspondingly, the Rayleigh scattering vanishes.

The optics interpretation is more obscure as it is a cancellation of the diffraction on the defect due the non-linear effects.
We see here for the first time, why the hydrodynamic interpretation is useful to analyze and study optical phenomena that are difficult to predict with the non-linear optics usual toolbox.
I will follow this approach in various experiments presented in this chapter.\\

\begin{figure}
    \centering
    \includegraphics[width=\textwidth]{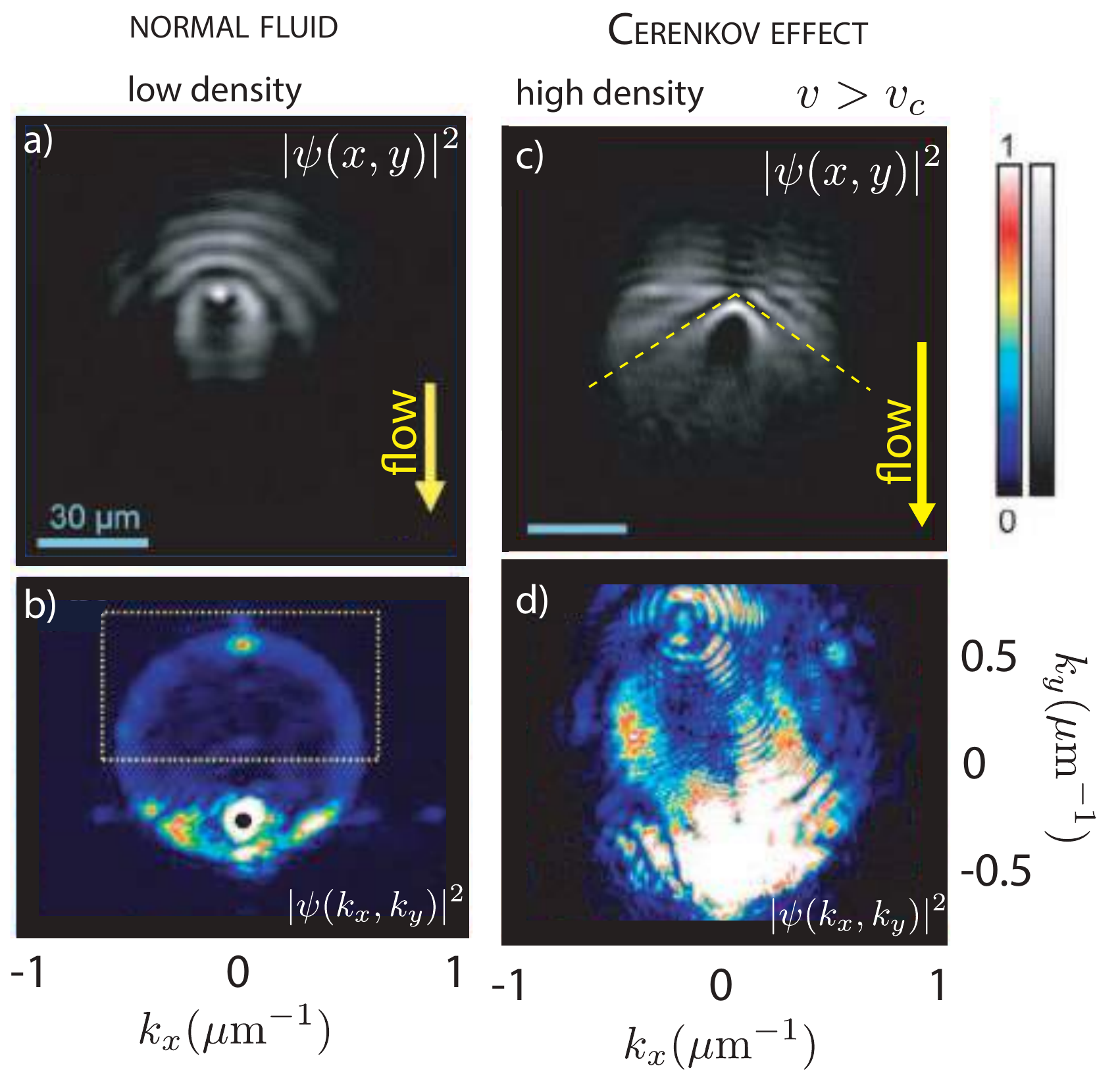}
    \caption{Comparison between a normal fluid (at low density a. and b. ) and a superfluid (at high density c. and d.) above the critical speed.
    The flow speed $v$ is downwards in real space images. For c) $v>v_c$.
    We can observe the break of superfluidity with the apparition of a Cerenkov cone at the defect position in c). The angle of the cone gives the critical velocity (which is the speed of sound in this case).
    In k-space images, the Rayleigh scattering is clearly visible in b). For the Cerenkov case d) polaritons are scattered at impulsion larger than the $k_in$ as expected by the Landau theory. Adapted from \cite{amo2009superfluidity}.}
    \label{fig:amoCerenkov}
\end{figure}

Finally we can break superfluidity by increasing the velocity above the critical speed $v_c$.
This experiment is reported in Fig. \ref{fig:amoCerenkov} c) and d).
We will see in the next section that the dispersion elementary excitations in a polariton fluid is linear at low wavevector, and this allows to define a speed of sound $c_s$ which is the critical velocity.
In this case, we observe an analogue of the Cerenkov effect. 
Indeed the excitations created in the fluid have a speed greater than the sound velocity in the medium, and we can observe the characteristic cone upstream of the defect.

\clearpage

\section[Injection of angular momentum in superfluids]{Injection of angular momentum in polariton superfluids}\label{polariton}

A rich variety of photon hydrodynamical effects  have been observed, from the unperturbed superfluid stream flowing around a defect \cite{amo2009superfluidity}, to the appearance of a Cerenkov cone in a supersonic flow \cite{amo2009superfluidity} presented in the previous section.
Other works from the LKB group have focused on the formation of topological excitations such as quantized vortices and dark solitons \cite{boulier2016injection,amo2011polariton}.
In this section I give the details about recent experiments done at LKB, where we have achieved the controlled injection of orbital angular momentum in a polariton superfluid using two different techniques.\\

We have discussed in paragraph \ref{rotatingfluid} a method proposed by Landau to probe superfluidity 
using a rotating bucket. 
We have revisited this idea by forcing the rotation of a polariton fluid and study the apparition of quantized vortices.
Quantized vortices are topological excitations characterized by the vanishing of the field density at a given point (the vortex core) and the quantized winding of the field phase from 0 to $2\pi$ around it.\\

\subsubsection{Injection of angular momentum using a Laguerre-Gauss pump}
The first method we have studied was to pump the microcavity with a Laguerre-Gauss mode of increasing charge.
We have reported the formation of a ring-shaped array of same sign vortices \cite{boulier2015vortex}.
As previously discussed in the linear regime, an interference pattern is visible. 
This pattern has a spiral shape and contains phase defects are visible. 
In the nonlinear (superfluid) regime, the interference disappears and up to eight vortices appear, minimizing the energy while conserving the quantized angular momentum (see figure 4.4).

\begin{figure}[h!]
    \centering
    \includegraphics[width=\textwidth]{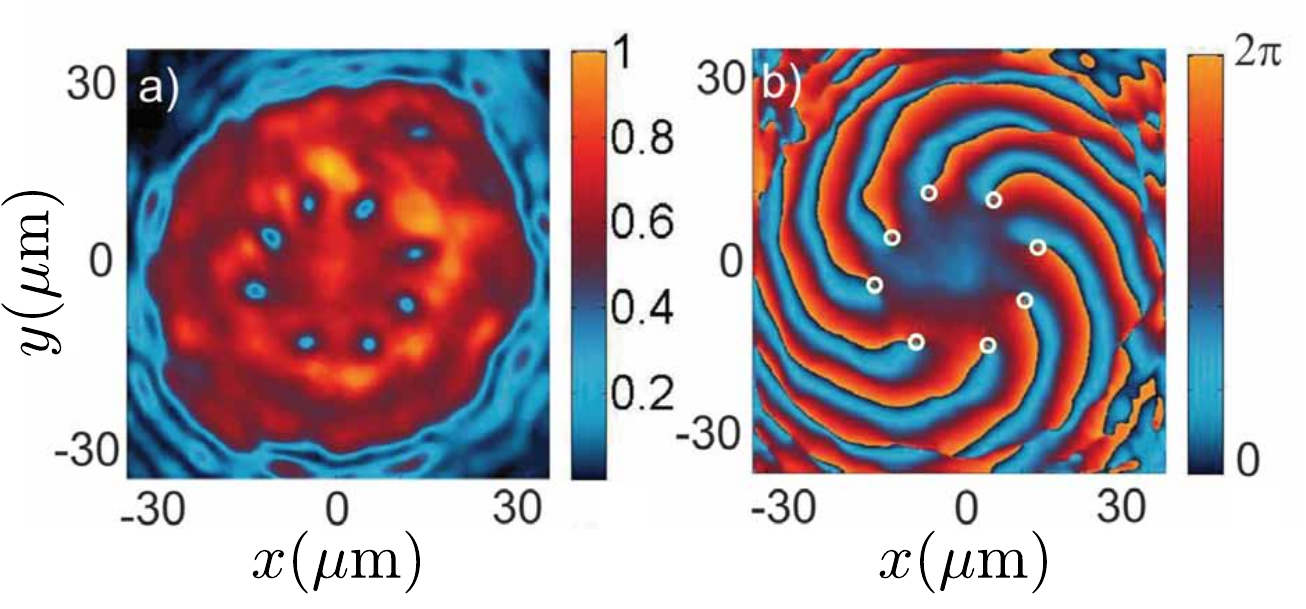}
    \caption{Nucleation of same sign vortices. 
    Experimental real space images of the polariton field in the nonlinear regime. a) Polariton density and b) phase pattern. Figure from Ref. \cite{boulier2015vortex}}
    \label{fig:vortexring}
\end{figure}

\subsubsection{Injection of angular momentum using a 4 tilted pumps}
The second experiment studies the propagation of 4 superfluids created with the same pump field at 4 different positions arranged in a square, all with an inward flow.
The pump is chosen to be resonant to allow for a fine tuning of the polariton density without generating excitonic reservoir and, consequently, a precise control of the non-linearities in the system.

If we send these fluids toward the center we simply observe the merging of the 4 converging fluids.
However, if we tilt the in-plane flow as presented in the figure~1 of PRL 116, 116402 (see below), we can inject an orbital angular momentum in the system and we see the apparition of quantized vortices.
We can evaluate the injected orbital angular momentum per photon $L$ in unit of $\hbar$ as:
\begin{equation}
    \frac{L}{\hbar}=R|\mathbf{k}|\sin \phi,
\end{equation}
with $R$ is the pump distance to the square center, $\mathbf{k}$ is the in-plane pump wavevector (choose identical for the 4 pumps) and $\phi$ is the tilt angle between the pump in-plane direction and the square center. 

The separation between the pumps is small enough to ensure the four polariton populations can meet, resulting in significant density at the square center. 
However to ensure that in the central region polaritons are free to evolve, we have cut the beam Gaussian tails to reduce direct illumination.

We use two different techniques for detection: the direct density measurement and a phase measurement.
 The polariton phase is measured with an off-axis interferometry setup: a beam splitter divides the real space image into two parts, one of which is expanded to generate a flat phase reference beam, used to make an off-axis interference pattern.
 With this method, the vortex position on the image is independent of the phase of the reference beam. The actual phase map is then numerically reconstructed with a phase retrieval algorithm.\\

What is striking in this experiment is that even though the orbital angular momentum injected is a continuous quantity, the number of topological defects remains a quantized quantity.
We have verified that the number of injected vortices (up to 5) matches a simple model including only the tilt angle (see figure 4 of PRL 116, 116402).\\

In conclusion we have observed the injection of angular momentum and the storage of topological charges in a non-equilibrium superfluid of light  \cite{boulier2016injection}. \\

\chapter*{Article 5: Injection of Orbital Angular Momentum and Storage of Quantized Vortices in Polariton Superfluids.}

\section*{Preprint}
\noindent The preprint version of this article is accessible on \href{https://arxiv.org/abs/1509.02680}{arXiv:1509.02680}.

\section*{Published version}

\href{http://journals.aps.org/prl/abstract/10.1103/PhysRevLett.116.116402}{\textbf{Injection of Orbital Angular Momentum and Storage of Quantized Vortices in Polariton Superfluids.}}
    T. Boulier, E. Cancellieri, N. D. Sangouard, \textbf{Q. Glorieux}, A. V. Kavokin, D. M. Whittaker, E. Giacobino, and A. Bramati.  \href{http://journals.aps.org/prl/abstract/10.1103/PhysRevLett.116.116402}{Phys. Rev. Lett. \textbf{116}, 116402} (2016).

\addcontentsline{toc}{section}{\textbf{Article 5}: Injection of orbital angular momentum and storage of quantized vortices in polariton superfluids. \textbf{PRL 116}, 402 (2016) }

\clearpage

\ifpapers
\fi

\clearpage

\section{Fluid of light in the propagating geometry}\label{section:propagation}

As we have seen in the previous section, fluids of light allow to revisit quantum gases experiments with the advantage of the precise control in state preparation and detection of optical systems.
So far I have presented results about cavity-based platforms which have intrinsic losses and therefore are restricted to driven-dissipative dynamics.
In 2016, I have started a new thematic in the LKB group to overcome this limitation with a different way of doing quantum simulation with light, by removing the need for a Fabry-Perot cavity. 
This approach brings complementary results to the exciton-polaritons experiments and gives the possibility for our group to compare  driven-dissipative and conservative dynamics with fluids of light.
Similarly to exciton-polaritons, it relies on the formal mapping between the dynamics of the problem we wish to simulate and the quantum system we will use to do it.

The system we propose to use is light propagating through a warm atomic vapor and I show that this system can be theoretically described by a non-linear Schr\"odinger equation similar to Eq. \ref{GPE1}.
To validate experimentally this approach, I present recent results about the dispersion relation of elementary excitations in this system and show that they follow a Bogoliubov spectrum similar to atomic BEC or photon BEC\cite{weitz} experiments.

\subsection{Bogoliubov dispersion relation}
Before going in the details of our system, I review briefly the physics of elementary excitations in the Gross-Pitaevskii equation (Eq. \ref{GPE1}).
More specifically, I derive the Bogoliubov dispersion relation for a uniform Bose gas in the absence of external potential.\\

We start by writing the quantum state as a mean field plus a small perturbation to linearize the Gross-Pitaevskii equation:
\begin{equation}
    \Psi(\mathbf{r},t)=[\psi_0(\mathbf{r},t)+\delta\psi(\mathbf{r},t)].
\end{equation}
Here $\psi_0(\mathbf{r},t)$ is the condensate wavefunction in the unperturbed state. 
It can then be written:
\begin{equation}
\psi_0=\sqrt{n(\mathbf{r})}\text{e}^{-i\mu t/\hbar},
\end{equation}
where $n(\mathbf{r}$ is the equilibrium density of particles and $\mu$ is the chemical potential.
For a uniform system $n(\mathbf{r}$ is independent of $\mathbf{r}$ and $\mu$ is given by $\mu=|\psi_0|^2g=ng$.
For $\delta\psi(\mathbf{r},t)$, we are interested in solution of the form:
\begin{equation}
    \delta\psi(\mathbf{r},t)=\left[u(\mathbf{r})\text{e}^{-i\omega t}-v^*(\mathbf{r})\text{e}^{i\omega t}\right]\text{e}^{-i\mu t/\hbar} .
\end{equation}
with $\omega$ a real quantity.
Inserting this ansatz in the linearized Gross-Pitaevskii equation we obtain two coupled equations for $u(\mathbf{r})$ and $v(\mathbf{r})$ known as the Bogoliubov equations:
\begin{eqnarray}\label{eq:couple}
    \left[-\frac{\hbar^2}{2m}\nabla^2+gn-\hbar \omega\right]u(\mathbf{r})-gn\ v(\mathbf{r})=0\\
    \nonumber
     \left[-\frac{\hbar^2}{2m}\nabla^2+gn+\hbar \omega\right]v(\mathbf{r})-gn\ u(\mathbf{r})=0
\end{eqnarray}
Because of the translational invariance in a uniform Bose gas we can take $u(\mathbf{r})$ et $v(\mathbf{r})$ to be of the form of plane waves:
\begin{equation}\label{uvq}
    u(\mathbf{r})=u_qe^{i\mathbf{q}\mathbf{r}} \text{ and }  v(\mathbf{r})=v_qe^{i\mathbf{q}\mathbf{r}}. 
\end{equation}
By injecting these expressions in Eq. \ref{eq:couple} we obtain two equations that are consistent only if the determinant of the coefficients vanishes.
Finally we obtain:
\begin{equation}
    (\hbar \omega)^2=\left[\frac{\hbar^2q^2}{2m}+gn\right]^2-\left(gn\right)^2
\end{equation}
We take the positive energy solutions and we obtain the Bogoliubov dispersion relation:
\begin{equation}\label{bogo1}
    \epsilon(q)= \hbar \omega(q)=\sqrt{\varepsilon_0(q)}\cdot\sqrt{  \varepsilon_0(q)+2gn}.
\end{equation}
Here we have introduced the notation $\varepsilon_0(q)=\frac{\hbar^2q^2}{2m}$ for the energy of the free-particle without interaction.

We can see from this expression that the Bogoliubov dispersion differs from the massive particle parabolic dispersion only at small $q$ when $2gn\leq \varepsilon_0(q)$.
In the limit $2gn/ \varepsilon_0(q)\gg 1$, we can approximate Eq. \ref{bogo1} by 
\begin{equation}
    \epsilon(q)=\hbar \sqrt{\frac{gn}{m}}q.
\end{equation}
In this limit, the spectrum become linear (sound-like) and we can define a speed of sound $c_s$ that scales with the square root of the density:
\begin{equation}
    c_s=\sqrt{\frac{gn}{m}}.
\end{equation}
If we use the Landau criteria for superfluidity (Eq. \ref{LandauCriteria}), we can see that this dispersion allows for a minimum of the quantity $\frac{\epsilon(q)}{q}$ which is precisely equal to $c_s$.
\begin{figure}
    \centering
    \includegraphics[width=9.5cm]{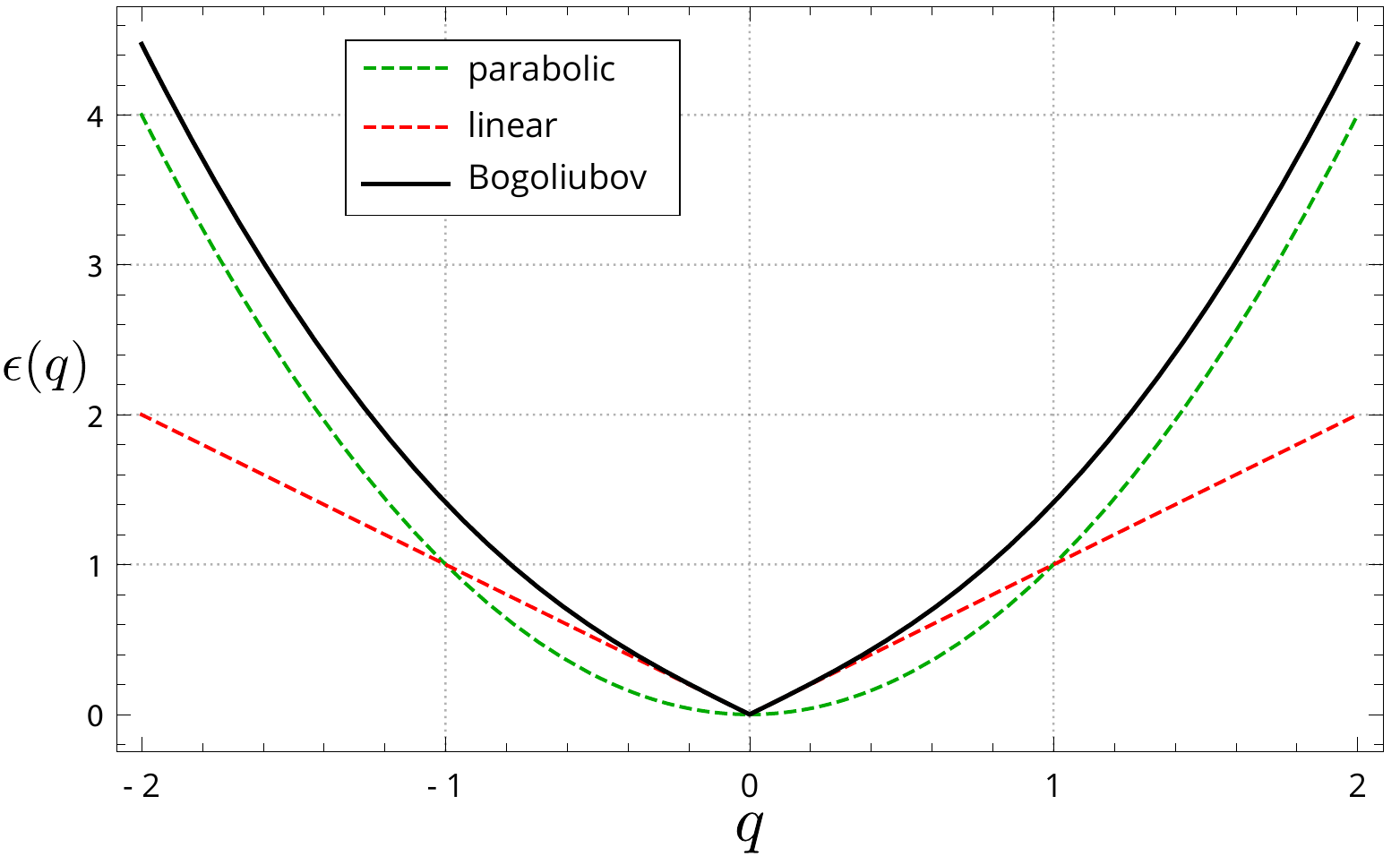}
    \caption{Dispersion relation for massive particles (dashed green), for sound waves (dashed red) and for Eq. \ref{bogo1} (black). $c_s=1$ and $\epsilon(q)$ in $\hbar$ unit. }
    \label{fig:bogo}
\end{figure}
\clearpage

\subsection{Non-linear propagation in a $\chi^{(3)}$ medium}
In the absence of sources, the propagation of light in a medium of refractive index $n$ is given by the Helmholtz equation:
\begin{equation}\label{eq:Helmholtz}
    \nabla^2E-\frac{n^2}{c^2}\frac{\partial^2E}{\partial t^2}=0,
\end{equation}
where $E$ is the electric field amplitude and $c$ is the speed of light in vacuum.
To take into account the propagation in a non-linear medium of we need to a non-linear polarization term $P_{NL}$ in the form:
\begin{equation}\label{eq:HelmholtzNL}
    \nabla^2E-\frac{n^2}{c^2}\frac{\partial^2E}{\partial t^2}=\frac{1}{\varepsilon_0 c^2}\frac{\partial^2P_{NL}}{\partial t^2}.
\end{equation}
$n$ is the linear part of the refractive index.\\
For a $\chi^{(3)}$ non-linear medium we have $P_{NL}=\varepsilon_0\chi^{(3)}E^3$.
We assume the electric field to be monochromatic (at frequency $\omega_0$) and to propagate along $z$. 
We introduce  $k_0$, the wavenumber of $E$, and we decompose the envelope $E_0$ from the oscillating part:
\begin{equation}
    E(x,y,z,t)=E_0(x,y,z) e^{i(k_0 z-\omega_0 t)}.
\end{equation}
We can inject this expression of $E$ into Eq. \ref{eq:HelmholtzNL} to get
\begin{equation}\label{440}
    \nabla^2_{\perp}E_0+\frac{\partial^2E_0}{\partial z^2}+2ik_0\frac{\partial E_0}{\partial z}-\left(k_0^2-n^2\frac{\omega_0^2}{c^2}\right)E_0=-\chi^{(3)}|E_0|^2\frac{\omega_0^2}{c^2}E_0,
\end{equation}
where $\nabla^2_{\perp}$ is the Laplacian in the transverse ($x,y$) plane. 
We then take into account that $E_0$ is slowly varying along $z$, known as the paraxial approximation, to eliminate the second derivative of $E_0$ along $z$. 
We can rewrite Eq. \ref{440} with this simplification:
\begin{equation}\label{x123}
    \nabla^2_{\perp}E_0+2ik_0\frac{\partial E_0}{\partial z}-\left(k_0^2-n^2\frac{\omega_0^2}{c^2}\right)E_0+\chi^{(3)}|E_0|^2\frac{\omega_0^2}{c^2}E_0=0.
\end{equation}
We can decompose the linear part of the refractive index as $n=n_0+\delta n(\mathbf{r})$ with a mean index plus a local perturbation (supposed weak).
As $k_0= n_0\omega_0/c$, we can rewrite Eq. \ref{x123} as:
\begin{equation}\label{eq:nlo1}
    i\frac{\partial E_0}{\partial z}=-\frac{1}{2k_0}\nabla^2_{\perp}E_0- \frac{\delta n k_0}{n_0}E_0-\frac{k_0\chi^{(3)}}{2n_0^2}|E_0|^2E_0.
\end{equation}
This equation is analogue to the Gross-Pitaevskii equation (Eq. \ref{GPE1}). We will therefore be able to use the same formalism, and in particular we can expect a Bogoliubov dispersion relation for the small perturbation on the electric field.
We introduce the light intensity $I$ and the non-linear index $n_2$.
We have:
\begin{equation}
    n_2=\frac{3\chi^{(3)}}{2n_0^2\varepsilon_0c}\text{ and } I=|E_0|^2  n_0\varepsilon_0 c/2.
\end{equation}
This allows to write an important quantity experimentally: the non-linear contribution to the index of refraction: $\Delta n=n_2I.$\\


To precise the analogy with the Gross-Pitaevskii equation (GPE), I have rewritten Eq. \ref{GPE1} and compared it to Eq. \ref{x123}.
\begin{eqnarray}
    \nonumber
    i \hbar \frac{\partial \psi(\mathbf{r},t)}{\partial t} &=& \left( -\frac{\hbar^2}{2m } \nabla^2 \ +\hspace{0.3cm} V(\mathbf{r})\hspace{0.3cm} +\  g\ \vert  \psi(\mathbf{r},t) \vert^2 \right) \psi(\mathbf{r},t)\\
    \nonumber  i\frac{\partial E_0(\mathbf{r_{\perp}},z)}{\partial z}&=&\left(-\frac{1}{2k_0}\nabla^2_{\perp}- \frac{\delta n(\mathbf{r_{\perp}}) k_0}{n_0}-\frac{k_0\chi^{(3)}}{2n_0^2}|E_0|^2\right)E_0(\mathbf{r_{\perp}},z).
\end{eqnarray}

The first remark is that the left term in GPE is a time derivative as it is a spatial derivative along the propagation direction for  Eq. \ref{x123}.
The dynamics in  Eq. \ref{x123} is then 2D in the transverse plane written $\mathbf{r_{\perp}}=(x,y)$.
Each slice at a given $z$ is equivalent to a time-snapshot in the GPE.
More precisely we can fix the initial state by designing a given field at the input of the medium $z=0$ and measure the evolution after an effective time by imaging the intensity and phase at the output of the medium $z=L$.

We can now propose a mapping of the different quantity to directly write the expected Bogoliubov dispersion for small perturbation in Eq. \ref{x123}.
\begin{eqnarray}
\nonumber \text{GPE} &\leftrightarrow& \text{Non-linear Optics}\\
    \nonumber  \psi(\mathbf{r},t) &\leftrightarrow& E_0(\mathbf{r_{\perp}},z) \\
    \nonumber m/\hbar &\leftrightarrow&  k_0\\
    \nonumber V(\mathbf{r}) /\hbar &\leftrightarrow& -\delta n(\mathbf{r_{\perp}}) k_0/n_0 \\
    \nonumber g/\hbar&\leftrightarrow& - \frac{k_0\chi^{(3)}}{2n_0^2}
\end{eqnarray}
We also need to \textit{translate} the frequency $\omega$ and the wavenumber $q$ of the Bogoliubov excitations. 
As time is mapped to $z$, $\omega$ is mapped to a spatial frequency in the propagation direction. It is denoted $\Omega_B$ to avoid confusion. We have:
\begin{eqnarray}
\nonumber \mathbf{q} &\leftrightarrow& \mathbf{k_{\perp}}\\
    \nonumber  \omega( \mathbf{q})&\leftrightarrow& \Omega_B(\mathbf{k_{\perp}})
\end{eqnarray}
We can now rewrite the Bogoliubov dispersion in the Non-linear Optics analogy:
\begin{equation}\label{bogo2}
    \epsilon(\kperp)=  \Omega_B(\kperp)=\sqrt{\frac{\kperp^2}{2k_0}}\cdot\sqrt{  \frac{\kperp^2}{2k_0}-\frac{k_0\chi^{(3)}}{n_0^2}|E_0|^2}.
\end{equation}
Similarly to Bose gases, we can define two limits depending on the value of the wavevector $\kperp$ compared to the inverse of the healing length $\xi$ defined as:
\begin{equation}
    \xi=-\frac{1}{k_0}\sqrt{\frac{2n_0^2}{\chi^{(3)}|E_0|^2}}.
\end{equation}
For $\kperp\gg 1/\xi$, we have a parabolic dispersion. For $\kperp\ll 1/\xi$ we observe a sonic dispersion with a speed of sound $c_s$ defined as:
\begin{equation}
    c_s=\sqrt{-\frac{\chi^{(3)}|E_0|^2}{2n_0^2}}.
\end{equation}

\subsubsection{Measure of the Bogoliubov dispersion}
So far, I have shown theoretically that we can consider the propagation in a $\chi^{(3)}$ analogue to the dynamic of a Bose gas in 2D \footnote{A fully quantum theoretical framework for fluids of light in the same configuration has been recently proposed \cite{larre2015propagation}.}.
We have recently pushed forward this effort with the experimental verification of this proposal.
Using the vocabulary of Bose gases,  in this context, offers modern perspectives on non--linear and quantum optics experiments.\\

To date, experimental implementations are still in their infancy with experiments in photo--refractive crystals \cite{wan2007dispersive,wan2010diffraction,khamis2008nonlinear}, and thermo-optic non--linear medium \cite{vocke2015experimental}.
In a non-local $\chi^{(3)}$  thermo-optic medium (methanol), the Bogoliubov dispersion of a fluid of light has been studied  \cite{vocke2015experimental} and indication of superfluid behaviour has been observed  \cite{vocke2016role}. 
Similarly, an iodine solution has been used to create an event horizon in light fluid acoustic black hole analogue \cite{elazar2012all}.
Even though these results highlight the potential of the quantum fluid interpretation, the  non-locality of these media (thermo-optic non-linearity is affected by heat diffusion) is detrimental for simulating system with local interactions.
By using warm atomic vapors as $\chi^{(3)}$ medium we have access to tunable and local interactions.  
Interestingly, a few works on the elimination of diffraction in a slow light atomic medium \cite{firstenberg2009elimination,firstenberg2009elimination2}, are using similar ideas without explicitly connecting to  fluid of light.\\

In our implementation, we use the following protocol:
\begin{itemize}
    \item We create a uniform background fluid with a pump laser propagating in a near resonance atomic vapor.
    \item We add a small perturbation by interfering a localize probe beam at the same frequency at an angle with respect to the background wavevector.
    \item We study the propagation of this perturbation by measuring the distance propagated in the transverse plane by these narrow wavepackets at the end of the medium.
    \item From this measurement of the group velocity, we simply integrate with respect to the wavevector and obtain the dispersion relation.
\end{itemize}
We have obtained a dispersion fitting a Bogoliubov spectrum confirming the theoretical prediction.
For more details the results are presented in the article attached.
I just add two comments (please read the paper first): one to provide an intuitive explanation of this surprising effect and the second to connect this effect to four-wave-mixing.\\

\begin{figure}[h!]
    \centering
    \includegraphics[width=1\textwidth]{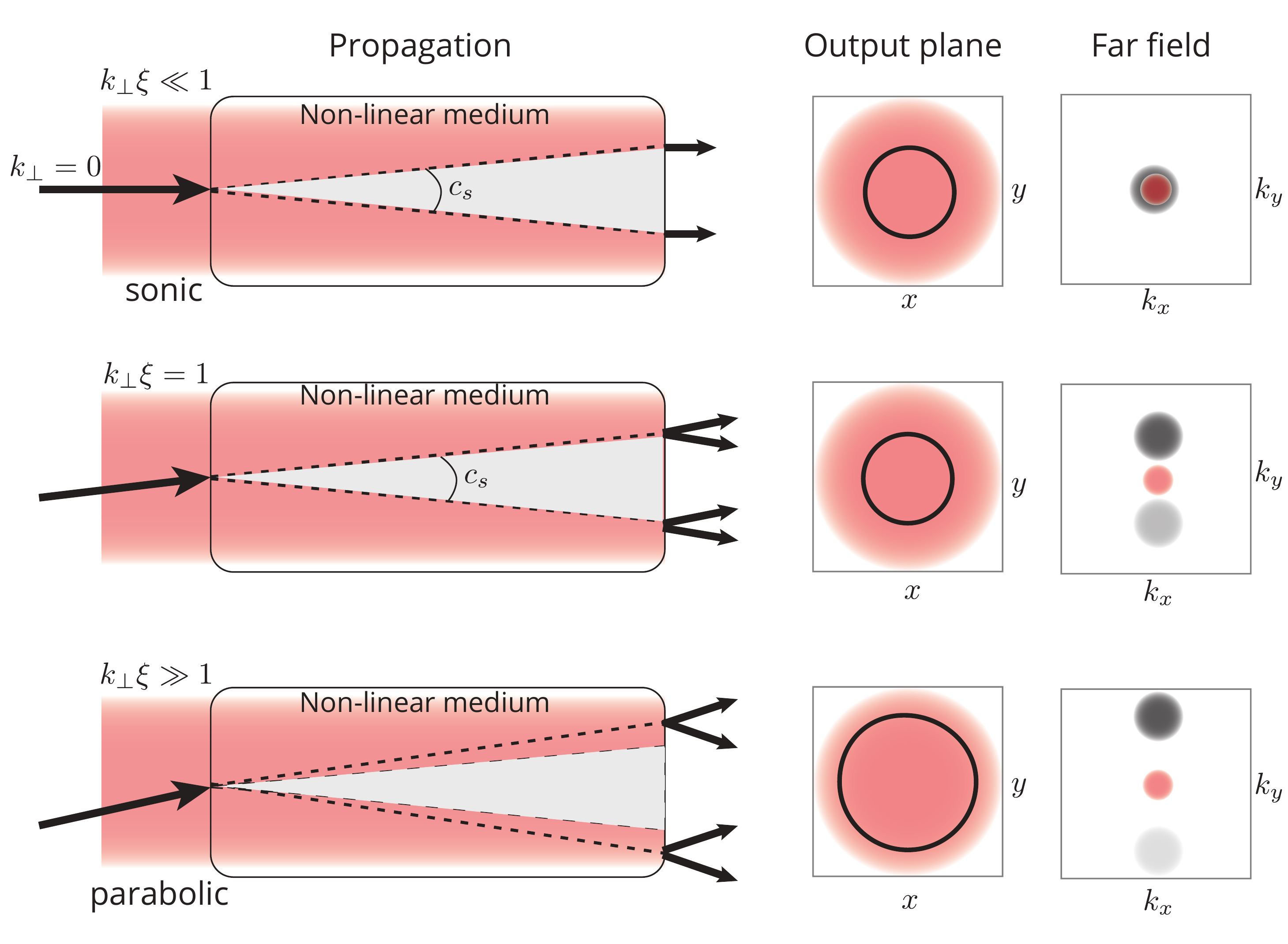}
    \caption{I represent here three limit cases for the propagation of a weak probe in a non-linear medium. 
    On the first line I draw the case $\kperp=0$. 
    We can see that even for $\kperp=0$, the Bogoliubov excitations (in dashed lines inside the medium) propagate at a finite speed $c_s$. 
    If we increase the incident angle ($\kperp$) while staying in the limit $\kperp<1/\xi$ the ring in the output plane does not change as the Bogoliubov modes always propagate
    at $c_s$ (shown on the second line).
    If we break this condition $\kperp\gg1/\xi$, Bogoliubov excitations behave like free particles and therefore an increasing $\kperp$ results in an increasing diameter of the ring (shown on the bottom line)
     At the output of the medium two modes are emitted to the far field at $\kperp$ and $-\kperp$. 
    These modes are usually called probe and conjugate in the language of four-wave-mixing.
    Due to phase matching condition, increasing $\kperp$ lead to a reduction of the power in the  $-\kperp$ mode. }
    \label{f46}
\end{figure}

\noindent\textbf{Anomalous refraction}\hspace{0.5cm}
In our experiment \cite{fontaine} the refraction law is indeed strongly modified compared to the linear case of Snell law (See figure \ref{f46}).
It is well known that imaging inside a non-linear medium is not a recommended task for this reason. What I want to show here, is that the quantum fluid language provide an intuitive vision of this phenomenon.
For small angles ($\kperp<\xi$), the probe beam light emerges always at the same place in the output plan independently of the angle.
This is strange because changing the incident angle should normally change the refracted angle.
This invariance is understood in the fluid context as a simple consequence of the constant group velocity or similarly of the linear dispersion.
Moreover, the probe position in the output plane depends on the pump intensity. 
Again, if one thinks in terms of fluids, it modifies the speed of sound and therefore the position in the output plane.

For larger angle  ($\kperp\gg\xi$), we found again a more usual behaviour with the position of the beam at the output directly linked to the incident angle.
We are in the limit of parabolic dispersion, which means that the group velocity is proportional to $\kperp$ and therefore the position in the output plane ($z=L$) is $\kperp L$. We can switch back to a simple geometric explanation as long as we stay in the paraxial approximation.
Coming with this explanation using only the non-linear optics phenomenon is not possible, especially the means to estimate the angle where the two regimes changes ($\xi$).
This is one of the strength of fluid of light language: to be able to bring new tools to explain non-linear and quantum optics experiment.\\

\noindent\textbf{Four-wave-mixing}\hspace{0.5cm}
This experiment is "nothing else than" degenerate four-wave-mixing.
One takes two photons from the pump and transfers one to the probe and one to the conjugate.
However, we need to be careful when using this vocabulary.
What the fluid of light language taught us, is that inside the medium we should not talk about probe and conjugate but more precisely about two Bogoliubov modes $u_q$ and $v_q$ propagating at $+\mathbf{q}$ and $-\mathbf{q}$ (using the notation of \ref{uvq}).
The most striking case is when $\kperp=0$. In this situation of colinear four-wave-mixing, probe and conjugate are at  $\kperp=0$, however we still have two Bogoliubov modes at $+\mathbf{q}$ and $-\mathbf{q}$ inside the medium as demonstrated by imaging the output (see figure \ref{f46}).

Probe and conjugate, as commonly use in four-wave-mixing, are actually far field denomination only.
What I present in the  figure \ref{f46}, is the incident probe beam, which splits in pairs of mode at  $+\mathbf{q}$ and $-\mathbf{q}$ inside the medium forming a ring.
After the medium, all these modes will combine and contribute to create a probe and a conjugate at $+\kperp$ and $-\kperp$ in the far field.
There is no reason, in general, for $\kperp$ and $\mathbf{q}$ to be the same.
When, we inject at mode at $\kperp\neq 0$, we break the symmetry between  for the ring of Bogoliubov modes. 
In the far field we observed a difference in intensity between probe and conjugate which is known in non-linear optics as phase matching.
We note that if the probe and pump have different frequency the phase matching condition would not be at $k_{\perp}=0$ anymore as in chapter 3.\\

An important prediction from the Bose gases physics is that the creation of Bogoliubov modes results from a quench: a sudden (non-adiabatic) change of one parameter of the system.
Here, the interactions appears suddenly at the medium input.
What this prediction tells us is that four-wave-mixing should not occur if we were able to ramp adiabatically the interactions.
This interesting (and surprising) prediction remains to be verified experimentally.
\clearpage

\clearpage

\chapter*{Article 6: Observation of the Bogoliubov dispersion in a fluid of light}

\section*{Preprint}
\noindent The preprint version of this article is accessible on \href{https://arxiv.org/abs/1807.10242}{arXiv:1807.10242}.

\section*{Published version}

  \href{https://journals.aps.org/prl/accepted/dd076Yd5H9214c7c92b385f9f1b8d57b2c746dd87 }{\textbf{Observation of the Bogoliubov dispersion in a fluid of light}}.  \\
   \begingroup
    \fontsize{9.5pt}{9.5pt}\selectfont
    Q. Fontaine,  T. Bienaim\'e,  S. Pigeon, E. Giacobino,  A. Bramati,   \textbf{Q. Glorieux}.\\
\endgroup
   \href{https://journals.aps.org/prl/accepted/dd076Yd5H9214c7c92b385f9f1b8d57b2c746dd87 }{{\it Phys. Rev. Lett.}},  \textbf{121}, 183604 , (2018).

\addcontentsline{toc}{section}{\textbf{Article 6}: Observation of the Bogoliubov dispersion  in a fluid of light.  PRL \textbf{121}, 183604 , (2018) }
\ifpapers
\fi


\chapter{Outlooks and future projects}\label{outlooks}
\vspace{-0.5cm}
In this work we have travelled through several implementations of light-matter interaction experiments from the generation of squeezed light in four-wave-mixing to the storage of light in an atomic vapor.
The future projects I want to develop are along the line of combining quantum optics and quantum memory with the concept of fluids of light.
Indeed,the experimental realizations of fluid of light presented in this manuscript are based on the Bogoliubov mean-field theory. 
An important leap forward in this field will consist in bridging the gap with quantum optics type measurements such as intensity noise, homodyne detection or  entanglement.
In this chapter I give a (partial) overview of the essential ideas we will explore in the coming years.

\section{Fluid of light in the propagating geometry}
\subsection[Shockwaves dynamics]{Shockwaves dynamics in 1D and 2D\\Collaboration with N. Pavloff and A. Kamchatnov}
\begin{figure}[]
    \centering
    \includegraphics[width=\textwidth]{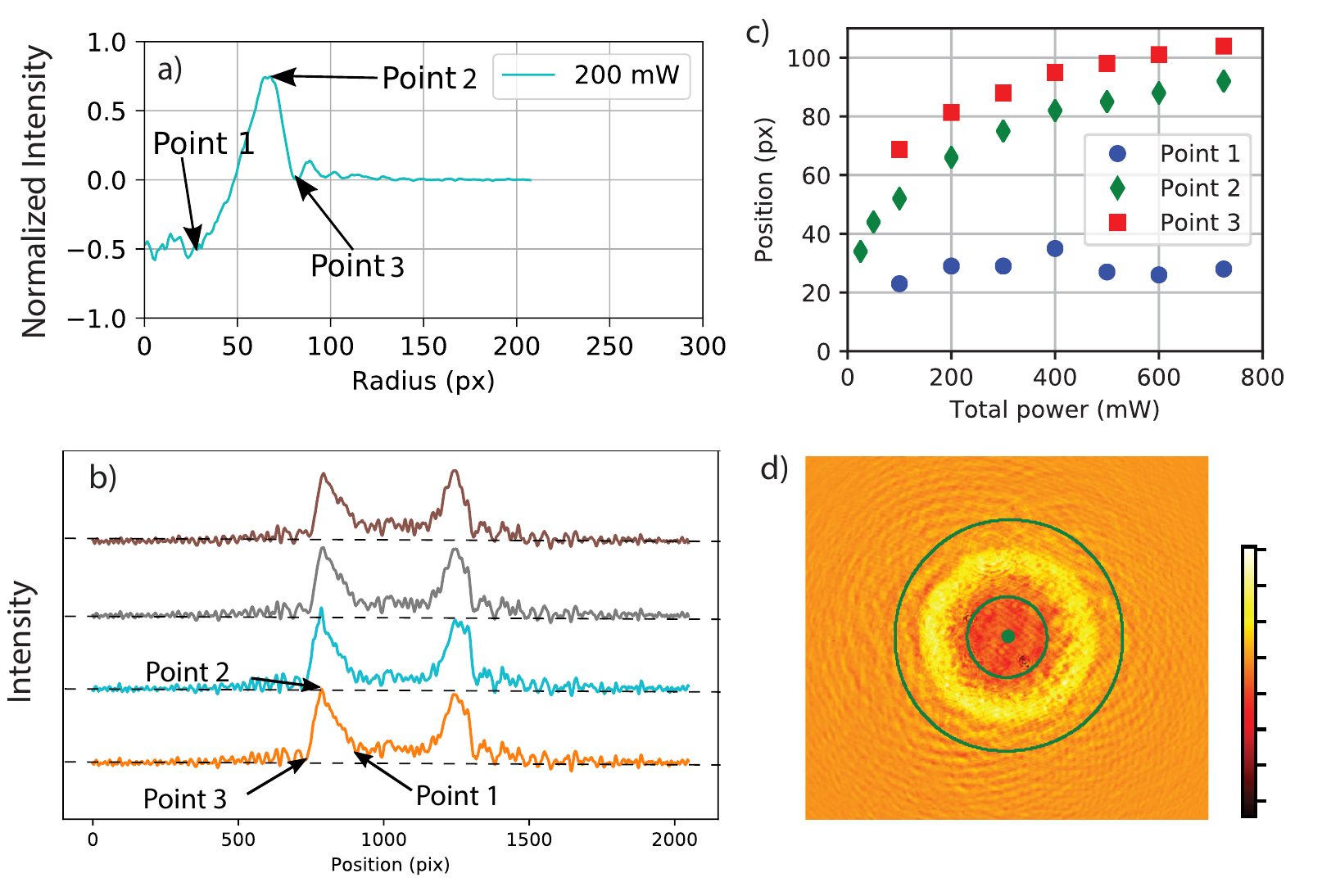}
    \caption{Preliminary results on shockwaves. a) 2D shock profile. Intensity is normalized to 0 far away from the shock. The negative value in the center is characteristic of 2D shockwaves. b) 1D configuration for various intensity. c) Scaling of points 1,2 and 3 as function of power. d) Typical intensity image.}
    \label{fig:shock}
\end{figure}

Before exploring the quantum world with fluids of light, some highly interesting experiments are still to be conducted in the classical regime.
One of them is focused on the generation of shockwaves in a fluid of light.
Shockwaves appear when the Bogoliubov approximation of a small perturbation propagating in a fluid breaks down.
Pioneering work about shockwaves in optics has been done by the Fleischer's group in photo-refractive crystal. 
Recently, we have started a collaboration with the group of Nicolas Pavloff in LPTMS to explore this field in detail and to understand the role of the geometry (1D and 2D) in shockwaves propagation.\\
The setup is identical to the setup used for Bogoliubov dispersion in chapter 4, however in this case, the magnitude of the perturbation is as large as the mean density.
We have studied the dynamic of the shock propagation for 3 characteristic points of the  shockwave (see Fig. \ref{fig:shock}):
\begin{itemize}
\item Point 1: the beginning of the linear slope
\item Point 2: the maximum
\item Point 3: the first minimum of the shockwave oscillations
\end{itemize}
The work in progress we are conducting with Nicolas Pavloff is to find theoretically a universal exponent for the propagation velocity of each of these three points and to verify it experimentally.
It is understood that these exponents should vary strongly depending on the dimensionality of the system.
Our implementation is intrinsically 2D (in the transverse plane) however, we can also explore 1D physics by making the system much longer in one direction (virtually infinite) and therefore invariant along this dimension.\\

A major difference between 1D and 2D for shockwaves consists in the prediction of the density in the center after the shock front has passed. 
In 1D the density of the fluid (here the intensity of light) should go back to average fluid density immediately after the shock front has passed, while for 2D (and 3D) the density should remain lower than the average density.
This effect is well known in fluid dynamics and has an immediate consequence when for example an explosion occurs inside a building. Indeed, when shockwaves due to the explosion reach a window, glass breaks and falls inside the building and not outside, a consequence of the lower density inside following the shock front.
We have reproduced this effect with light and observed a reduction of the density after the shock in 2D and no reduction in 1D as seen in figure \ref{fig:shock}a.

\subsection[Superfluid flow around a defect]{Superfluid flow around a defect \\
Collaboration with C. Michel and M. Bellec}

\begin{figure}[h!]
    \centering
    \includegraphics[width=0.7\textwidth]{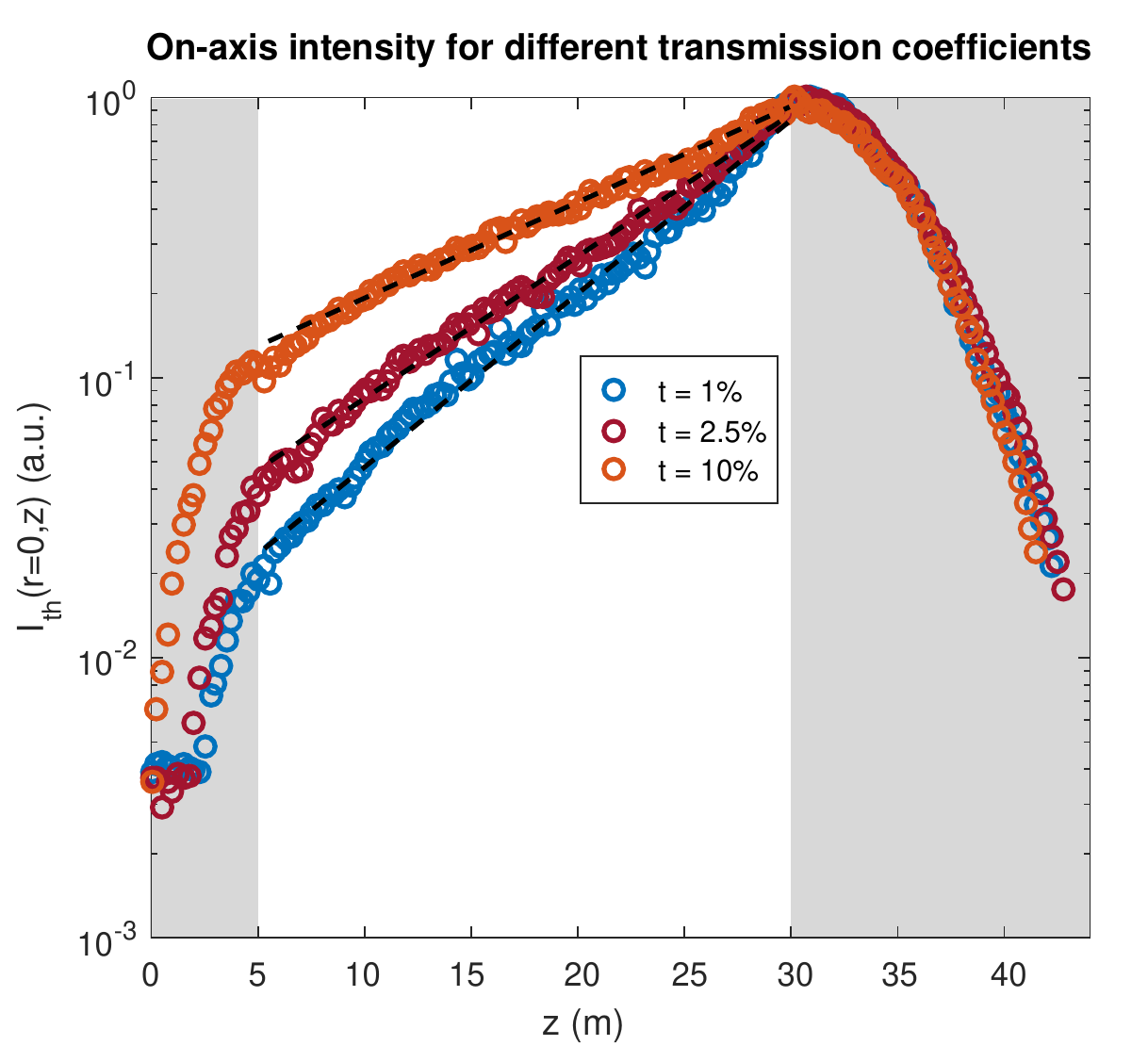}
    \caption{On axis intensity of a Bessel beam shaped to have an exponential increase along the propagation. Various attenuation coefficients are used. The dashed lines are the fits with linear slopes in log-scale. The region where the intensity is designed to follow an exponential increase is shown in white. }
    \label{fig:bessel}
\end{figure}

We have recently observed the Bogoliubov dispersion for light propagating into an atomic medium.
The linear part of this dispersion is, following Landau criteria, an indication of possible superfluid flow.
The smoking gun of superfluidity is the observation of  Rayleigh scattering cancellation when the fluid hits a defect as seen in figure \ref{fig:amosuperfluid} for a polariton fluid.
To settle our platform, based on light propagating through atomic vapor, as a realistic alternative to the study of superfluidity, we need to perform a similar experiment \cite{Bellec_Defect}.\\

The defect here should be a refractive index change around the mean value as described in Eq. \ref{eq:nlo1}. As we want the defect to be constant in (effective) time, one needs to make it constant in the (real) space along the propagation direction.
Local modifications of the refractive index can be created optically by designing spatially dependant optical pumping.
Experimentally, this requires an external defect beam, tuned near resonance, which propagates with constant shape and amplitude through the atomic medium.
This seems a difficult task, but works in progress in my group have shown this to be achievable.
Two steps have to be considered.

The first step is the design of a non-diffracting beam. Bessel beams can be created using an optical axicon or with more control using a spatial light modulator.
A distance of 10~cm for a beam size of 50~$\mu$m is achievable at 800~nm, to be compared to a Rayleigh length of 9~mm for a Gaussian beam of the same size.

The second stage is the control of the amplitude along the propagation direction to compensate the absorption by the atomic medium, which will be large as the defect beam is set near resonance for maximal efficiency.
The intensity of the beam follows a Beer law along the medium, so we must design a beam with an exponentially rising intensity inside the first maxima of the Bessel function.
This is done using the SLM and an approach similar to \cite{vcivzmar2008generation}.
I present in figure \ref{fig:bessel} the preliminary results of this technique.\\

In the near future, we will study the effect of scattering on this optically designed defect in an experiment similar to \cite{Bellec_Defect,amo2009superfluidity}.
A strong asset of our system is that the defect can be of any arbitrary shape, size and depth by simple optical control.
This opens the way, for example, to the study of propagation in a disorder media in the presence of interactions, by sending a non-diffracting speckle pattern as defect.

\subsection[Optomechanical signature of superfluidity]{Optomechanical signature of superfluidity \\
Collaboration with P.E. Larr\'e and I. Carusotto}

\begin{figure}[h]
    \centering
    \includegraphics[width=\textwidth]{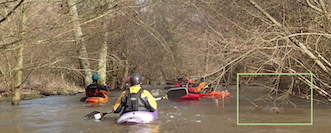}
    \caption{Kayaking in a normal fluid. Kayaking in a superfluid would be a tricky task as the paddle will not help the boat to move as it will not feel any drag force.}
    \label{fig:kayak}
\end{figure}

An exciting prospect of the experiments described in the previous section is to replace the optically induced defect by a real object and observe the optomechanical signature of superfluidity.
Let us start by the analogy with a matter fluid.
Imagine a flexible tree branch touching a river as highlighted in green in figure \ref{fig:kayak}. 
When  flowing water  hits the branch, it  moves it in the flow direction by applying a drag force on it.
But with a superfluid the drag force vanishes and the branch comes back to its initial position.
We want to produce an analogous experiment with light.

A defect can be created by any dielectric object with a refractive index different from the medium. 
We propose to use a tapered optical nanofiber in glass as a defect.
The setup consists in the introduction of a nanofiber inside a rubidium vapor cell, along the cell direction.
Sending near resonant light in the cell with a small angle with respect to the nanofiber will slightly move it thanks to the radiation pressure (or the drag force in the hydrodynamics language).
By increasing the light intensity (while staying below superfluidity threshold) one can increase the radiation pressure and in consequence the displacement of the nanofiber should be larger.
However, when the light becomes superfluid the drag force no longer exists and the nanofiber should come back to its initial position as in absence of light.
This effect has been predicted in Ref. \cite{larre2015optomechanical}. 
The interpretation in non-linear optics language is more convoluted as it means that radiation pressure is cancelled by non-linearities above a certain threshold. Once again the hydrodynamic language provides us with an intuitive interpretation of a novel phenomena that is virtually impossible to predict using non-linear optics intuition.

In the implementation I propose here, we have a crucial tool to achieve this
challenging experiment: we know how to precisely monitor the displacement of a nanofiber with a resolution below 2~nm.
To do so, we use the fact that a nanoparticle deposited on a nanofiber scatters light inside the guided mode of the fiber.
By placing a nanofiber within a standing wave (in the transverse plane), and monitoring the intensity of the scattered light, we can monitor fiber displacement with a resolution of 2~nm, well below diffraction limit.

For us, the next experimental step is to combine the nanofiber and the non-linear medium into one experiment and probe the optomechanical signature of superfluidity.
        
\begin{figure}[h!]
    \centering
    \includegraphics[width=0.8\textwidth]{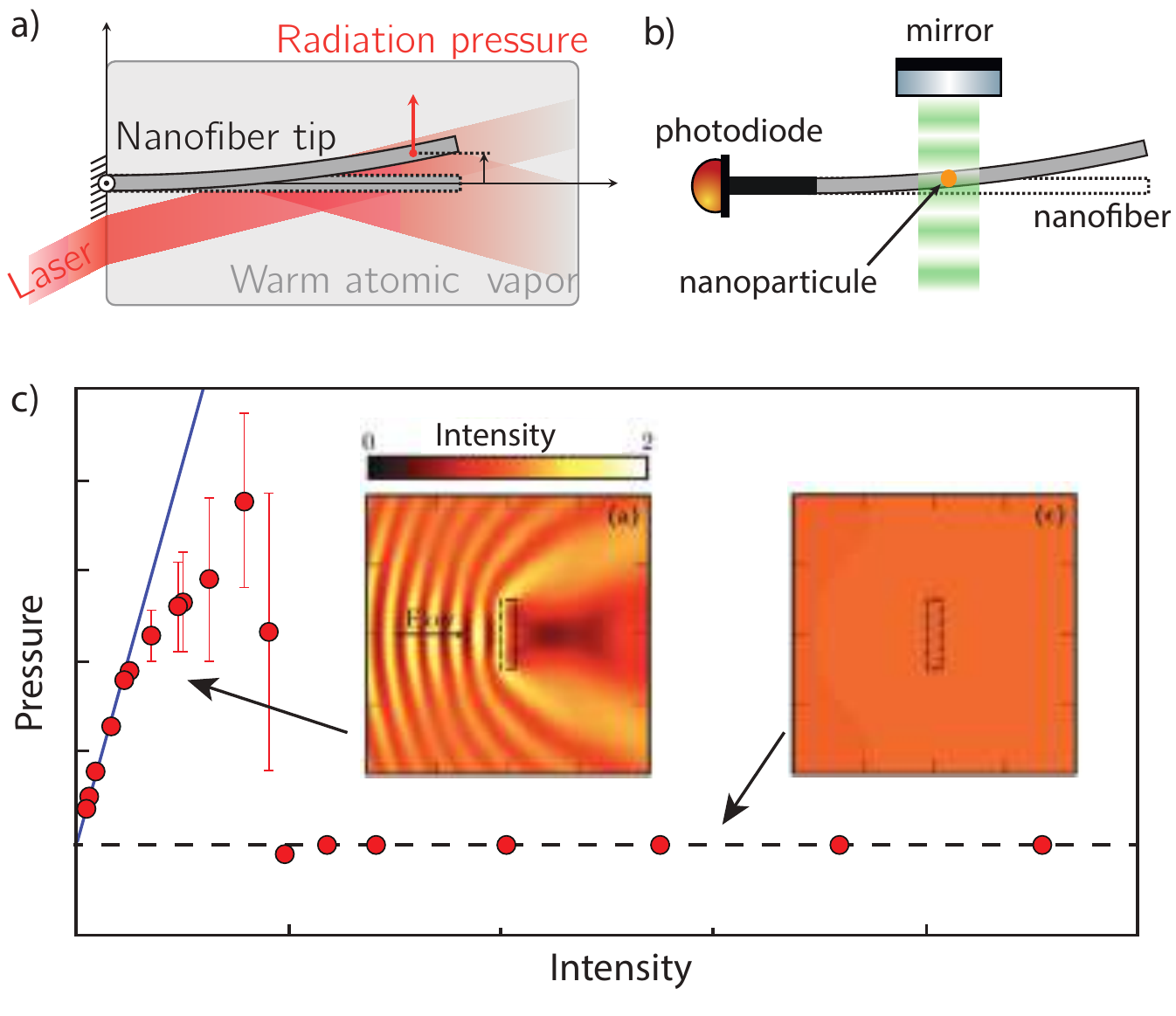}
    \caption{Optomechanical signature of superfluidity. a) Proposed implementation. b) Displacement measurement setup. The standing wave (green laser) is scattered inside the guided mode of the fiber by a nanoparticle. c) Pressure force as function of intensity. At the threshold of superfluidity the pressure force becomes zero and the frag force vanishes. Insets are numerical simulations of normal fluid and superfluids regimes. Adapted from \cite{larre2015optomechanical}}
    \label{fig:larre}
\end{figure}

\subsection[Analogue gravity]{Analogue gravity\\
Collaboration with D. Faccio}
A vast topic that fluids of light can help to investigate is analogue gravity, i.e. how to mimic gravitational effects in a non-linear optics experiment.
The underlying idea behind hydrodynamical model for gravity has been introduced by Unruh.
In 1981, he showed that sound waves in an accelerated flowing medium mimic the space-time geometry of a black hole \cite{unruh1981experimental}.
As a consequence it is possible to look at quantum effects in the vicinity of the horizon and for example study the analogue of Hawking radiation, in a non-linear quantum system.

We can get an intuition about this effect using once again the analogy with water.
Let us imagine a river flowing towards a few meters waterfall with a faster and faster flow speed.
A fish swimming in the river (with a maximum velocity) will be irrevocably sent through the waterfall if he crosses the point where water flow speed becomes larger than its maximum velocity.
This point is analogous to a black hole horizon (for fish).
It is also true for (density) waves propagating on the river at the speed of sound. 
The black hole becomes a dumb hole (for sound waves instead of light).\\

Another familiar implementation is the hydraulic jump which appears in a sink.
When water flows from the tap (at a given flow rate) and hits the bottom of the sink a circular region with laminar flow appears in the center and after an hydraulic jump one can observe turbulent flow.
Because the flow rate is constant, water velocity is much faster in the center than at larger radius.
In consequence, no waves can penetrate inside the central region because, there, the sound speed is slower than the water velocity.
A region of space where no wave can enter is the time reversal opposite of a black hole also known as a white hole.\\

In optics, a similar idea can be implemented using the Madelung transformations.
The density of the fluid is analogous to the intensity and therefore the speed of sound scales with the square root of the intensity.
A sudden change in intensity in the transverse plane is then suitable to create the analogue of a black hole horizon \cite{nguyen2015acoustic}.
Interesting theoretical proposals \cite{larre2012quantum} investigate this configuration to observe the optical analogue of the Hawking radiation, and a potential implementation in our platform can be envisioned.\\

Another striking effect is known as the Zel'dovich effect i.e. the amplification of radiation scattering off rotating absorbing surfaces.
The first demonstration of this effect dates back to 1971 when Zel'dovich showed that electromagnetic waves can be amplified when reflected on a rotating conducting object \cite{zel1971generation}.
This effect has been extended to cosmology by Misner who showed, in 1972, that this reflection can also occur near rotating Kerr black holes \cite{fackerell1972weak}.
Observation of this type of super-radiance in hydrodynamics has also been possible and is known as \textit{over-reflection effect} \cite{richartz2015rotating}.
Surprisingly, no experiment in optics have studied this effect so far.
Our platform can be suitable for the observation of this effect in the optical domain. Indeed, analogue of rotating black holes can be created by injecting a fluid with an orbital angular momentum using a Laguerre-Gauss beam.
The general topic of analogue gravity (and even analogue physics) is a good match with our approach which aims to find novel effects in non-linear and quantum optics through the perspective of other physical phenomena.

\subsection[Disorder and interactions]{Disorder and interactions\\Collaboration with N. Cherroret}

Quantum fluids of light are characterized by long-range coherence. 
This not only favors superfluid behaviors but also localization phenomena in disordered potentials. 
In our platform, we can shape the spatial distribution $n_1(x, y, z)$ of the linear refractive index (i.e. the external potential) by applying a control field with the intensity profile modeled with a spatial light modulator.
In particular, a disorder potential can be generated by illuminating the atomic medium with a random speckle intensity pattern. 
We plan to investigate the interplay between localization features, which correspond to freezing the motion in the disordered landscape, and perfect transmission, corresponding to a super-flow regime. 
In particular, the competition between localization and superfluid transport is envisaged as a novel route for guiding light in a disordered non-linear environment. \\

\begin{figure}[h]
\center
\includegraphics[width=0.5\columnwidth]{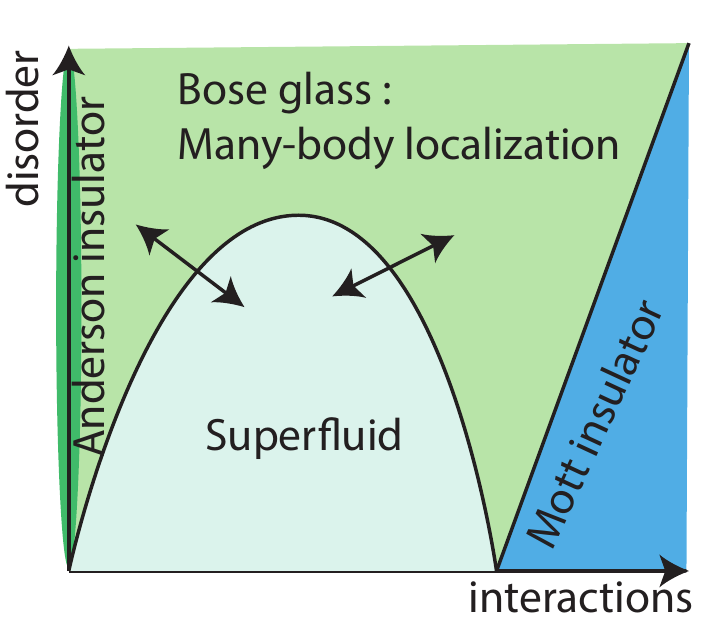}
\caption{Phase diagram of a 1D system at T=0 in presence of disorder and interactions. We propose to study the phase transitions marked with the black arrows. }
\label{MBL}
\end{figure}

At zero temperature, a 1D phase diagram can be calculated in the presence of interactions as shown schematically  in Figure \ref{MBL}.
While the non--interacting regime can exhibit an Anderson localization, slightly repulsive interactions between the photons tend, on the contrary, to suppress this effect. 
A transition in favor of superfluidity should then occur, although the critical threshold for light superfluidity is expected to be substantially affected by the disorder. 
I propose to study the phase transitions marked with black arrows in Figure \ref{MBL}, at the quantum level in 1D and 2D. 
This comes along the lines of the complex but very exciting problem of many-body localization (MBL) which extends the concept of localization beyond the mean field approximation.
Our system benefits of crucial assets to study this effect: a full quantum description is available \cite{Larre_Quench}, temperature, kinetic energy and interactions are tunable, disorder can be increased adiabatically and the dynamics of thermalization can be studied at various times and various spatial scales.

\clearpage
\section{Quantum simulation with photons}
I have tried in this manuscript to describe the tools which make the connection between quantum optics and quantum simulation with light.
Before concluding, I want to explain how quantum memories or four-wave-mixing configurations can be used in the context of quantum simulation.
Ultra-cold atoms in optical lattices are the current gold-standard to simulate Hamiltonian (conservative) dynamics.
Let me briefly discuss why photonics quantum simulators are an exciting direction to explore !

\subsection{From ultracold atoms to quantum fluids of light }
The design of an analog quantum simulator relies on three steps \cite{bloch2012quantum}. 
First, we should prepare an input state relevant for the physical problem of interest (either a well defined quantum state or an equilibrium state at non-zero temperature).
Next, we need to realize the proper mapping of the Hamiltonian to be simulated, including  both the single-particle physics (external potential, effective mass) and the interaction between the constituents.
Finally, measurements are performed on the output state to extract information on the simulated dynamics with the highest-possible precision. 
The cold-atoms community has achieved tremendous progresses in the recent years on these three tasks.
We will briefly review how they are implemented on ultra-cold atoms and optical platforms.\\

\vspace{-3mm}
\textbf{Initial state preparation}\hspace{0.5cm}
Preparing a state for quantum simulation implies to control the density distribution in the position or the momentum space, the energy or the temperature.
Because ultracold atoms are prepared in a trap, it is somehow complicated to control precisely the local density for example.
For light-based platforms, state preparation is more straightforward as it relies on optics. 
Cavity-setups are effective 2D systems in the transverse direction, and therefore, the density distribution in the position space is set by imaging a given intensity pattern onto a defined plane of the system (typically the cavity plane).
Momentum distribution relies on the angular distribution of the photons and can then be tuned in the Fourier plane.
Shifting away from zero, the mean of this momentum distribution injects a kinetic energy in the system and induces an effective flow in the transverse plane.  
On the other hand, the momentum distribution variance (the spread in momentum) defines an effective temperature of the photon fluid. 
For a variance of $1/l_\perp^2$, we can write the gaussian distribution of momentum exp$[{-k^2l_\perp^2}]$ and compare it to the thermodynamics expression exp$[{-\frac{\hbar^2k^2}{2m}/\frac{k_B T}{4 \pi}]}$ to identify $T\propto 1/l_\perp^2$.
For quantum fluids of light, thermalization occurs through the non-linear interaction mediated by the coupling with matter. 
An interesting feature of these setups is the possibility to isolate a subregion of the entire fluid by imaging it and therefore study the thermalization locally which is impossible with cold atoms.\\

\vspace{-3mm}
\textbf{Engineering the simulated Hamiltonians.}\hspace{0.5cm} 
An example of the Hamiltonian design with ultracold atoms is the simulation of condensed-matter systems of electrons moving on an array of atom cores.
This configuration can be mapped to interacting ultra-cold atoms in an optical lattice.
The interactions between atoms are tuned with the use of Feshbach resonances and the energy landscape at the single-particle level is controlled with an optical lattice, created using the interference pattern of far-detuned overlapping laser beams, thanks to the dipole force.
This precise control allows to simulate a quantum phase transition from a superfluid to a Mott insulator when the on-site interaction energy becomes much larger than the hoping energy due to tunneling between neighboring sites. 
For quantum fluids of light, the same level of control can be reached.
The effective photon-photon interaction is mediated by the non-linearity $\chi^{(3)}$ of the medium and the potential landscape is created by auxiliary laser beams.
\noindent Another example in the control of the Hamiltonian for an atomic quantum fluid is the development of artificial gauge fields, opening the way to the simulation of synthetic magnetic fields for neutral atoms, quantum Hall effect and topological phases.
 Similarly, fluids of light can also implement artificial gauge fields and synthetic magnetism \cite{otterbach2010effective,schine2016synthetic} by using for example photon orbital angular momentum \cite{westerberg2016synthetic}. \\

\vspace{-3mm}
\textbf{Detection of the final state}\hspace{0.5cm}
On the detection side, quantum optics techniques offer genuine advantages when working with a photonics platform compared to quantum simulations with ultra-cold gases. 
Typical ultra-cold atoms experiments use time-of-flight images after releasing the atoms from the trapping potential  to provide insight into the momentum distribution and coherence properties \cite{bloch2012quantum}. 
Only recently, the trapped density distribution of a quantum gas has become available to experimentalists, at the cost of a complex in situ imaging setup called a quantum gas microscope \cite{bakr2009quantum}.
With quantum fluids of light, detection turns out to be a much easier task, consisting in measuring the light exiting the system.
Similar to state preparation, we can access the momentum distribution by imaging the Fourier plane and the density distribution by imaging  the real plane. 
Moreover, coherence properties are available through interferometric techniques  ($g^{(1)}$ and  $g^{(2)}$ measurements) which are parts of the standard  quantum optics toolbox. 
This ease of preparation and detection is actually one of the main benefits of quantum fluids of light and one attractive aspect of photonic quantum simulation in respect to quantum simulation with ultra-cold atoms.

\subsection{Double-$\Lambda$ configuration}
The goal is to refine the EIT configuration to allow for more complex equations to be simulated.
By adding a forth level and a second coupling beam it is possible to do so.
I will not go into the details of the derivation, the interested reader can use \cite{angelakis2011luttinger}, but I will just mention the main result that we are planning to implement in our setup.

With two coupling beams (the EIT beam is noted $\Omega_c$ and the second coupling beam $\Omega_s$) the evolution equation for the probe Rabi frequency $\Omega_p$ is:
\begin{equation}
   - i\frac{\partial}{\partial z}\Omega_p=\left[\frac 12 \nabla_{\perp}^2+V(r)-G(r)|\Omega_p|^2\right] \Omega_p,
\end{equation}
with the potentiel given by 
\begin{equation}
    V(r)=-\frac{\kappa |\Omega_s|^2}{2\Delta (|\Omega_c|^2+|\Omega_s|^2)},
\end{equation}
and the non-linear coupling term:
\begin{equation}
   G(r)=-\frac{\kappa |\Omega_s|^2}{2\Delta (|\Omega_c|^2+|\Omega_s|^2)^2}.
\end{equation}
$\kappa$ is a constant given by $\kappa=N \mu^2/(\varepsilon_0\hbar \Gamma)$. 
Obviously this two coefficients are coupled, but this opens the way to control of potential and non-linearity with much more degrees of liberty that what we currently achieve with two-level atoms.

\subsection{Quantum optics}
Going \textit{quantum} is certainly a fundamental landmark in the field of photonics quantum simulators.
Because light in warm atomic vapor is a well controlled system which has already proved to be an excellent source of quantum correlated beams (see chapter \ref{chap:3} for example) I anticipate that it could be the first system to lay the groundwork for achieving valuable results .

\begin{figure}
    \centering
    \includegraphics[width=\textwidth]{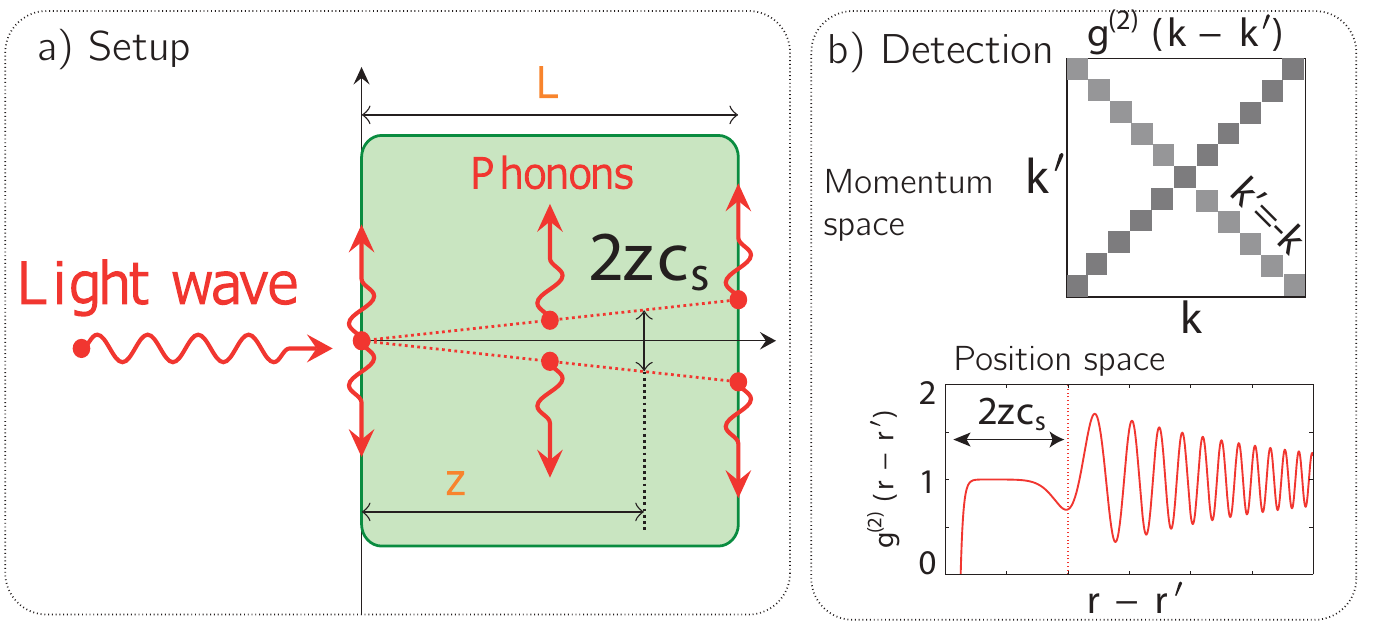}
    \caption{a) Schematic representation of the dynamical Casimir effect in a non-linear medium. Pairs of quantum correlated phonons (or intensity modulation) are generated at the input of the cell by the non-linearity quench. They propagate in the transverse direction at $c_s$ the speed of sound and can be detected by homodyne detection at the output plane. b) Expected correlations in the momentum space and position space. Adapted from \cite{Larre_Quench} }
    \label{fig:casimir}
\end{figure}

One step in this direction could be done by investigating the dynamical Casimir effect \cite{Larre_Quench}.
In the same experimental configuration as in section \ref{section:propagation}, but without the probe beam to seed an excited Bogoliubov mode, we have an interesting test-bed for quantum simulation.
Indeed, because of the quench of the interaction at the entrance of the medium, quantum correlated pairs of Bogoliubov modes are generated even in the absence of the probe beam analogous to the dynamical Casimir effect.
This is the same difference as between bright squeezing and squeeze vacuum as discussed in chapter \ref{chap:3}.
The main challenge is now to find a technique to measure the correlations presented in figure \ref{fig:casimir}.

This experiment is basically a four-wave-mixing setup with a detection scheme in the plane of the cell output and not in the far field.
Homodyne detection seems to be the suitable tool, however this means going to non-zero frequency to be realistic.
The step that remains to be done is therefore to include frequency dependance inside the evolution equation and see how the analogy with the non-linear Schr\"odinger equation holds.

\begin{figure}[h!]
    \centering
    \includegraphics[width=0.87\textwidth]{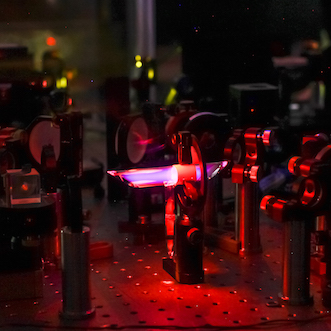}
    \caption{A nice picture of a rubidium cell to conclude...}
    \label{fig:rb}
\end{figure}

\clearpage
\section{Nano-optics}
An important part of my work since I joined the LKB in 2013 has been dedicated to the development of a novel experimental apparatus for nano-optics.
In writing this manuscript, I want to show the possible links that still remain to be tied between quantum optics and quantum memory on one hand and fluids of light and quantum simulation with light in the other.
Therefore nanooptics experiments did not fit well in this effort.
However, the interested reader can learn more about these works in the following references \cite{PRAjoos,Manceau18,Geng16,vezzo15,PhysRevB.90.035311}:

\begin{itemize}

\item  \href{https://doi.org/10.1103/PhysRevApplied.9.064035}{Polarization Control of Linear Dipole Radiation Using an Optical Nanofiber.}\\
   M. Joos, C. Ding, \ V. Loo, G. Blanquer, E. Giacobino, A. Bramati, V. Krachmalnicoff, \ \textbf{Q Glorieux}. \\
 \href{https://doi.org/10.1103/PhysRevApplied.9.064035}{Phys. Rev. Applied}, \textbf{9}, 064035 (2018).

\item 
\href{https://doi.org/10.1002/cphc.201800694}{CdSe/CdS dot--in--rods nanocrystals fast blinking dynamics.}\\
M. Manceau,  S. Vezzoli,  \textbf{Q Glorieux},  E. Giacobino  L. Carbone, M. De Vittorio  J-P. Hermier,  A. Bramati. \\
 \href{https://doi.org/10.1002/cphc.201800694}{ChemPhysChem}, Accepted. (2018).

\item \href{http://www.nature.com/articles/srep19721}{Localised excitation of a single photon source by a nanowaveguide.}\\
W. Geng, M. Manceau, N, Rahbany, V. Sallet, M. De Vittorio, L. Carbone, \textbf{Q. Glorieux}, A. Bramati, C. Couteau. \\
\href{http://www.nature.com/articles/srep19721}{Scientific Reports \textbf{6}, 19721} (2016).

\item \href{http://pubs.acs.org/doi/abs/10.1021/acsnano.5b01354}{Exciton Fine Structure of CdSe/CdS Nanocrystals Determined by Polarization Microscopy at Room Temperature.}\\
S. Vezzoli, M. Manceau, G. Lem\'{e}nager, \textbf{Q. Glorieux}, E. Giacobino, L. Carbone, M. De Vittorio, A. Bramati. \\ \href{http://pubs.acs.org/doi/abs/10.1021/acsnano.5b01354}{ACS Nano \textbf{9}, 7992} (2015).

\item \href{http://journals.aps.org/prb/abstract/10.1103/PhysRevB.90.035311}{Effect of charging on CdSe/CdS dot-in-rods single-photon emission.}\\
M. Manceau, S. Vezzoli, \textbf{Q. Glorieux}, F. Pisanello, E. Giacobino, L. Carbone, M. De Vittorio, A. Bramati.\\  \href{http://journals.aps.org/prb/abstract/10.1103/PhysRevB.90.035311}{Phys. Rev. B \textbf{90}, 035311} (2014).

\end{itemize}

\clearpage

\addcontentsline{toc}{chapter}{Bibliography}  
\bibliographystyle{IEEEtran} 
\bibliography{biblio} 


\end{document}